\documentclass[a4paper,10pt]{article}
\usepackage{abstract}
\usepackage{adjustbox}
\usepackage{algorithm}
\usepackage{amsfonts}
\usepackage{amsmath}
\usepackage{longtable}
\usepackage[algo2e]{algorithm2e}
\usepackage{bm}

\usepackage{booktabs}
\usepackage{breqn}
\usepackage{amssymb}
\usepackage{amsthm}
\usepackage[titletoc,toc,title]{appendix}
\usepackage{array}
\usepackage{authblk}
\DeclareMathOperator*{\argmin}{arg\,min}
\usepackage{bbm}
\usepackage{caption}
\usepackage{dsfont}
\usepackage{esvect}
\usepackage{xcolor}
\usepackage[round]{natbib}
\usepackage{animate}
\usepackage{float}
\usepackage{setspace}
\usepackage{geometry}
\usepackage{graphicx}
\usepackage{layout}
\usepackage{multirow}
\usepackage{neuralnetwork}
\usepackage{rotating}
\usepackage{leftidx}
\usepackage{enumerate}
\newcommand{\verteq}{\rotatebox{90}{$\,=$}}
\usepackage{subcaption}

\usepackage{tikz}
\usetikzlibrary{automata,fit,calc,shapes,shadows,arrows,positioning,decorations.markings,decorations.pathreplacing,matrix,shapes,arrows,positioning,chains,shapes.multipart}
\usetikzlibrary{shapes.multipart}

\DeclareMathOperator*{\argmax}{arg\,max}
\usepackage{pgfplots}

\def\cdf(#1)(#2)(#3){0.5*(1+(erf((#1-#2)/(#3*sqrt(2)))))}%

\tikzset{
declare function={
normcdf(\x,\m,\s)=1/(1 + exp(-0.07056*((\x-\m)/\s)^3 - 1.5976*(\x-\m)/\s));
}
}

\pgfmathdeclarefunction{gauss}{2}{%
\pgfmathparse{1/(#2*sqrt(2*pi))*exp(-((x-#1)^2)/(2*#2^2))}%
}

\tikzset{
partial ellipse/.style args={#1:#2:#3}{
insert path={+ (#1:#3) arc (#1:#2:#3)}
}
}

\makeatletter
\providecommand\phantomcaption{\caption@refstepcounter\@captype}
\makeatother

\makeatletter
\def\namedlabel#1#2{\begingroup
#2%
\def\@currentlabel{#2}%
\phantomsection\label{#1}\endgroup
}

\makeatother

\tikzset{
->-/.style={decoration={
markings,\theoremstyle{plain}
\newlist{casenv}{enumerate}{4}
\setlist[casenv]{leftmargin=*,align=left,widest={iiii}}
\setlist[casenv,1]{label={{\itshape\ \casename} \arabic*.},ref=\arabic*}
\setlist[casenv,2]{label={{\itshape\ \casename} \roman*.},ref=\roman*}
\setlist[casenv,3]{label={{\itshape\ \casename\ \alph*.}},ref=\alph*}
\setlist[casenv,4]{label={{\itshape\ \casename} \arabic*.},ref=\arabic*}
mark=at position .5 with {\arrow{>}}},postaction={decorate}},
-<-/.style={decoration={
markings,
mark=at position .5 with {\arrow{<}}},postaction={decorate}},
}

\usepackage{hyperref}
\title{\bf{Pseudo-Model-Free Hedging for Variable Annuities via Deep Reinforcement Learning\footnote{This work was first initiated by the authors at the Illinois Risk Lab in January 2020. This work was presented at the 2020 Actuarial Research Conference in August 2020, the United As One: 24th International Congress on Insurance: Mathematics and Economics in July 2021, the 2021 Actuarial Research Conference in August 2021, Heriot-Watt University in November 2021, University of Amsterdam in June 2022, and the 2022 Insurance Data Science Conference in June 2022. The authors thank the participants for fruitful comments. This work utilizes resources supported by the National Science Foundation's Major Research Instrumentation program, grant \#1725729, as well as the University of Illinois at Urbana-Champaign. The authors are grateful to anonymous reviewers for their careful reading and insightful comments.}}}
\author[$\star$]{Wing Fung Chong}
\affil[$\star$]{Maxwell Institute for Mathematical Sciences and Department of Actuarial Mathematics and Statistics, Heriot-Watt
University, Edinburgh, United Kingdom. alfred.chong@hw.ac.uk.}
\author[$\sharp$]{Haoen Cui}
\affil[$\sharp$]{School of Computer Science, Georgia Institute of Technology, Atlanta, United States. haoen.cui@gatech.edu.}
\author[$\ddagger$]{Yuxuan Li\footnote{Corresponding author.}}
\affil[$\ddagger$]{Department of Mathematics, University of Illinois at Urbana-Champaign, Urbana, United States. yuxuanl9@illinois.edu.}
\date{\today}
\begin{document}

\allowdisplaybreaks
\sloppy

\theoremstyle{definition}
\newtheorem{theorem}{Theorem}[section]
\newtheorem{corollary}[theorem]{Corollary}
\newtheorem{lemma}[theorem]{Lemma}
\newtheorem{proposition}[theorem]{Proposition}

\newtheorem{definition}{Definition}[section]
\newtheorem{problem}{Problem}[section]
\newtheorem{remark}{Remark}[section]
\newtheorem{example}{Example}[section]
\setcounter{section}{0}

\maketitle
\begin{abstract}
This paper proposes a two-phase deep reinforcement learning approach, for hedging variable annuity contracts with both GMMB and GMDB riders, which can address model miscalibration in Black-Scholes financial and constant force of mortality actuarial market environments. In the training phase, an infant reinforcement learning agent interacts with a pre-designed training environment, collects sequential anchor-hedging reward signals, and gradually learns how to hedge the contracts. As expected, after a sufficient number of training steps, the trained reinforcement learning agent hedges, in the training environment, equally well as the correct Delta while outperforms misspecified Deltas. In the online learning phase, the trained reinforcement learning agent interacts with the market environment in real time, collects single terminal reward signals, and self-revises its hedging strategy. The hedging performance of the further trained reinforcement learning agent is demonstrated via an illustrative example on a rolling basis to reveal the self-revision capability on the hedging strategy by online learning.

\end{abstract}

\section{Introduction}\label{sec:intro}

Variable annuities are long-term life products, in which policyholders participate in financial investments for profit sharing with insurers. Various guarantees are embedded in these contracts, such as guaranteed minimum maturity benefit (GMMB), guaranteed minimum death benefit (GMDB), guaranteed minimum accumulation benefit (GMAB), guaranteed minimum income benefit (GMIB), and guaranteed minimum withdrawal benefit (GMWB). According to the Insurance Information Institute in 2020, the sales of variable annuity contracts in the United States have amounted to, on average, $100.7$ billion annually, from 2016 to 2020.

Due to their popularity in the market and their dual-risk bearing nature, valuation and risk management of variable annuities have been substantially studied in the literature. By the risk-neutral option pricing approach, to name a few, \cite{Milevsky_2001} studied the valuation of the GMDB rider; valuation and hedging of the GMMB rider under the Black-Scholes (BS) financial market model were covered in \cite{Hardy_2003}; the GMWB rider was extensively investigated by \cite{Milevsky_2006}, \cite{Dai_2008}, and \cite{Chen_2008}; valuation and hedging of the GMMB rider were studied in \cite{Cui_2017} under the Heston financial market model; valuation of the GMMB rider, together with the feature that a contract can be surrendered before its maturity, was examined by \cite{Jeon_2018}, in which optimal surrender strategies were also provided. For a comprehensive review of this approach, see \cite{Feng_2018}.

Valuation and risk management of variable annuities have recently been advanced via various approaches as well. \cite{Trottier_2018} studied the hedging of variable annuities in the presence of basis risk based on a local optimization method. \cite{Chong_2019} revisited the pricing and hedging problem of equity-linked life insurance contracts utilizing the so-called principle of equivalent forward preferences. \cite{Feng_Yi_2019} compared the dynamic hedging approach to the stochastic reserving approach for the risk management of variable annuities. \cite{Moenig_2021_a} investigated the valuation and hedging problem of a portfolio of variable annuities via a dynamic programming method. \cite{Moenig_2021_b} explored the impact of market incompleteness on the policyholder's behavior. \cite{Wang_Zou_2021} solved the optimal fee structure for the GMDB and GMMB riders. \cite{Dang_2020} and \cite{Dang_2022} proposed and analyzed efficient simulation methods for measuring the risk of variable annuities.

Recently, state-of-the-art machine learning methods have been deployed to revisit the valuation and hedging problems of variable annuities at a portfolio level. \cite{Gan_2013} proposed a three-step technique, by (i) selecting representative contracts with clustering method, (ii) pricing these contracts with Monte Carlo (MC) simulation, and (iii) predicting the value of the whole portfolio based on the values of representative contracts with kriging method. To further boost the efficiency and the effectiveness of selecting and pricing the representative contracts, as well as valuating the whole portfolio, various methods at each of these three steps have been proposed. For instance, \cite{Gan_2015} extended the ordinary kriging method to the universal kriging method; \cite{Hejazi_2016} used a neural network as the predictive model to valuate the whole portfolio; \cite{Gan_2018} implemented the generalized beta of the second kind method instead of the kriging method to capture the non-Gaussian behavior of the market price of variable annuities. See also, \cite{Gan_2018_2}, \cite{Gan_2020}, \cite{Gweon_2020}, \cite{Liu_2020}, \cite{Lin_2020}, \cite{Feng_2020}, and \cite{Quan_2021} for recent developments in this three-step technique.
Similar idea has also been applied to the calculation of Greeks and risk measures of a portfolio of variable annuities; see \cite{Gan_2017}, \cite{Gan_2017_2}, and \cite{Xu_2018}. 
All of the above literature applying the machine learning methods involve the supervised learning, which requires a pre-labelled dataset (in this case, it is the set of fair prices of the representative contracts) to train a predictive model.

Other than valuating and hedging variable annuities, supervised learning methods have also been applied to different actuarial contexts. \cite{Wuthrich_2018} used a neural network for the chain-ladder factors in the chain-ladder claim reserving model to include heterogeneous individual claim features. \cite{Gao_2019} applied a convolutional neural network to classify drivers using their telematics data. \cite{Cheridito_2020} estimated the risk measures of a portfolio of assets and liabilities with a feedforward neural network. \cite{Richman_2021} and \cite{Perla_2021} studied the mortality rate forecasting problem, where \cite{Richman_2021} extended the traditional Lee-Carter model to multiple populations using a neural network, while \cite{Perla_2021} applied deep learning techniques directly on a time-series data of mortality rate. \cite{Hu_2022} modified the loss function in tree-based models to improve the predictive performance when applying to imbalanced datasets which are common in the insurance practice.

Meanwhile, a flourishing sub-field in machine learning, called the reinforcement learning (RL), has been skyrocketing and has proved its powerfulness in various tasks; see \cite{Silver_2017}, and the references therein. Contrary to the supervised learning, the RL does not require a pre-labelled dataset for training. Instead, in the RL, an {\it agent interacts} with an {\it environment}, by sequentially {\it observing states}, {\it taking}, as well as {\it revising}, {\it actions}, and {\it collecting rewards}. Without possessing any prior knowledge of the environment, the agent needs to, {\it explore} the environment while {\it exploit} the collected reward signals, for learning. For a representative monograph of RL, see \cite{Sutton_2018}; for its broad applications in economics, game theory, operations research, and finance, see the recent survey paper by \cite{Charpentier_2021}.

The mechanism of RL resembles how a hedging agent hedges any contingent claim dynamically. Indeed, the hedging agent could not know any specifics of the market environment, but could only observe states from the environment, take a hedging strategy, and learn from reward signals to progressively improve the hedging strategy. However, in the context of hedging, if an insurer builds a hedging agent based on a certain RL method, called RL agent hereafter, and allows this infant RL agent to interact and learn from the market environment right away, the insurer could bear enormous financial loss while the infant RL agent is still exploring the environment before it could effectively exploit the reward signals. Moreover, provided that the insurer could not know any specifics of the market environment as well, she could not supply any information derived from theoretical models to the infant RL agent, and thus the agent could only obtain the reward signals via the realized terminal profit and loss, based on the realized net liability and hedging portfolio value; these signals should not be effective for an infant RL agent to learn from the market environment.

To resolve these two issues above, we propose a {\it two-phase (deep) RL approach}, which is composed of a {\it training phase} and an {\it online learning phase}. In the training phase, based on her best knowledge of the market, the insurer constructs a training environment. An infant RL agent is then designated to interact and learn from this training environment for a period of time. Comparing to putting the infant RL agent in the market environment right away, the infant RL agent could be supplied by more information derived from the constructed training environment, such as the net liabilities before any terminal times. In this paper, we propose that the RL agent collects {\it anchor-hedging reward signals} during the training phase. After the RL agent is experienced with the training environment, in the online learning phase, the insurer finally designates the trained RL agent in the market environment. Again, since no theoretical model for the market environment is available to the insurer, the trained RL agent could only collect {\it single terminal reward signals} in this phase. In this paper, an illustrative example is provided to demonstrate the hedging performance using this approach.

All RL methods can be classified into either MC or temporal-difference (TD) learning. As a TD method shall be employed in this paper, in both the training and online learning phases, the following RL literature review focuses on the latter method. \cite{Sutton_1984} and \cite{Sutton_1988} first introduced the TD method for prediction of value function. Based upon their works, \cite{Watkins_1989} and \cite{Watkins_1992} proposed the well-known Q-learning for finite state and action spaces. Since then, the Q-learning has been improved substantially, in \cite{Hasselt_2010} for the Double Q-learning, and in \cite{Mnih_2013}, as well as \cite{Mnih_2015}, for the deep Q-learning which allows infinite state space. Any Q-learning approaches, or in general tabular solution methods and value function approximation methods, are only applicable to finite action space. However, in the context of hedging, the action space is infinite. Instead of discretizing the action space, {\it proximal policy optimization} (PPO) by \cite{Schulman_2017}, which is a {\it policy gradient method}, shall be applied in this paper; our Section~\ref{sec:PPO} shall provide its self-contained review.

To the best of our knowledge, this paper is the first work to implement the RL algorithms with online learning to hedge contingent claims, particularly variable annuities. Contrary to \cite{Xu_2020} and \cite{Carbonneau_2021}, in which both adapted the state-of-the-art DH approach in \cite{Buhler_2019}, this paper is in line with the recent works by \cite{Kolm_2019} and \cite{Cao_2021}, while extends with actuarial components. We shall outline the differences between the RL and DH approaches throughout Sections \ref{sec:RL_approach} and \ref{sec:baseline_results}, as well as Appendices \ref{app:dh_agent_1} and \ref{sec:REINFORCE}. \cite{Kolm_2019} discretized the action space and implemented RL algorithms for finitely many possible actions; however, as mentioned above, this paper does not discretize the action space but adapts the recently advanced policy gradient method, namely, the PPO. Comparing with \cite{Cao_2021}, in addition to the actuarial elements, this paper puts forward online learning to self-revise the hedging strategy.

In the illustrative example, we assume that the market environment is the BS financial and constant force of mortality (CFM) actuarial markets, and the focus is on contracts with both GMMB and GMDB riders. Furthermore, we assume that the model of the market environment being presumed by the insurer, which shall be supplied as the training environment, is also the BS and the CFM, but with a different set of parameters. That is, while the insurer constructs correct dynamic models of the market environment for the training environment, the parameters in the model of the market environment are not the same as those in the market environment. Section \ref{sec:pit_revisit} shall set the stage of this illustrative example, and shall show that, if the insurer forwardly implements, in the market environment, the incorrect Delta hedging strategy based on her presumed model of the market environment, then its hedging performance for the variable annuities is worse than that by the correct Delta hedging strategy based on the market environment. In Sections \ref{sec:baseline_results} and \ref{sec:illustrative}, this illustrative example shall be revisited using the two-phase RL approach. As we shall see in Section \ref{sec:illustrative}, the hedging performance of the RL agent is even worse than that of the incorrect Delta, at the very beginning of hedging in real time. However, delicate analysis shows that, with a fair amount of future trajectories (which are different from simulated scenarios, with more details in Section \ref{sec:illustrative}), the hedging performance of the RL agent becomes comparable with that of the correct Delta within a reasonable amount of time. Therefore, the illustrative example addresses model miscalibration issue in hedging variable annuity contracts with GMMB and GMDB riders in BS financial and CFM actuarial market environments, which is common in practice.



This paper is organized as follows. Section \ref{sec:prob_formulation} formulates the continuous hedging problem for variable annuities, reformulates it to the discrete and Markov setting, and motivates as well as outlines the two-phase RL approach. Section \ref{sec:RL_approach} discusses the RL approach in hedging variable annuities and provides a self-contained review of RL, particularly the PPO, which is a TD policy gradient method, while Section \ref{sec:online_learning} presents the implementation details of the online learning phase. Sections \ref{sec:baseline_results} and \ref{sec:illustrative} revisit the illustrative example in the training and online learning phases respectively. Section \ref{sec:assumption_practice} collates the assumptions of utilizing the two-phase RL approach for hedging contingent claims, as well as their implications in practice. This paper finally concludes and comments on future directions in Section~\ref{sec:conclusion}.

\section{Problem Formulation and Motivation}\label{sec:prob_formulation}

\subsection{Classical Hedging Problem and Model-Based Approach}
We first review the classical hedging problem for variable annuities and its model-based solution to introduce some notations and to motivate the RL approach.

\subsubsection{Actuarial and Financial Market Models}\label{sec:models}

Let $(\Omega, \mathcal{F}, \mathbb{P})$ be a rich enough complete probability space. Consider the current time $t=0$ and fix $T>0$ as a deterministic time in the future. Throughout this paper, all time units are in year.


There are one risk-free asset and one risky asset in the financial market. 
Let $B_t$ and $S_t$, for $t\in\left[0,T\right]$, be the time-$t$ values of the risk-free asset and the risky asset respectively. Let $\mathbb{G}^{\left(1\right)}=\left\{\mathcal{G}^{\left(1\right)}_t\right\}_{t\in\left[0,T\right]}$ be the filtration which contains all financial market information; in particular, both processes $B=\left\{B_t\right\}_{t\in\left[0,T\right]}$ and $S=\left\{S_t\right\}_{t\in\left[0,T\right]}$ are $\mathbb{G}^{\left(1\right)}$-adapted.


There are $N$ policyholders in the actuarial market. For each policyholder $i=1,2,\dots,N$, denote $T_{x_i}^{\left(i\right)}$ as her random future lifetime, who is of age $x_i$ at the current time $0$. Define, for each $i=1,2,\dots,N$, and for any $t\geq 0$, $J_t^{\left(i\right)}=\mathds{1}_{\left\{T_{x_i}^{\left(i\right)}>t\right\}}$, be the corresponding time-$t$ jump value generated by the random future lifetime of the $i$-th policyholder; that is, if the $i$-th policyholder survives at some time $t\in\left[0,T\right]$, $J_t^{\left(i\right)}=1$; otherwise, $J_t^{\left(i\right)}=0$. Let $\mathbb{
G}^{\left(2\right)}=\left\{\mathcal{G}^{\left(2\right)}_t\right\}_{t\in\left[0,T\right]}$ be the filtration which contains all actuarial market information; in particular, all single-jump processes $J^{\left(i\right)}=\left\{J_t^{\left(i\right)}\right\}_{t\in\left[0,T\right]}$, for $i=1,2,\dots,N$, are $\mathbb{G}^{\left(2\right)}$-adapted.


Let $\mathbb{F}=\left\{\mathcal{F}_t\right\}_{t\in\left[0,T\right]}$ be the filtration which contains all actuarial and financial market information; that is, $\mathbb{F}=\mathbb{G}^{\left(1\right)}\vee\mathbb{G}^{\left(2\right)}$. Therefore, the filtered probability space is given by $(\Omega, \mathcal{F}, \mathbb{F}, \mathbb{P})$.


\subsubsection{Variable Annuities with Guaranteed Minimum Maturity Benefit and Guaranteed Minimum Death Benefit Riders}

At the current time $0$, an insurer writes a variable annuity contract to each of these $N$ policyholders. Each contract is embedded with both GMMB and GMDB riders. Assume that all these $N$ contracts expire at the same fixed time $T$. In the following, fix a generic policyholder $i=1,2,\dots,N$.

At the current time $0$, the policyholder deposits $F_0^{\left(i\right)}$ into her segregated account to purchase $\rho^{\left(i\right)}>0$ shares of the risky asset; that is, $F_0^{\left(i\right)}=\rho^{\left(i\right)}S_0$. Assume that the policyholder does not revise the number of shares $\rho^{\left(i\right)}$ throughout the effective time of the contract.

For any $t\in\left[0,T_{x_i}^{\left(i\right)}\wedge T\right]$, the time-$t$ segregated account value of the policyholder is given by $F_t^{\left(i\right)}=\rho^{\left(i\right)}S_te^{-m^{\left(i\right)}t}$, where $m^{\left(i\right)}\in\left(0,1\right)$ is the continuously compounded annualized rate at which the asset-value-based fees are deducted from the segregated account by the insurer. For any $t\in\left(T_{x_i}^{\left(i\right)}\wedge T,T\right]$, the time-$t$ segregated account value $F_t^{\left(i\right)}$ must be $0$; indeed, if the policyholder dies before the maturity, i.e. $T_{x_i}^{\left(i\right)}<T$, then, due to the GMDB rider of a minimum guarantee $G_D^{\left(i\right)}>0$, the beneficiary inherits $\max\left\{F^{\left(i\right)}_{T_{x_i}^{\left(i\right)}},G_D^{\left(i\right)}\right\}$, which can be decomposed into $F_{T_{x_i}^{\left(i\right)}}^{\left(i\right)}+\left(G_D^{\left(i\right)}-F_{T_{x_i}^{\left(i\right)}}^{\left(i\right)}\right)_+$, at the policyholder's death time $T_{x_i}^{\left(i\right)}$ right away. Due to the GMMB rider of a minimum guarantee $G_M^{\left(i\right)}>0$, if the policyholder survives beyond the maturity, i.e. $T_{x_i}^{\left(i\right)}>T$, the policyholder acquires $\max\left\{F_T^{\left(i\right)},G_M^{\left(i\right)}\right\}$ at the maturity, which can be decomposed into $F_T^{\left(i\right)}+\left(G_M^{\left(i\right)}-F_T^{\left(i\right)}\right)_+$.

\subsubsection{Net Liability of Insurer}\label{sec:net_lia}
The liability of the insurer thus has two parts. The liability from the GMMB rider at the maturity for the $i$-th policyholder, where $i=1,2,\dots,N$, is given by  $\left(G_M^{\left(i\right)}-F_T^{\left(i\right)}\right)_+$ if the $i$-th policyholder survives beyond the maturity, and is $0$ otherwise. The liability from the GMDB rider at the death time $T_{x_i}^{\left(i\right)}$ for the $i$-th policyholder, where $i=1,2,\dots,N$, is given by $\left(G_D^{\left(i\right)}-F_{T_{x_i}^{\left(i\right)}}^{\left(i\right)}\right)_+$ if the $i$-th policyholder dies before the maturity, and is $0$ otherwise. Therefore, at any time $t\in\left[0,T\right]$, the future gross liability of the insurer accumulated to the maturity for these $N$ contracts is given by
\begin{equation*}
\sum_{i = 1}^{N}\left(\left(G_M^{\left(i\right)}-F_T^{\left(i\right)}\right)_+J_T^{\left(i\right)} + \frac{B_T}{B_{T_{x_i}^{\left(i\right)}}}\left(G_D^{\left(i\right)}-F_{T_{x_i}^{\left(i\right)}}^{\left(i\right)}\right)_+\mathds{1}_{\{T_{x_i}^{\left(i\right)} < T\}}J_t^{\left(i\right)}\right).
\end{equation*}
Denote $V_t^{\text{GL}}$, for $t\in\left[0,T\right]$, as the time-$t$ value of the discounted (via the risk-free asset $B$) future gross liability of the insurer; if the liability is $0$, the value will be $0$.

From the asset-value-based fees collected by the insurer, a portion, known as the rider charge, is used to fund the liability due to the GMMB and GMDB riders; the remaining portion is used to cover overhead, commissions, and any other expenses. From the $i$-th policyholder, where $i=1,2,\dots,N$, the insurer collects $m_e^{\left(i\right)}F_t^{\left(i\right)}J_t^{\left(i\right)}$ as the rider charge at any time $t\in\left[0,T\right]$, where $m_e^{\left(i\right)}\in\left(0,m^{\left(i\right)}\right]$. Therefore, the cumulative future rider charge to be collected, from any time $t\in\left[0,T\right]$ onward, till the maturity, by the insurer from these $N$ policyholders, is given by $\sum_{i=1}^{N}\int_{t}^{T}m_e^{\left(i\right)}F_s^{\left(i\right)}J_s^{\left(i\right)}\left(B_T/B_s\right)ds$. Denote $V_t^{\text{RC}}$, for $t\in\left[0,T\right]$, as its time-$t$ discounted (via the risk-free asset $B$) value; if the cumulative rider charge is $0$, the value will be $0$.

Hence, due to these $N$ variable annuity contracts with both GMMB and GMDB riders, for any $t\in\left[0,T\right]$, the time-$t$ net liability of the insurer for these $N$ contracts is given by $L_t=V_t^{\text{GL}}-V_t^{\text{RC}}$, which is $\mathcal{F}_t$-measurable.

One of the many ways to set the rate $m^{\left(i\right)}\in\left(0,1\right)$ for the asset-value-based fees, and the rate $m_e^{\left(i\right)}\in\left(0,m^{\left(i\right)}\right]$ for the rider charge, for $i=1,2,\dots,N$, is based on the time-$0$ net liability of the insurer for the $i$-th policyholder. More precisely, $m^{\left(i\right)}$ and $m_e^{\left(i\right)}$ are determined via $L_0^{\left(i\right)}=V_0^{\text{GL},\left(i\right)}-V_0^{\text{RC},\left(i\right)}=0$, where $V_0^{\text{GL},\left(i\right)}$ and $V_0^{\text{RC},\left(i\right)}$ are the time-$0$ values of, respectively, the discounted future gross liability and the discounted cumulative future rider charge, of the insurer for the $i$-th policyholder.

\subsubsection{Continuous Hedging and Hedging Objective}\label{hedging_va}
The insurer aims to hedge this dual-risk bearing net liability via investing in the financial market. To this end, let $\tilde{T}$ be the death time of the last policyholder; that is, $\tilde{T}=\max_{i=1,2,\dots,N}T_{x_i}^{\left(i\right)}$, which is random.

While the net liability $L_t$ is defined for any time $t\in\left[0,T\right]$, as the difference between the values of discounted future gross liability and discounted cumulative future rider charge, $L_t=0$ for any $t\in\left(\tilde{T}\wedge T,T\right]$. Indeed, if $\tilde{T}<T$, then, for any $t\in\left(\tilde{T}\wedge T,T\right]$, one has $T_{x_i}^{\left(i\right)}<t\leq T$ for all $i = 1,2,\dots, N$, and hence, the future gross liability accumulated to the maturity, and the cumulative rider charge from time $\tilde{T}$ onward, are both $0$, so are their values. Therefore, the insurer only hedges the net liability $L_t$, for any $t\in\left[0,\tilde{T}\wedge T\right]$.


Let $H_t$ be the hedging strategy, i.e. the number of shares of the risky asset being held by the insurer, at time $t\in\left[0,T\right)$. Hence, $H_t=0$, for any $t\in\left[\tilde{T}\wedge T,T\right)$. Let $\mathcal{H}$ be the admissible set of hedging strategies, which is defined by
\begin{align*}
\mathcal{H}=&\;\left\{H=\left\{H_t\right\}_{t\in\left[0,T\right)}:(\text{i})\;H\text{ is }\mathbb{F}\text{-adapted, }(\text{ii})\;H\in\mathbb{R},\;\mathbb{P}\times\mathcal{L}\text{-a.s., and }(\text{iii})\;\text{for any }t\in\left[\tilde{T}\wedge T,T\right),\;H_t=0\right\},
\end{align*}
where $\mathcal{L}$ is the Lebesgue measure on $\mathbb{R}$.

Let $P_t$ be the time-$t$ value, for $t\in\left[0,T\right]$, of the insurer's hedging portfolio. Then $P_0=0$, and together with the rider charges collected from the $N$ policyholders, as well as the withdrawal for paying the liabilities due to the beneficiaries' inheritance from those policyholders who have already been dead, for any $t\in\left(0,T\right]$,
\begin{equation*}
P_t=\int_{0}^{t}\left(P_s-H_sS_s\right)\frac{dB_s}{B_s}+\int_{0}^{t}H_sdS_s+\sum_{i=1}^{N}\int_{0}^{t}m_e^{\left(i\right)}F_s^{\left(i\right)}J_s^{\left(i\right)}ds -\sum_{i=1}^{N}\left(G_D^{\left(i\right)}-F_{T_{x_i}^{\left(i\right)}}^{\left(i\right)}\right)_+\mathds{1}_{\{T_{x_i}^{\left(i\right)}\leq t < T\}},
\end{equation*}
which obviously depends on $\left\{H_s\right\}_{s\in\left[0,t\right)}$.

As in \cite{Bertsimas_2000}, the insurer's hedging objective function at the current time $0$ should be given by the root-mean-square error (RMSE) of the terminal profit and loss (P\&L), 
which is, for any $H\in\mathcal{H}$,
\begin{equation*}
\sqrt{\mathbb{E}^{\mathbb{P}}\left[\left(P_{\tilde{T}\wedge T}-L_{\tilde{T}\wedge T}\right)^2\right]}.
\end{equation*}
If the insurer has full knowledge of the objective probability measure $\mathbb{P}$, and hence the correct dynamics of the risk-free asset and the risky asset in the financial market, as well as the correct mortality model in the actuarial market, the optimal hedging strategy, being implemented forwardly, is given by minimizing the RMSE of the terminal P\&L:
\begin{equation*}
H^*=\argmin_{H\in\mathcal{H}}\sqrt{\mathbb{E}^{\mathbb{P}}\left[\left(P_{\tilde{T}\wedge T}-L_{\tilde{T}\wedge T}\right)^2\right]}.
\end{equation*}

\subsection{Pitfall of Model-Based Approach}
However, having correct model is usually not the case in practice. Indeed, the insurer, who is the hedging agent above, usually has little information regarding the objective probability measure $\mathbb{P}$, and hence easily misspecifies the financial market dynamics and the mortality model, which will in turn yield a poor performance from the supposedly optimal hedging strategy when it is implemented forwardly in the future. Section \ref{sec:pit_revisit} outlines such an illustrative example which shall be discussed throughout the remaining of this paper.

To rectify this, we propose a {\it two-phase} {\it (deep) RL} {\it approach} to solve an optimal hedging strategy. In this approach, an RL agent, which is not the insurer herself but is built by the insurer to hedge on her behalf, does not have any knowledge of the objective probability measure $\mathbb{P}$, the financial market dynamics, and the mortality model; Section \ref{sec:motivation_revisited} shall explain this approach in details. Before that, in the following Section \ref{sec:dis_markov_hedging}, the classical hedging problem shall first be reformulated with a Markov decision process (MDP) in a discrete-time setting so that RL methods can be implemented. The illustrative example outlined in Section \ref{sec:pit_revisit} shall be revisited using the proposed two-phase RL approach in Sections \ref{sec:baseline_results} and \ref{sec:illustrative}.

In the remaining of this paper, unless otherwise specified, all expectation operators shall be taken with respect to the objective probability measure $\mathbb{P}$, and denoted simply as $\mathbb{E}\left[\cdot\right]$.

\subsection{Discrete and Markov Hedging}\label{sec:dis_markov_hedging}
\subsubsection{Discrete Hedging and Hedging Objective}\label{sec:dis_hedging}
Let $t_0,t_1,\dots,t_{n-1}\in\left[0,T\right)$, for some $n\in\mathbb{N}$, be the time when the hedging agent decides the hedging strategy, such that $0=t_0<t_1<\dots<t_{n-1}<T$. Denote also $t_n=T$.

Let $t_{\tilde{n}}$ be the first time (right) after the last policyholder dies or all contracts expire, for some $\tilde{n}=1,2,\dots,n$, which is random; that is, $t_{\tilde{n}}=\min\left\{t_k,\;k=1,2,\dots,n:t_k\geq\tilde{T}\right\}$, and when $\tilde{T}>T$, by convention, $\min\emptyset=t_n$.
Therefore, $H_t=0$, for any $t=t_{\tilde{n}},t_{\tilde{n}+1},\dots,t_{n-1}$. With a slight abuse of notation, the admissible set of hedging strategies in discrete time is
\begin{align*}
\mathcal{H}=&\;\left\{H=\left\{H_t\right\}_{t=t_0,t_1,\dots,t_{n-1}}:(\text{i})\;\text{for any }t=t_{0},t_{1},\dots,t_{n-1},\;H_t\text{ is }\mathcal{F}_t\text{-measurable, }\right.\\&\;\left.\quad\quad\quad\quad\quad\quad\quad\quad\quad\quad\;\;\;\;(\text{ii})\;\text{for any }t=t_{0},t_{1},\dots,t_{n-1},\;H_t\in\mathbb{R},\;\mathbb{P}\text{-a.s., and}\right.\\&\;\left.\quad\quad\quad\quad\quad\quad\quad\quad\quad\quad\;\;\;\;(\text{iii})\;\text{for any }t=t_{\tilde{n}},t_{\tilde{n}+1},\dots,t_{n-1},\;H_t=0\right\}.
\end{align*}

While the hedging agent decides the hedging strategy at the discrete time points, the actuarial and financial market models are continuous. Hence, the net liability $L_t=V_t^{\text{GL}}-V_t^{\text{RC}}$ is still defined for any time $t\in\left[0,T\right]$ as before. Moreover, if $t\in\left[t_k,t_{k+1}\right)$, for some $k=0,1,\dots,n-1$, $H_t=H_{t_k}$; thus, $P_0=0$, and, if $t\in\left(t_k,t_{k+1}\right]$, for some $k=0,1,\dots,n-1$,
\begin{equation}\label{eq:hedging_port_value}
P_t=\left(P_{t_k}-H_{t_k}S_{t_k}\right)\frac{B_{t}}{B_{t_k}}+H_{t_k}S_{t}+\sum_{i=1}^{N}\int_{t_k}^{t}m_e^{\left(i\right)}F_s^{\left(i\right)}J_s^{\left(i\right)}\frac{B_{t}}{B_{s}}ds  -\sum_{i = 1}^{N}\frac{B_t}{B_{T_{x_i}^{\left(i\right)}}}\left(G_D^{\left(i\right)}-F_{T_{x_i}^{\left(i\right)}}^{\left(i\right)}\right)_+\mathds{1}_{\{t_k<T_{x_i}^{\left(i\right)}\leq t<T\}}.
\end{equation}
For any $H\in\mathcal{H}$, the hedging objective of the insurer at the current time $0$ is $\sqrt{\mathbb{E}\left[\left(P_{t_{\tilde{n}}}-L_{t_{\tilde{n}}}\right)^2\right]}$. Hence, the optimal discrete hedging strategy, being implemented forwardly, is given by
\begin{equation}
H^*=\argmin_{H\in\mathcal{H}}\sqrt{\mathbb{E}\left[\left(P_{t_{\tilde{n}}}-L_{t_{\tilde{n}}}\right)^2\right]}=\argmin_{H\in\mathcal{H}}\mathbb{E}\left[\left(P_{t_{\tilde{n}}}-L_{t_{\tilde{n}}}\right)^2\right].
\label{eq:hedging}
\end{equation}

\subsubsection{Markov Decision Process}\label{sec:MDP}
An MDP can be characterized by its state space, action space, Markov transition probability, and reward signal. In turn, these derive the value function and the optimal value function, which are equivalently known as, respectively, the objective function and the value function, in optimization as in the previous sections. In the remaining of this paper, we shall adapt the MDP language.
\begin{itemize}
\item ({\it State}) Let $\mathcal{X}$ be the state space in $\mathbb{R}^p$, where $p\in\mathbb{N}$. Each state in the state space represents a possible observation with $p$ features in the actuarial and financial markets. Denote $X_{t_k}\in\mathcal{X}$ as the observed state at any time $t_k$, where $k=0,1,\dots,n$; the state should minimally include an information related to the number of surviving policyholders $\sum_{i=1}^{N}J_{t_k}^{\left(i\right)}$, and the term to maturity $T-t_k$, in order to terminate the hedging at time $t_{\tilde{n}}$, which is the first time when $\sum_{i=1}^{N}J_{t_{\tilde{n}}}^{\left(i\right)}=0$, or which is when $T-t_{\tilde{n}}=0$. The states (space) shall be specified in Sections \ref{sec:baseline_results} and \ref{sec:online_learning}.
\item ({\it Action}) Let $\mathcal{A}$ be the action space in $\mathbb{R}$. Each action in the action space is a possible hedging strategy. Denote $H_{t_k}\left(X_{t_k}\right)\in\mathcal{A}$ as the action at any time $t_k$, where $k=0,1,\dots,n-1$, which is assumed to be Markovian with respect to the observed state $X_{t_k}$; that is, given the current state $X_{t_k}$, the current action $H_{t_k}\left(X_{t_k}\right)$ is independent of the past states $X_{t_0},X_{t_1},\dots,X_{t_{k-1}}$. In the sequel, for notational simplicity, we simply write $H_{t_k}$ to represent $H_{t_k}\left(X_{t_k}\right)$, for $k=0,1,\dots,n-1$. If the feature of the number of surviving policyholders $\sum_{i=1}^{N}J_{t_k}^{\left(i\right)}=0$, for $k=0,1,\dots,n-1$, in the state $X_{t_k}$, then $H_{t_k}=0$; in particular, for any $t_k$, where $k=\tilde{n},\tilde{n}+1,\dots,n-1$, the hedging strategy $H_{t_k}=0$.
\item (Markov property) At any time $t_k$, where $k=0,1,\dots,{n-1}$, given the current state $X_{t_k}$ and the current hedging strategy $H_{t_k}$, the transition probability distribution of the next state $X_{t_{k+1}}$ in the market is independent of the past states $X_{t_0},X_{t_1},\dots,X_{t_{k-1}}$ and the past hedging strategies $H_{t_0},H_{t_1},\dots,H_{t_{k-1}}$; that is, for any Borel set $\overline{B}\in\mathcal{B}\left(\mathcal{X}\right)$,
\begin{equation}
\mathbb{P}\left(X_{t_{k+1}}\in\overline{B}\vert H_{t_{k}},X_{t_{k}},H_{t_{k-1}},X_{t_{k-1}},\dots,H_{t_{1}},X_{t_{1}},H_{t_{0}},X_{t_{0}}\right)=\mathbb{P}\left(X_{t_{k+1}}\in\overline{B}\vert H_{t_{k}},X_{t_{k}}\right).
\label{eq:markov_property}
\end{equation}
\item ({\it Reward}) At any time $t_k$, where $k=0,1,\dots,{n-1}$, given the current state $X_{t_k}$ in the market and the current hedging strategy $H_{t_k}$, a reward signal $R_{t_{k+1}}\left(X_{t_k},H_{t_k},X_{t_{k+1}}\right)$ is received, by the hedging agent, as a result of transition to the next state $X_{t_{k+1}}$. The reward signal shall be specified after introducing the (optimal) value function below. In the sequel, occasionally, for notational simplicity, we simply write $R_{t_{k+1}}$ to represent $R_{t_{k+1}}\left(X_{t_k},H_{t_k},X_{t_{k+1}}\right)$, for $k=0,1,\dots,n-1$.
\item (State, action, and reward sequence) The states, actions, and reward signals form an {\it episode}, which is sequentially given by:
\begin{equation*}
\left\{X_{t_0},H_{t_0},X_{t_1},R_{t_1},H_{t_1},X_{t_2},R_{t_2},H_{t_2},\dots,X_{t_{\tilde{n}-1}},R_{t_{\tilde{n}-1}},H_{t_{\tilde{n}-1}},X_{t_{\tilde{n}}},R_{t_{\tilde{n}}}\right\}.
\end{equation*}
\item (Optimal value function) Based on the reward signals, the value function, at any time $t_k$, where $k=0,1,\dots,n-1$, with the state $x\in\mathcal{X}$, is defined by, for any hedging strategies $H_{t_k},H_{t_{k+1}},\dots,H_{t_{n-1}}$,
\begin{equation}
V\left(t_k,x;H_{t_k},H_{t_{k+1}},\dots,H_{t_{n-1}}\right)=\mathbb{E}\left[\sum_{l=k}^{n-1}\gamma^{t_{l+1}-t_k}R_{t_{l+1}}\Big\vert X_{t_k}=x\right],
\label{eq:value_function}
\end{equation}
where $\gamma\in\left[0,1\right]$ is the discount rate; the value function, at the time $t_n=T$ with the state $x\in\mathcal{X}$, is defined by $V\left(t_n,x\right)=0$. Hence, the optimal discrete hedging strategy, being implemented forwardly, is given by
\begin{equation}
H^*=\argmax_{H\in\mathcal{H}}\mathbb{E}\left[\sum_{k=0}^{n-1}\gamma^{t_{k+1}}R_{t_{k+1}}\Big\vert X_{0}=x\right].
\label{eq:hedging2}
\end{equation}
In turn, the optimal value function, at any time $t_k$, where $k=0,1,\dots,n-1$, with the state $x\in\mathcal{X}$, is
\begin{equation}
V^*\left(t_k,x\right)=V\left(t_k,x;H^*_{t_k},H^*_{t_{k+1}},\dots,H^*_{t_{n-1}}\right),\text{ and }V^*\left(t_n,x\right)=0.
\label{eq:optimal_value_function}
\end{equation}
\item (Reward engineering) To ensure the hedging problem being reformulated with the MDP, the value functions, given by that in \eqref{eq:hedging2}, and the negative of that in \eqref{eq:hedging}, should coincide; that is,
\begin{equation}\label{eq:rd_engin}
\mathbb{E}\left[\sum_{k=0}^{n-1}\gamma^{t_{k+1}}R_{t_{k+1}}\Big\vert X_{0}=x\right]=-\mathbb{E}\left[\left(P_{t_{\tilde{n}}}-L_{t_{\tilde{n}}}\right)^2\right].
\end{equation}

Hence, two possible constructions for the reward signals are proposed as follows; each choice of the reward signals shall be utilized in one of the two phases in the proposed RL approach.
\begin{itemize}
\item (Single terminal reward) An obvious choice is to only have a reward signal from the negative squared terminal P\&L; that is, for any time $t_k$, 
\begin{equation}\label{eq:reward_2}
R_{t_{k+1}} = 
\begin{cases*}
-\left(P_{t_{\tilde{n}}}-L_{t_{\tilde{n}}}\right)^2 & if  $k= \tilde{n}-1$, \\
0 & otherwise.
\end{cases*} 
\end{equation}
Necessarily, the discount rate is given as $\gamma=1$.
\item (Sequential anchor-hedging reward) A less obvious choice is via telescoping the RHS of Equation \eqref{eq:rd_engin}, that
\begin{equation*}
-\mathbb{E}\left[\left(P_{t_{\tilde{n}}}-L_{t_{\tilde{n}}}\right)^2\right]=-\mathbb{E}\left[\sum_{k=0}^{\tilde{n}-1}\left(\left(P_{t_{k+1}}-L_{t_{k+1}}\right)^2-\left(P_{t_{k}}-L_{t_{k}}\right)^2\right)+\left(P_0-L_0\right)^2\right].
\end{equation*}
Therefore, when $L_0=P_0$, another possible construction for the reward signal is, for any time $t_k$,
\begin{equation}\label{eq:reward_1}
R_{t_{k+1}} = 
\begin{cases*}
\left(P_{t_{k}}-L_{t_{k}}\right)^2-\left(P_{t_{k+1}}-L_{t_{k+1}}\right)^2 & if  $k=0,1,\dots,\tilde{n}-1$, \\
0 & otherwise.
\end{cases*} 
\end{equation}
Again, the discount rate is necessarily given as $\gamma=1$. The constructed reward in \eqref{eq:reward_1} outlines an {\it anchor-hedging} scheme. First, note that, at the current time $0$, when $L_0=P_0$, there is no local hedging error. Then, at each future hedging time before the last policyholder dies and before the maturity, the hedging performance is measured by the local squared P\&L, i.e. $\left(P_{t_{k}}-L_{t_{k}}\right)^2$, which serves as an anchor. At the next hedging time, if the local squared P\&L is smaller than the anchor, it will be rewarded, i.e. $R_{t_{k+1}}>0$; however, if the local squared P\&L becomes larger, it will be penalized, i.e. $R_{t_{k+1}}<0$.
\end{itemize}
\end{itemize}

\subsection{Illustrative Example}\label{sec:pit_revisit}
The illustrative example below demonstrates the poor hedging performance by the Delta hedging strategy when the insurer miscalibrates the parameters in the market environment. We consider that the insurer hedges a variable annuity contract, with both GMMB and GMDB riders, of a single policyholder, i.e. $N=1$, with the contract characteristics given in Table \ref{table:params_contrac}.


\begin{table}[!htb]
\centering
\begin{tabular}{@{}lc@{}}
\toprule
Parameter & Value 
\\ \midrule
Expiration date $T$ & $1$ \\
Minimum guarantee at maturity $G_M$ & $100$\\
Minimum guarantee at death $G_D$ & $100$\\
\bottomrule   
\end{tabular}
\caption{Contract Characteristics}
\label{table:params_contrac}
\end{table}

The market environment follows the Black-Scholes (BS) in the financial part and the constant force of mortality (CFM) in the actuarial front. The risk-free asset earns a constant risk-free interest rate $r>0$ that, for any $t\in\left[0,T\right]$, $dB_t=rB_tdt$, while the value of the risky asset evolves as a geometric Brownian motion that, for any $t\in\left[0,T\right]$, $dS_t=\mu S_tdt+\sigma S_tdW_t$, where $\mu$ is a constant drift, $\sigma>0$ is a constant volatility, and $W=\left\{W_t\right\}_{t\in\left[0,T\right]}$ is the standard Brownian motion. The random future lifetime of the policyholder $T_x$ has a CFM $\nu>0$; that is, for any $0\leq t\leq s\leq T$, the conditional survival probability $\mathbb{P}\left(T_x >s\vert T_x>t\right)=e^{-\nu\left(s-t\right)}$. Moreover, the Brownian motion $W$ in the financial market and the future lifetime $T_x$ in the actuarial market are independent. Table \ref{table:param_financial_actuarial_market} summarizes the parameters in the market environment. Note that the risk-free interest rate, the risky asset initial price, the initial age of the policyholder, and the investment strategy of the policyholder, are observable by the insurer.

\begin{table}[!htb]
\begin{subtable}{.4\linewidth}
\centering
\caption{Black-Scholes Financial Market}
\begin{tabular}{@{}lc@{}}
\toprule
Parameter & Value  \\
\midrule
Risk-free interest rate $r$ & $0.02$\\
Risky asset initial price $S_0$ & $100$\\
Risky asset drift $\mu$ & $-0.2$\\
Risky asset volatility $\sigma$ & $0.4$\\
\bottomrule   
\end{tabular}
\label{sub_table:params_fin_market}
\end{subtable}%
\begin{subtable}{.65\linewidth}
\centering
\caption{Constant Force of Mortality Actuarial Market}
\begin{tabular}{@{}lc@{}}
\toprule
Parameter & Value 
\\ \midrule
Initial number of policyholders $N$ & $1$\\
Initial age of policyholders $x$ & $20$\\
Constant force of mortality $\nu$ & $0.03$\\
Investment strategy of policyholders $\rho$ & $1.19$\\
\bottomrule   
\end{tabular}
\label{sub_table:params_act_market}
\end{subtable}


\caption{Parameters setting of market environment}
\label{table:param_financial_actuarial_market}
\end{table}

Based on her best knowledge of the market, the insurer builds a model of the market environment. Suppose that the model happens to be the BS and the CFM as the market environment, but the insurer {\it miscalibrates} the parameters. Table \ref{table:param_financial_actuarial} lists these parameters in the model of the market environment. In particular, the risky asset drift and volatility, as well as the force of mortality constant are different from those in the market environment. For the observable parameters, they are the same as those in the market environment.

\begin{table}[!htb]
\begin{subtable}{.4\linewidth}
\centering
\caption{Black-Scholes Financial Market}
\begin{tabular}{@{}lc@{}}
\toprule
Parameter & Value  \\
\midrule
Risk-free interest rate $r$ & $0.02$\\
Risky asset initial price $S_0$ & $100$\\
{\bf Risky asset drift} $\mu$ & $0.08$\\
{\bf Risky asset volatility} $\sigma$ & $0.2$\\
\bottomrule   
\end{tabular}
\label{sub_table:params_fin}
\end{subtable}%
\begin{subtable}{.65\linewidth}
\centering
\caption{Constant Force of Mortality Actuarial Market}
\begin{tabular}{@{}lc@{}}
\toprule
Parameter & Value 
\\ \midrule
Initial number of policyholders $N$ & $1$\\
Initial age of policyholders $x$ & $20$\\
{\bf Constant force of mortality} $\nu$ & $0.02$\\
Investment strategy of policyholders $\rho$ & $1.19$\\
\bottomrule   
\end{tabular}
\label{sub_table:params_act}
\end{subtable} 


\caption{Parameters setting of model of market environment, with bolded parameters being different from those in market environment}
\label{table:param_financial_actuarial}
\end{table}

At any time $t\in\left[0,T\right]$, the value of the hedging portfolio of the insurer is given by \eqref{eq:baseline_hedging_portfolio}, with $N=1$, in which the values of the risky asset and the single-jump process follow the market environment with the parameters in Table \ref{table:param_financial_actuarial_market}. At any time $t\in\left[0,T\right]$, the value of the net liability of the insurer is given by \eqref{eq:net_liability}, with $N=1$, in both the market environment and its model; for its detailed derivations, we defer it to Section \ref{sec:MDP_training}, as the model of the market environment, with multiple homogeneous policyholders for effective training, shall be supplied as the training environment. Since the parameters in the model of the market environment (see Table \ref{table:param_financial_actuarial}) are different from those in the market environment (see Table \ref{table:param_financial_actuarial_market}), the net liability evaluated by the insurer using the model is different from that of the market environment. There are two implications. Firstly, the Delta hedging strategy of the insurer using the parameters in Table \ref{table:param_financial_actuarial} is incorrect, while the correct Delta hedging strategy should use the parameters in Table \ref{table:param_financial_actuarial_market}. 
Secondly, the asset-value-based fee $m$ and the rider charge $m_e$ given in Table \ref{sub_table:params_fee}, which are determined by the insurer based on the time-$0$ value of her net liability by Table \ref{table:param_financial_actuarial} via the method in Section \ref{sec:net_lia}, are mispriced. They would not lead to zero time-$0$ value of her net liability in the market environment which is based on Table \ref{table:param_financial_actuarial_market}.


\begin{table}[!htb]
\centering
\begin{tabular}{@{}lc@{}}
\toprule
Parameter & Value 
\\ \midrule
Rate for asset-value-based fee $m$ & $0.02$ \\
Rate for rider charge $m_e$ & $0.019$ \\
\bottomrule   
\end{tabular}
\caption{Fee structures derived from model of market environment}
\label{sub_table:params_fee}
\end{table} 

\begin{figure}[H]  
\centering
\begin{subfigure}[t]{.5\textwidth}
\centering
\includegraphics[width=1\linewidth]{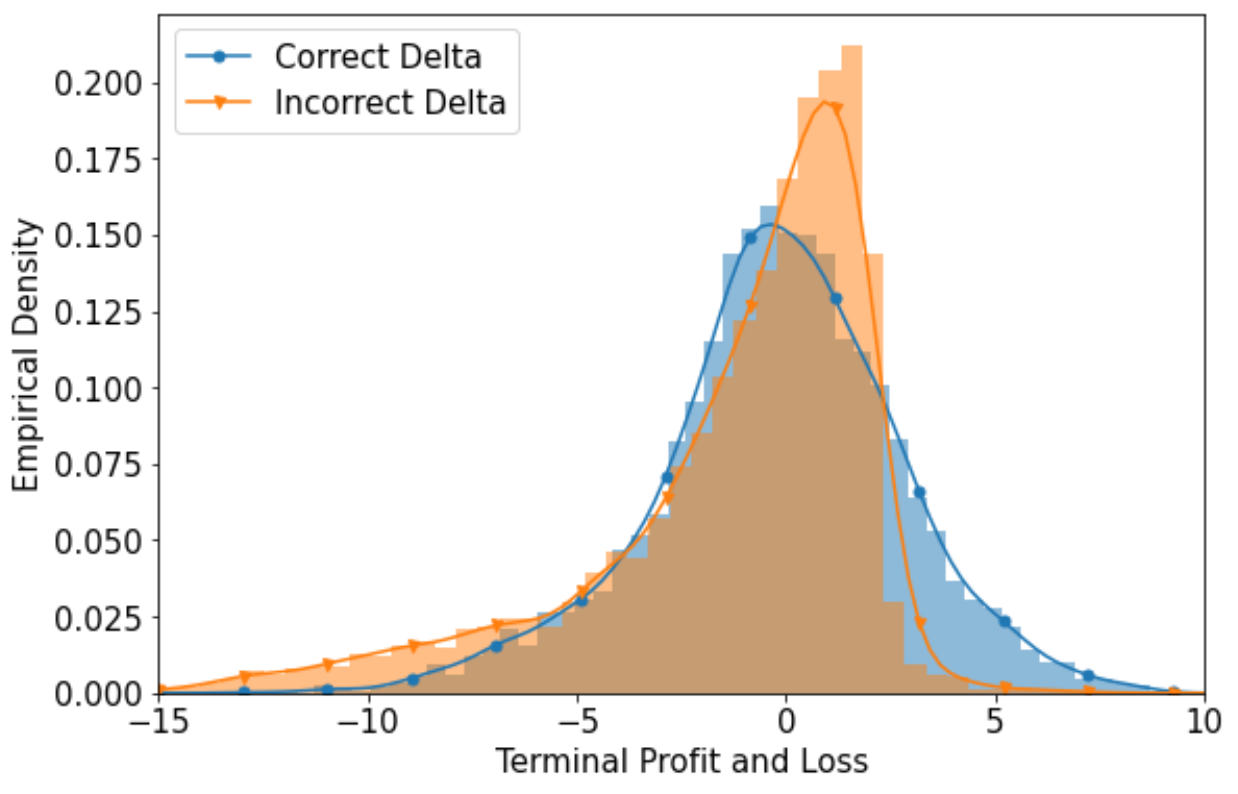}  
\caption{Empirical density}
\label{fig:mf_dist_motivating}
\end{subfigure}%
\begin{subfigure}[t]{.5\textwidth}
\centering
\includegraphics[width=1\linewidth]{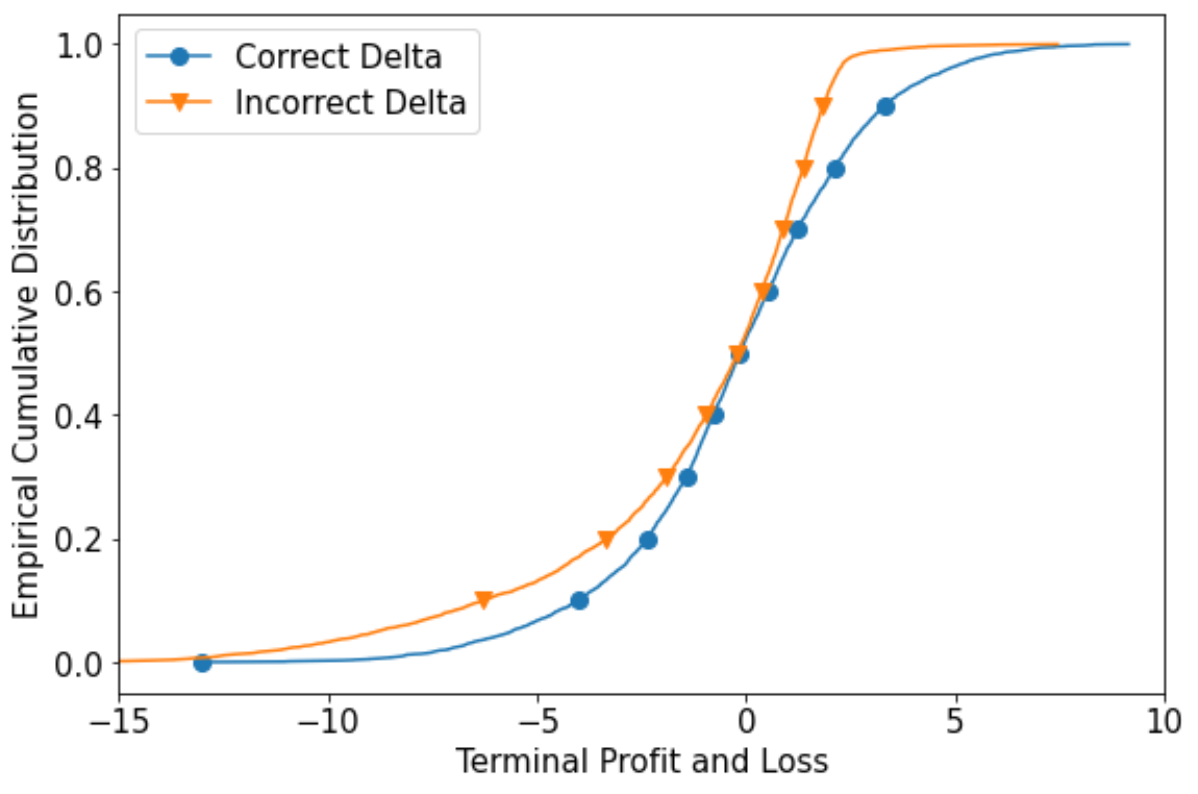}
\caption{Empirical cumulative distribution}
\label{fig:mf_cdf_motivating}
\end{subfigure}%
\caption{Empirical density and cumulative distribution functions of realized terminal P\&Ls by different Delta strategies}
\label{fig:delta_compares}
\end{figure}

To evaluate the hedging performance of the incorrect Delta strategy by the insurer in the market environment for the variable annuity of contract characteristics in Table \ref{table:params_contrac}, $5000$ market scenarios using the parameters in Table \ref{table:param_financial_actuarial_market} are simulated to realize terminal P\&Ls. For comparison, the terminal P\&Ls by the correct Delta hedging strategy are also obtained. Figure \ref{fig:delta_compares} shows the empirical density and cumulative distribution functions of the $5000$ realized terminal P\&Ls by each Delta hedging strategy, while Table \ref{table:only_delta_stats} outlines the summary statistics of the empirical distributions, in which $\widehat{\text{RMSE}}$ is the estimated RMSE of the terminal P\&L similar to \eqref{eq:hedging}.

In Figure \ref{fig:mf_dist_motivating}, the empirical density function of realized terminal P\&Ls by the incorrect Delta hedging strategy is depicted to be more heavy-tailed on the left than that by the correct Delta strategy. In fact, the terminal P\&L by the incorrect Delta hedging strategy is stochastically dominated by that by the correct Delta strategy in the first-order; see Figure \ref{fig:mf_cdf_motivating}. Table \ref{table:only_delta_stats} shows that the terminal P\&L by the incorrect Delta hedging strategy has a mean and a median farther from zero, a higher standard deviation, larger left-tail risks in terms of Value-at-Risk and Tail Value-at-Risk, and a larger RMSE than that by the correct Delta strategy.

These observations conclude that, even in a market environment as simple as the BS and the CFM, the incorrect Delta hedging strategy based on the miscalibrated parameters by the insurer does not perform well when it is being implemented forwardly. In general, the hedging performance of model-based approaches depends crucially on the calibration of parameters for the model of the market environment.



\begin{table}[H]
\centering
\begin{tabular}{@{}rccccccccc@{}}
\toprule
Terminal P\&L of & \multirow{2}{*}{Mean} & \multirow{2}{*}{Median} & \multirow{2}{*}{Std. Dev.} & \multirow{2}{*}{$\text{VaR}_{90}$} & \multirow{2}{*}{$\text{VaR}_{95}$} & \multirow{2}{*}{$\text{TVaR}_{90}$} & \multirow{2}{*}{$\text{TVaR}_{95}$} & \multirow{2}{*}{$\widehat{\text{RMSE}}$}
\\
Hedging Strategy  & &  &  &  &  &  &  & 
\\\midrule
Correct Delta  & $-0.24$ &  $-0.14$ & $2.96$ & $-4.00$ & $-5.59$ & $-5.99$ & $-7.22$ & $2.97$ \\
Incorrect Delta  & $-1.25$ &  $-0.22$ & $3.41$ & $-6.27$ & $-8.80$ & $-9.24$ & $-11.05$ & $3.63$ \\
\bottomrule                     
\end{tabular}
\caption{Summary statistics of empirical distributions of realized terminal P\&Ls by different Delta strategies}
\label{table:only_delta_stats}
\end{table}

\subsection{Two-Phase Reinforcement Learning Approach}\label{sec:motivation_revisited}

In an RL approach, at the current time $0$, the insurer builds an RL agent to hedge on her behalf in the future. The {\it agent interacts} with a market {\it environment}, by sequentially {\it observing states}, {\it taking}, as well as {\it revising}, {\it actions}, which are the hedging strategies, and {\it collecting rewards}. Without possessing any prior knowledge of the market environment, the agent needs to, {\it explore} the environment while {\it exploit} the collected reward signals, for effective learning.



An intuitive proposition would be allowing an infant RL agent to learn directly from such market environment, like the one in Section \ref{sec:pit_revisit}, moving forward. However, recall that the insurer actually does not know any exact market dynamics in the environment and thus is not able to provide any theoretical model for the net liability to the RL agent. In turn, the RL agent could not receive any sequential anchor-hedging reward signal in \eqref{eq:reward_1} from the environment, but instead receives the single terminal reward signal in \eqref{eq:reward_2}. Since the rewards, except the terminal one, are all zero, the infant RL agent would learn ineffectively from such sparse rewards, i.e. the RL agent shall take a tremendous amount of time to finally learn a nearly optimal hedging strategy in the environment. Most importantly, while the RL agent is exploring and learning from the environment, which is not a simulated one, the insurer could suffer from huge financial burden due to any sub-optimal hedging performances.


In view of this, we propose that the insurer should first designate the infant RL agent to interact and learn from a training environment, which is constructed by the insurer based on her best knowledge of the market, for example, the model of the market environment in Section \ref{sec:pit_revisit}. Since the training environment is known to the insurer (but is unknown to the RL agent), the RL agent can be supplied by a net liability theoretical model, and consequently learn from the sequential anchor-hedging reward signal in \eqref{eq:reward_1} of the training environment. Therefore, the infant RL agent would be guided by the net liability to learn effectively from the local hedging errors. After interacting and learning from the training environment for a period of time, in order to gauge the effectiveness, the RL agent shall be tested for its hedging performance in simulated scenarios from the same training environment. This first phase is called the {\it training phase}.

\vspace{3mm}
\noindent
{\bf Training Phase}:
\vspace{0.8mm}
\begin{itemize}
\item[(i)] The insurer constructs the MDP training environment.
\vspace{-2mm}
\item[(ii)] The insurer builds the infant RL agent which uses the PPO algorithm.
\vspace{-2mm}
\item[(iii)] The insurer assigns the RL agent in the MDP training environment to interact and learn for a period of time, during which the RL agent collects the anchor-hedging reward signal in \eqref{eq:reward_1}.
\vspace{-2mm}
\item[(iv)] The insurer deploys the trained RL agent to hedge in simulated scenarios from the same training environment and documents the baseline hedging performance.
\end{itemize}

If the hedging performance of the trained RL agent in the training environment is satisfactory, the insurer should then proceed to assign it to interact and learn from the market environment. Since the training and market environments are usually different, such as having different parameters as in Section \ref{sec:pit_revisit}, the initial hedging performance of the trained RL agent in the market environment is expected to diverge from the fine baseline hedging performance in the training environment. However, different from an infant RL agent, the trained RL agent is experienced so that the sparse reward signal in \eqref{eq:reward_2} should be sufficient for the agent to revise the hedging strategy, from the nearly optimal one in the training environment to that in the market environment, within a reasonable amount of time. This second phase is called the {\it online learning phase}.

\vspace{3mm}
\noindent
{\bf Online Learning Phase}:
\vspace{0.8mm}
\begin{itemize}
\item[(v)] The insurer assigns the RL agent in the market environment to interact and learn in real time, during which the RL agent collects the single terminal reward signal in \eqref{eq:reward_2}.
\end{itemize}

\begin{figure}[ht!]
\begin{subfigure}{1\linewidth}
\centering
\begin{tikzpicture}[scale=0.95, transform shape]
\node [draw,
    minimum width=2.5cm,
    minimum height=1.5cm,
]  (insurer) {Insurer};
\node [draw,
    minimum width=2.5cm, 
    minimum height=1.5cm,
    right=2.5cm of insurer
] (agent) {RL Agent};
\node [draw,
    minimum width=2.5cm, 
    minimum height=1.5cm, 
    below right= 1.5cm and -1.7cm of insurer
]  (MDP) {MDP Training Environment};
\draw [-latex](insurer) -- (MDP)
node[midway,left]{(i) construct};
\draw [-latex](insurer) -- (agent)
node[midway,above]{(ii) build};
\draw [-latex](MDP) -- ([xshift=-0.5cm]agent.south);
\node[] at (5.6,-1.4){(iii) interact and};
\node[] at (5.35,-1.8){learn};
\draw [-latex](agent) -- ([xshift=0.5cm]MDP.north);

\draw[dotted] (7.5,-4) -- (7.5,1);

\node [draw,
    minimum width=2.5cm, 
    minimum height=1.5cm,
    right=3cm of agent
] (agent_2) {RL Agent};
\node [draw,
    minimum width=2.5cm, 
    minimum height=1.5cm, 
    below = 1.5cm of agent_2
]  (MDP_e) {MDP Training Environment};
\draw [-latex]([xshift=-0.1cm]MDP_e.north) -- ([xshift=-0.1cm]agent_2.south);

\draw [-latex]([xshift=0.1cm]agent_2.south) -- ([xshift=0.1cm]MDP_e.north)
node[midway, right]{{\parbox{5cm}{(iv) hedge and \\ {\color{white}\;\;\;\;\;\;\;}realize performance}}};

\node[] at (7.1,1.5)
{\textbf{Training Phase}};
\end{tikzpicture}
\caption{Training phase}
\label{fig:relation_a}
\end{subfigure}

\par\bigskip
\begin{subfigure}{1\linewidth}
\centering
\begin{tikzpicture}[scale=0.95, transform shape]
\node [draw,
    minimum width=2.5cm, 
    minimum height=1.5cm
] (agent_3) at (16.5, 0)
{RL Agent};
\node [draw,
    minimum width=2.5cm, 
    minimum height=1.5cm, 
    below = 1.5cm of agent_3
]  (MDP_e) {Market Environment};
\draw [-latex]([xshift=-0.1cm]MDP_e.north) -- ([xshift=-0.1cm]agent_3.south)
node[midway, left]{{\color{white}{\parbox{5cm}{(v) interact and \\
{\color{white}\;\;\;\;\;\;}learn in real time}}}};
\draw [-latex]([xshift=0.1cm]agent_3.south) -- ([xshift=0.1cm]MDP_e.north)
node[midway, right]{\parbox{5cm}{(v) interact and \\
{\color{white}\;\;\;\;\;\;}learn in real time}};
\node[above = 0.4cm of agent_3]
{\textbf{Online Learning Phase}};

\end{tikzpicture}
\caption{Online learning phase}
\label{fig:relation_b}
\end{subfigure}
\caption{The relationship among insurer, RL agent, MDP training environment, and market environment of the two-phase RL approach}
\label{fig:relation}
\end{figure}
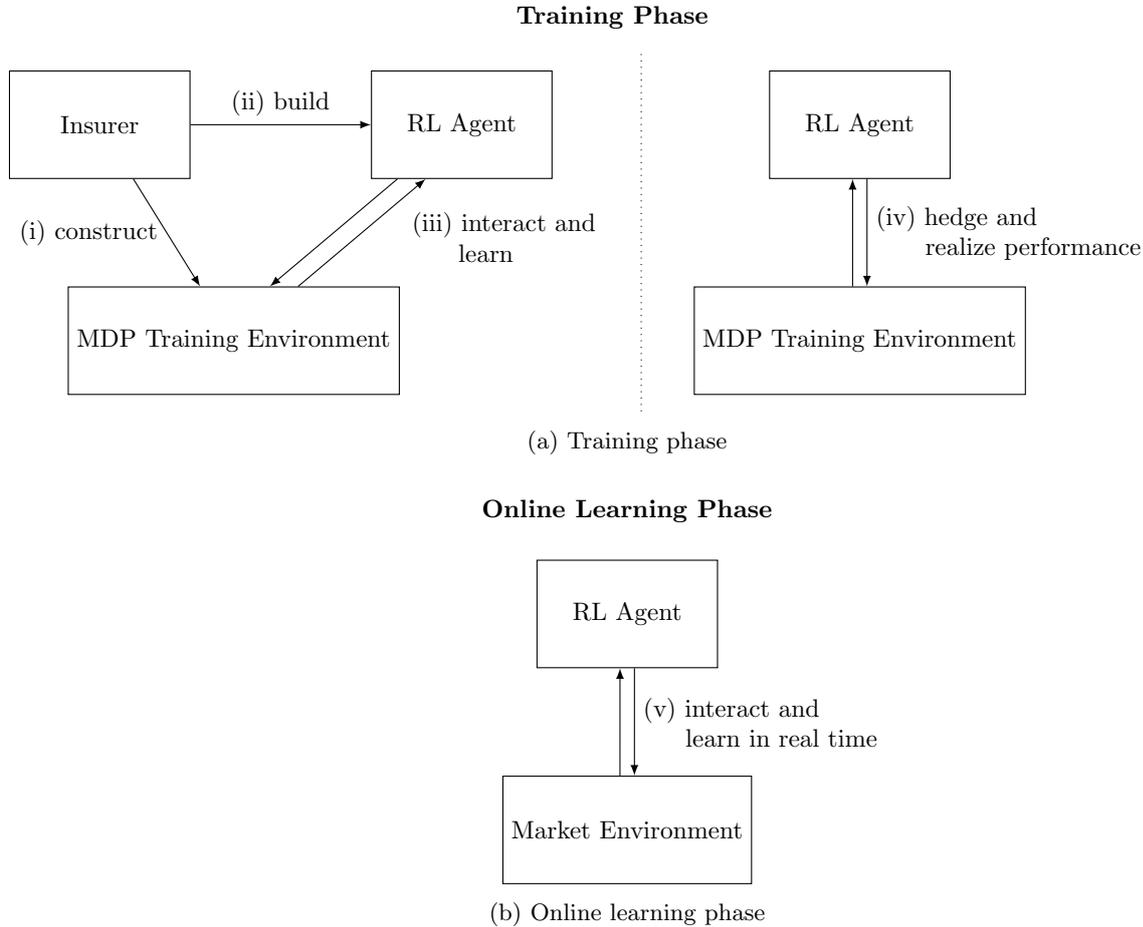

These summarize the proposed two-phase RL approach. Figure \ref{fig:relation} depicts the above sequence clearly. There are several assumptions underneath this two-phase RL approach in order to apply it effectively to a hedging problem of a contingent claim; as they involve specifics in later sections, we collate their discussions and elaborate their implications in practice in Section \ref{sec:assumption_practice}. In the following section, we shall briefly review the training essentials of RL in order to introduce the PPO algorithm. For the details of online learning phase, we defer them until Section \ref{sec:online_learning}.  




\section{Review of Reinforcement Learning}\label{sec:RL_approach}
\subsection{Stochastic Action for Exploration}\label{sec:stoc_action}
One of the fundamental ideas in RL is that, at any time $t_k$, where $k=0,1,\dots,{n-1}$, given the current state $X_{t_k}$, the RL agent does not take a deterministic action $H_{t_k}$ but extends it to a stochastic action, in order to {\it explore} the MDP environment and in turn learn from the reward signals. The stochastic action is sampled through a so-called {\it policy}, which is defined below.


Let $\mathcal{P}\left(\mathcal{A}\right)$ be a set of probability measures over the action space $\mathcal{A}$; each probability measure $\mu\left(\cdot\right)\in\mathcal{P}\left(\mathcal{A}\right)$ maps a Borel set $\overline{A}\in\mathcal{B}\left(\mathcal{A}\right)$ to $\mu\left(\overline{A}\right)\in\left[0,1\right]$. The policy $\pi\left(\cdot\right)$ is a mapping from the state space $\mathcal{X}$ to the set of probability measures $\mathcal{P}\left(\mathcal{A}\right)$; that is, for any state $x\in\mathcal{X}$, $\pi\left(x\right)=\mu\left(\cdot\right)\in\mathcal{P}\left(\mathcal{A}\right)$. The value function and the optimal value function, at any time $t_k$, where $k=0,1,\dots,\tilde{n}-1$, with the state $x\in\mathcal{X}$, are then generalized as, for any policy $\pi\left(\cdot\right)$,
\begin{equation}
V\left(t_k,x;\pi\left(\cdot\right)\right)=\mathbb{E}\left[\sum_{l=k}^{\tilde{n}-1}R_{t_{l+1}}\Big\vert X_{t_k}=x\right],\quad V^*\left(t_k,x\right)=\sup_{\pi\left(\cdot\right)}V\left(t_k,x;\pi\left(\cdot\right)\right);
\label{eq:pi_value_functions}
\end{equation}
at any time $t_k$, where $k=\tilde{n},\tilde{n}+1,\dots,n-1$, with the state $x\in\mathcal{X}$, for any policy $\pi\left(\cdot\right)$, $V\left(t_k,x;\pi\left(\cdot\right)\right)=V^*\left(t_k,x\right)=0$. In particular, if $\mathcal{P}\left(\mathcal{A}\right)$ contains only all Dirac measures over the action space $\mathcal{A}$, which is the case in the DH approach of \cite{Buhler_2019} (see Appendix \ref{app:dh_agent_1} for more details), the value function and the optimal value function reduce to \eqref{eq:value_function} and \eqref{eq:optimal_value_function}. With this relaxed setting, solving the optimal hedging strategy $H^*$ boils down to finding the optimal policy $\pi^*\left(\cdot\right)$.

\subsection{Policy Approximation and Parameterization}\label{sec:pc_approxi}
As the hedging problem has the infinite action space $\mathcal{A}$, tabular solution methods for problems of finite state space and finite action space (such as Q-learning), or value function approximation methods for problems of infinite state space and finite action space (such as deep Q-learning) are not suitable. Instead, a {\it policy gradient method} is employed.

To this end, the policy $\pi\left(\cdot\right)$ is approximated and parametrized by the weights $\theta_{\text{p}}$ in an artificial neural network (ANN); in turn, denote the policy by $\pi\left(\cdot;\theta_{\text{p}}\right)$. 
The ANN $\mathcal{N}_{\text{p}}\left(\cdot;\theta_{\text{p}}\right)$ (to be defined in \eqref{eq:policy_network} below) takes a state $x\in\mathcal{X}$ as the input vector, and outputs parameters of a probability measure in $\mathcal{P}\left(\mathcal{A}\right)$. In the sequel, the set $\mathcal{P}\left(\mathcal{A}\right)$ contains all Gaussian measures (see, for example, \cite{Wang_Thaleia_2020} and \cite{Wang_2020}), in which each has a mean $c$ and a variance $d^2$, which depend on the state input $x\in\mathcal{X}$ and the ANN weights $\theta_{\text{p}}$. Therefore, for any state $x\in\mathcal{X}$, 
\begin{equation*}
\pi\left(x;\theta_{\text{p}}\right)=\mu\left(\cdot;\theta_{\text{p}}\right)\sim\text{Gaussian}\left(c\left(x;\theta_{\text{p}}\right),d^2\left(x;\theta_{\text{p}}\right)\right),
\end{equation*}
where $\left(c\left(x;\theta_{\text{p}}\right),d^2\left(x;\theta_{\text{p}}\right)\right)=\mathcal{N}_{\text{p}}\left(x;\theta_{\text{p}}\right)$.

With such approximation and parameterization, solving the optimal policy $\pi^*$ further boils down to finding the optimal ANN weights $\theta^*_{\text{p}}$. Hence, denote the value function and the optimal value function in \eqref{eq:pi_value_functions} by $V\left(t_k,x;\theta_{\text{p}}\right)$ and $V\left(t_k,x;\theta^*_{\text{p}}\right)$, for any $t_k$, where $k=0,1,\dots,\tilde{n}-1$, with $x\in\mathcal{X}$. However, the (optimal) value function still depends on the objective probability measure $\mathbb{P}$, the financial market dynamics, and the mortality model, which are unknown to the RL agent. Before formally introducing the policy gradient methods to tackle this issue, we shall first explicitly construct the ANNs for the approximated policy, as well as for an estimate of the value function (to prepare the algorithm of policy gradient method to be reviewed below).

\subsection{Network Architecture}\label{sec:network_art}
As alluded above, in this paper, the ANN involves two parts, which are the policy network and the value function network.

\subsubsection{Policy Network}\label{sec:pc_network}
Let $N_{\text{p}}$ be the number of layers for the policy network. For $l=0,1,\dots,N_{\text{p}}$, let $d_{\text{p}}^{\left(l\right)}$ be the dimension of the $l$-th layer, where the $0$-th layer is the input layer; the $1,2,\dots,\left(N_{\text{p}}-1\right)$-th layers are hidden layers; the $N_{\text{p}}$-th layer is the output layer. In particular, $d_{\text{p}}^{\left(0\right)}=p$, which is the number of features in the actuarial and financial parts, and $d_{\text{p}}^{\left(N_{\text{p}}\right)}=2$, which outputs the mean $c$ and the variance $d^2$ of the Gaussian measure. The policy network $\mathcal{N}_{\text{p}}:\mathbb{R}^{p}\rightarrow\mathbb{R}^{2}$ is defined as, for any $x\in\mathbb{R}^{p}$,
\begin{equation}
\mathcal{N}_{\text{p}}\left(x\right)=\left(W_{\text{p}}^{\left(N_{\text{p}}\right)}\circ\psi\circ W_{\text{p}}^{\left(N_{\text{p}}-1\right)}\circ\psi\circ W_{\text{p}}^{\left(N_{\text{p}}-2\right)}\circ\dots\circ\psi\circ W_{\text{p}}^{\left(1\right)}\right)\left(x\right),
\label{eq:policy_network}
\end{equation}
where, for $l=1,2,\dots,N_{\text{p}}$, the mapping $W_{\text{p}}^{\left(l\right)}:\mathbb{R}^{d_{\text{p}}^{\left(l-1\right)}}\rightarrow\mathbb{R}^{d_{\text{p}}^{\left(l\right)}}$ is affine, and the mapping $\psi:\mathbb{R}^{d_{\text{p}}^{\left(l\right)}}\rightarrow\mathbb{R}^{d_{\text{p}}^{\left(l\right)}}$ is a componentwise activation function. Let $\theta_{\text{p}}$ be the parameter vector of the policy network; in turn, denote the policy network in \eqref{eq:policy_network} by $\mathcal{N}_{\text{p}}\left(x;\theta_{\text{p}}\right)$, for any $x\in\mathbb{R}^p$.

\subsubsection{Value Function Network}\label{sec:value_network}
The value function network is constructed similarly as in the policy network, except that all subscripts p (policy) are replaced by v (value). In particular, the value function network $\mathcal{N}_{\text{v}}:\mathbb{R}^{p}\rightarrow\mathbb{R}$ is defined as, for any $x\in\mathbb{R}^{p}$,
\begin{equation}
\mathcal{N}_{\text{v}}\left(x\right)=\left(W_{\text{v}}^{\left(N_{\text{v}}\right)}\circ\psi\circ W_{\text{v}}^{\left(N_{\text{v}}-1\right)}\circ\psi\circ W_{\text{v}}^{\left(N_{\text{v}}-2\right)}\circ\dots\circ\psi\circ W_{\text{v}}^{\left(1\right)}\right)\left(x\right),
\label{eq:value_function_network}
\end{equation}
which models an approximated value function $\hat{V}$ (see Section \ref{sec:PPO} below). Let $\theta_{\text{v}}$ be the parameter vector of the value function network; in turn, denote the value function network in \eqref{eq:value_function_network} by $\mathcal{N}_{\text{v}}\left(x;\theta_{\text{v}}\right)$, for any $x\in\mathbb{R}^p$.

\subsubsection{Shared Layers Structure}\label{sec:share_layer}
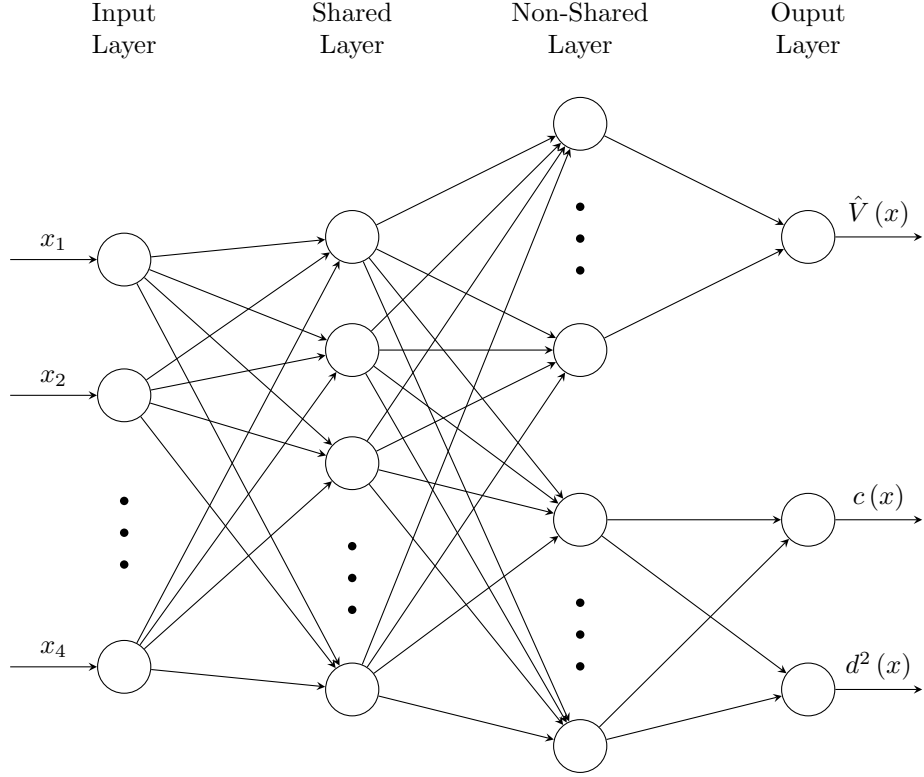
\begin{figure}[h!]
\centering
\tikzset{%
every neuron/.style={
circle,
draw,
minimum size=0.7cm
},
neuron missing/.style={
draw=none, 
scale=3,
text height=0.333cm,
execute at begin node=\color{black}$\vdots$
},
}

\begin{tikzpicture}[x=1.5cm, y=1.5cm, >=stealth]

\foreach \m [count=\y] in {1,2,missing,3}
\node [every neuron/.try, neuron \m/.try] (input-\m) at (0,2.5- 1.2*\y) {};

\foreach \m [count=\y] in {1,2,3,missing,4}
\node [every neuron/.try, neuron \m/.try ] (shidden-\m) at (2,2.5-\y) {};

\foreach \m [count=\y] in {1,missing,2}
\node [every neuron/.try, neuron \m/.try ] (i1hidden-\m) at (4,0-\y) {};

\foreach \m [count=\y] in {1,missing,2}
\node [every neuron/.try, neuron \m/.try ] (i2hidden-\m) at (4,3.5-\y) {};

\foreach \m [count=\y] in {1}
\node [every neuron/.try, neuron \m/.try ] (voutput-\m) at (6,2.5-\y) {};

\foreach \m [count=\y] in {1}
\node [every neuron/.try, neuron \m/.try ] (poutput-\m) at (6,0-\y) {};

\foreach \m [count=\y] in {1}
\node [every neuron/.try, neuron \m/.try ] (p2output-\m) at (6,-1.5-\y) {};
\foreach \l [count=\i] in {1, 2, 4}
\draw [<-] (input-\i) -- ++(-1,0)
node [above, midway] {$x_\l$};

\foreach \l [count=\i] in {1}
\draw [->] (voutput-\i) -- ++(1,0)
node [above, midway] {$\hat{V}\left(x\right)$};

\foreach \l [count=\i] in {1}
\draw [->] (poutput-\i) -- ++(1,0)
node [above, midway] {$c\left(x\right)$};

\foreach \l [count=\i] in {2}
\draw [->] (p2output-\i) -- ++(1,0)
node [above, midway] {$d^2\left(x\right)$};

\foreach \i in {1,...,3}
\foreach \j in {1,...,4}
\draw [->] (input-\i) -- (shidden-\j);

\foreach \i in {1,...,4}
\foreach \j in {1,...,2}
\draw [->] (shidden-\i) -- (i1hidden-\j);

\foreach \i in {1,...,4}
\foreach \j in {1,...,2}
\draw [->] (shidden-\i) -- (i2hidden-\j);

\foreach \i in {1,...,2}
\foreach \j in {1}
\draw [->] (i1hidden-\i) -- (poutput-\j);

\foreach \i in {1,...,2}
\foreach \j in {1}
\draw [->] (i1hidden-\i) -- (p2output-\j);

\foreach \i in {1,...,2}
\foreach \j in {1}
\draw [->] (i2hidden-\i) -- (voutput-\j);

\foreach \l [count=\x from 0] in {Input, Shared, Non-Shared, Ouput}
\node [align=center, above] at (\x*2,3) {\l \\ Layer};

\end{tikzpicture}
\caption{An example of policy and value function artificial neural networks with a shared hidden layer and a non-shared hidden layer}
\label{fig:structure}
\end{figure}

Since the policy and value function networks should extract features from the input state vector in a similar manner, they are assumed to share the first few layers. More specifically, let $N_{\text{s}}\left(<\min\left\{N_{\text{p}},N_{\text{v}}\right\}\right)$ be the number of shared layers for the policy and value function networks; for $l=1,2,\dots,N_{\text{s}}$, $W_{\text{p}}^{\left(l\right)}=W_{\text{v}}^{\left(l\right)}=W_{\text{s}}^{\left(l\right)}$, and hence, for any $x\in\mathbb{R}^{p}$,
\begin{equation*}
\mathcal{N}_{\text{p}}\left(x;\theta_{\text{p}}\right)=\left(W_{\text{p}}^{\left(N_{\text{p}}\right)}\circ\psi\circ W_{\text{p}}^{\left(N_{\text{p}}-1\right)}\circ\dots\circ\psi\circ W_{\text{p}}^{\left(N_{\text{s}}+1\right)}\circ\psi\circ W_{\text{s}}^{\left(N_{\text{s}}\right)}\circ\dots\circ\psi\circ W_{\text{s}}^{\left(1\right)}\right)\left(x\right),
\end{equation*}
\begin{equation*}
\mathcal{N}_{\text{v}}\left(x;\theta_{\text{v}}\right)=\left(W_{\text{v}}^{\left(N_{\text{v}}\right)}\circ\psi\circ W_{\text{v}}^{\left(N_{\text{v}}-1\right)}\circ\dots\circ\psi\circ W_{\text{v}}^{\left(N_{\text{s}}+1\right)}\circ\psi\circ W_{\text{s}}^{\left(N_{\text{s}}\right)}\circ\dots\circ\psi\circ W_{\text{s}}^{\left(1\right)}\right)\left(x\right).
\end{equation*}
Let $\theta$ be the parameter vector of the policy and value function networks. Figure \ref{fig:structure} depicts such a shared layers structure.



\subsection{Proximal Policy Optimization: A Temporal-Difference Policy Gradient Method}\label{sec:PPO}
A policy gradient method entails that, starting from initial ANN weights $\theta^{\left(0\right)}$, and via interacting with the MDP environment to observe the states and collect the reward signals, the RL agent gradually updates the ANN weights, by the {\it (stochastic) gradient ascent} on a certain {\it surrogate performance measure} defined for the ANN weights. That is, at each update step $u=1,2,\dots$,
\begin{equation}\label{eq:sga}
\theta^{\left(u\right)}=\theta^{\left(u-1\right)}+\alpha\widehat{\nabla_{\theta}\mathcal{J}^{\left(u-1\right)}\left(\theta^{\left(u-1\right)}\right)},
\end{equation}
where the hyperparameter $\alpha\in\left[0,1\right]$ is the learning rate of the RL agent, and, based on the experienced episode(s), $\widehat{\nabla_{\theta}\mathcal{J}^{\left(u-1\right)}\left(\theta^{\left(u-1\right)}\right)}$ is the estimated gradient of the surrogate performance measure $\mathcal{J}^{\left(u-1\right)}\left(\cdot\right)$ evaluating at $\theta=\theta^{\left(u-1\right)}$.

REINFORCE, which is pioneered by \cite{Williams_1992}, is a Monte Carlo policy gradient method, which updates the ANN weights by each episode. As this paper applies a temporal-difference (TD) policy gradient method, we relegate the review of REINFORCE to Appendix \ref{sec:REINFORCE}, where the {\it Policy Gradient Theorem}, the foundation of any policy gradient methods, is presented.

PPO, which is pioneered by \cite{Schulman_2017}, is a TD policy gradient method, which updates the ANN weights by a batch of $K\in\mathbb{N}$ realizations. At each update step $u=1,2,\dots$, based on the ANN weights $\theta^{\left(u-1\right)}$, and thus the policy $\pi\left(\cdot;\theta_{\text{p}}^{\left(u-1\right)}\right)$, the RL agent experiences $E^{\left(u\right)}\in\mathbb{N}$ realized episodes for the $K$ realizations.
\begin{itemize}
\item If $E^{\left(u\right)}=1$, the episode is given by
\begin{align*}
&\;\left\{\dots,x_{t_{K_s^{\left(u\right)}}}^{\left(u-1\right)},h_{t_{K_s^{\left(u\right)}}}^{\left(u-1\right)},x_{t_{K_s^{\left(u\right)}+1}}^{\left(u-1\right)},r_{t_{K_s^{\left(u\right)}+1}}^{\left(u-1\right)},h_{t_{K_s^{\left(u\right)}+1}}^{\left(u-1\right)},\right.\\&\;\left.\quad\dots,x_{t_{K_s^{\left(u\right)}+K-1}}^{\left(u-1\right)},r_{t_{K_s^{\left(u\right)}+K-1}}^{\left(u-1\right)},h_{t_{K_s^{\left(u\right)}+K-1}}^{\left(u-1\right)},x_{t_{K_s^{\left(u\right)}+K}}^{\left(u-1\right)},r_{t_{K_s^{\left(u\right)}+K}}^{\left(u-1\right)},\dots\right\},
\end{align*}
where $K_s^{\left(u\right)}=0,1,\dots,\tilde{n}-1$, such that the time $t_{K_s^{\left(u\right)}}$ is when the episode is initiated in this update, and $h_{t_k}^{\left(u-1\right)}$, for $k=0,1,\dots,\tilde{n}-1$, is the time-$t_k$ realized hedging strategy being sampled from the Gaussian distribution with the mean $c\left(x_{t_k}^{\left(u-1\right)};\theta_{\text{p}}^{\left(u-1\right)}\right)$ and the variance $d^2\left(x_{t_k}^{\left(u-1\right)};\theta_{\text{p}}^{\left(u-1\right)}\right)$; necessarily, $\tilde{n}-K_s^{\left(u\right)}\geq K$.
\item If $E^{\left(u\right)}=2,3,\dots$, the episodes are given by
\begin{align*}
&\;\left\{\dots,x_{t_{K_s^{\left(u\right)}}}^{\left(u-1,1\right)},h_{t_{K_s^{\left(u\right)}}}^{\left(u-1,1\right)},x_{t_{K_s^{\left(u\right)}+1}}^{\left(u-1,1\right)},r_{t_{K_s^{\left(u\right)}+1}}^{\left(u-1,1\right)},h_{t_{K_s^{\left(u\right)}+1}}^{\left(u-1,1\right)},\right.\\&\;\left.\quad\dots,x_{t_{\tilde{n}^{\left(1\right)}-1}}^{\left(u-1,1\right)},r_{t_{\tilde{n}^{\left(1\right)}-1}}^{\left(u-1,1\right)},h_{t_{\tilde{n}^{\left(1\right)}-1}}^{\left(u-1,1\right)},x_{t_{\tilde{n}^{\left(1\right)}}}^{\left(u-1,1\right)},r_{t_{\tilde{n}^{\left(1\right)}}}^{\left(u-1,1\right)}\right\},\\
&\;\left\{x_{t_0}^{\left(u-1,2\right)},h_{t_0}^{\left(u-1,2\right)},x_{t_1}^{\left(u-1,2\right)},r_{t_1}^{\left(u-1,2\right)},h_{t_1}^{\left(u-1,2\right)},\right.\\&\;\left.\quad\dots,x_{t_{\tilde{n}^{\left(2\right)}-1}}^{\left(u-1,2\right)},r_{t_{\tilde{n}^{\left(2\right)}-1}}^{\left(u-1,2\right)},h_{t_{\tilde{n}^{\left(2\right)}-1}}^{\left(u-1,2\right)},x_{t_{\tilde{n}^{\left(2\right)}}}^{\left(u-1,2\right)},r_{t_{\tilde{n}^{\left(2\right)}}}^{\left(u-1,2\right)}\right\},\\&\;\dots,\\
&\;\left\{x_{t_0}^{\left(u-1,E^{\left(u\right)}-1\right)},h_{t_0}^{\left(u-1,E^{\left(u\right)}-1\right)},x_{t_1}^{\left(u-1,E^{\left(u\right)}-1\right)},r_{t_1}^{\left(u-1,E^{\left(u\right)}-1\right)},h_{t_1}^{\left(u-1,E^{\left(u\right)}-1\right)},\right.\\&\;\left.\quad\dots,x_{t_{\tilde{n}^{\left(E^{\left(u\right)}-1\right)}-1}}^{\left(u-1,E^{\left(u\right)}-1\right)},r_{t_{\tilde{n}^{\left(E^{\left(u\right)}-1\right)}-1}}^{\left(u-1,E^{\left(u\right)}-1\right)},h_{t_{\tilde{n}^{\left(E^{\left(u\right)}-1\right)}-1}}^{\left(u-1,E^{\left(u\right)}-1\right)},x_{t_{\tilde{n}^{\left(E^{\left(u\right)}-1\right)}}}^{\left(u-1,E^{\left(u\right)}-1\right)},r_{t_{\tilde{n}^{\left(E^{\left(u\right)}-1\right)}}}^{\left(u-1,E^{\left(u\right)}-1\right)}\right\},\\
&\;\left\{x_{t_0}^{\left(u-1,E^{\left(u\right)}\right)},h_{t_0}^{\left(u-1,E^{\left(u\right)}\right)},x_{t_1}^{\left(u-1,E^{\left(u\right)}\right)},r_{t_1}^{\left(u-1,E^{\left(u\right)}\right)},h_{t_1}^{\left(u-1,E^{\left(u\right)}\right)},\right.\\&\;\left.\quad\dots,x_{t_{K_f^{\left(u\right)}-1}}^{\left(u-1,E^{\left(u\right)}\right)},r_{t_{K_f^{\left(u\right)}-1}}^{\left(u-1,E^{\left(u\right)}\right)},h_{t_{K_f^{\left(u\right)}-1}}^{\left(u-1,E^{\left(u\right)}\right)},x_{t_{K_f^{\left(u\right)}}}^{\left(u-1,E^{\left(u\right)}\right)},r_{t_{K_f^{\left(u\right)}}}^{\left(u-1,E^{\left(u\right)}\right)},\dots\right\},
\end{align*}
where $K_f^{\left(u\right)}=1,2,\dots,\tilde{n}^{\left(E^{\left(u\right)}\right)}$, such that the time $t_{K_f^{\left(u\right)}}$ is when the last episode is finished (but not necessarily terminated) in this update; necessarily, $\tilde{n}^{\left(1\right)}-K_s^{\left(u\right)}+\sum_{e=2}^{E^{\left(u\right)}-1}\tilde{n}^{\left(e\right)}+K_f^{\left(u\right)}=K$.
\end{itemize}
The surrogate performance measure of PPO consists of three components. In the following, fix an update step $u=1,2,\dots$.

Inspired by \cite{Schulman_2015}, in which the time-$0$ value function difference between two policies is shown to be equal to the expected advantage, together with importance sampling and KL divergence constraint reformulation, the first component in the surrogate performance measure of PPO is given by:
\begin{itemize}
\item if $E^{\left(u\right)}=1$,
\begin{equation*}
L_{\text{CLIP}}^{\left(u-1\right)}\left(\theta_{\text{p}}\right)=\mathbb{E}\left[\sum_{k=K_s^{\left(u\right)}}^{K_s^{\left(u\right)}+K-1}\min\left\{q_{t_k}^{\left(u-1\right)}\hat{A}^{\left(u-1\right)}_{\theta^{\left(u-1\right)}_{\text{p}},t_k},\text{clip}\left(q_{t_k}^{\left(u-1\right)},1-\epsilon,1+\epsilon\right)\hat{A}^{\left(u-1\right)}_{\theta^{\left(u-1\right)}_{\text{p}},t_k}\right\}\right],
\end{equation*}
where the importance sampling ratio $q_{t_k}^{\left(u-1\right)}=\frac{\phi\left(H^{\left(u-1\right)}_{t_k};X^{\left(u-1\right)}_{t_k},\theta_{\text{p}}\right)}{\phi\left(H^{\left(u-1\right)}_{t_k};X^{\left(u-1\right)}_{t_k},\theta^{\left(u-1\right)}_{\text{p}}\right)}$, in which $\phi\left(\cdot;X^{\left(u-1\right)}_{t_k},\theta_{\text{p}}\right)$ is the Gaussian density function with mean $c\left(X^{\left(u-1\right)}_{t_k};\theta_{\text{p}}\right)$ and variance $d^2\left(X^{\left(u-1\right)}_{t_k};\theta_{\text{p}}\right)$, the estimated advantage is evaluated at $\theta_{\text{p}}=\theta^{\left(u-1\right)}_{\text{p}}$ and bootstrapped through the approximated value function that
\begin{equation*}
\hat{A}^{\left(u-1\right)}_{\theta^{\left(u-1\right)}_{\text{p}},t_k}=
\begin{cases}
\sum_{l=k}^{K_s^{\left(u\right)}+K-1}R_{t_{l+1}}^{\left(u-1\right)}+\hat{V}\left(t_{K_s^{\left(u\right)}+K},X^{\left(u-1\right)}_{t_{K_s^{\left(u\right)}+K}};\theta_{\text{v}}^{\left(u-1\right)}\right)&\\\quad\quad\quad\quad\quad\quad\quad\quad\quad\quad\quad\quad\;-\hat{V}\left(t_k,X^{\left(u-1\right)}_{t_k};\theta_{\text{v}}^{\left(u-1\right)}\right)&\text{if }K_s^{\left(u\right)}+K<\tilde{n},\\
\sum_{l=k}^{\tilde{n}-1}R_{t_{l+1}}^{\left(u-1\right)}-\hat{V}\left(t_k,X^{\left(u-1\right)}_{t_k};\theta_{\text{v}}^{\left(u-1\right)}\right)&\text{if }K_s^{\left(u\right)}+K=\tilde{n},\\
\end{cases}
\end{equation*}
and the function $\text{clip}\left(q_{t_k}^{\left(u-1\right)},1-\epsilon,1+\epsilon\right)=\min\left\{\max\left\{q_{t_k}^{\left(u-1\right)},1-\epsilon\right\},1+\epsilon\right\}$. The approximated value function $\hat{V}$ is given by the output of the value network, i.e. $\hat{V}\left(t_k,X^{\left(u-1\right)}_{t_k};\theta_{\text{v}}^{\left(u-1\right)}\right)=\mathcal{N}_{\text{v}}\left(X^{\left(u-1\right)}_{t_k};\theta_{\text{v}}^{\left(u-1\right)}\right)$ as defined in \eqref{eq:value_function_network} for $k=0,1,\dots,\tilde{n}-1$.
\item if $E^{\left(u\right)}=2,3,\dots$,
\begin{align*}
&\;L_{\text{CLIP}}^{\left(u-1\right)}\left(\theta_{\text{p}}\right)=\mathbb{E}\left[\sum_{k=K_s^{\left(u\right)}}^{\tilde{n}^{\left(1\right)}-1}\min\left\{q_{t_k}^{\left(u-1,1\right)}\hat{A}^{\left(u-1,1\right)}_{\theta^{\left(u-1\right)}_{\text{p}},t_k},\text{clip}\left(q_{t_k}^{\left(u-1,1\right)},1-\epsilon,1+\epsilon\right)\hat{A}^{\left(u-1,1\right)}_{\theta^{\left(u-1\right)}_{\text{p}},t_k}\right\}\right.\\
&\;\left.+\sum_{e=2}^{E^{\left(u\right)}-1}\sum_{k=0}^{\tilde{n}^{\left(e\right)}-1}\min\left\{q_{t_k}^{\left(u-1,e\right)}\hat{A}^{\left(u-1,e\right)}_{\theta^{\left(u-1\right)}_{\text{p}},t_k},\text{clip}\left(q_{t_k}^{\left(u-1,e\right)},1-\epsilon,1+\epsilon\right)\hat{A}^{\left(u-1,e\right)}_{\theta^{\left(u-1\right)}_{\text{p}},t_k}\right\}\right.\\
&\;\left.+\sum_{k=0}^{K_f^{\left(u\right)}-1}\min\left\{q_{t_k}^{\left(u-1,E^{\left(u\right)}\right)}\hat{A}^{\left(u-1,E^{\left(u\right)}\right)}_{\theta^{\left(u-1\right)}_{\text{p}},t_k},\text{clip}\left(q_{t_k}^{\left(u-1,E^{\left(u\right)}\right)},1-\epsilon,1+\epsilon\right)\hat{A}^{\left(u-1,E^{\left(u\right)}\right)}_{\theta^{\left(u-1\right)}_{\text{p}},t_k}\right\}\right].
\end{align*}
\end{itemize}

Similar to REINFORCE in Appendix \ref{sec:REINFORCE}, the second component in the surrogate performance measure of PPO minimizes the loss between the bootstrapped sum of reward signals and the approximated value function. To this end, define:
\begin{itemize}
\item if $E^{\left(u\right)}=1$,
\begin{equation*}
L_{\text{VF}}^{\left(u-1\right)}\left(\theta_{\text{v}}\right)=\mathbb{E}\left[\sum_{k=K_s^{\left(u\right)}}^{K_s^{\left(u\right)}+K-1}\left(\hat{A}^{\left(u-1\right)}_{\theta^{\left(u-1\right)}_{\text{p}},t_k}+\hat{V}\left(t_k,X^{\left(u-1\right)}_{t_k};\theta_{\text{v}}^{\left(u-1\right)}\right)-\hat{V}\left(t_k,X^{\left(u-1\right)}_{t_k};\theta_{\text{v}}\right)\right)^2\right];
\end{equation*}
\item if $E^{\left(u\right)}=2,3,\dots$,
\begin{align*}
&\;L_{\text{VF}}^{\left(u-1\right)}\left(\theta_{\text{v}}\right)=\mathbb{E}\left[\sum_{k=K_s^{\left(u\right)}}^{\tilde{n}^{\left(1\right)}-1}\left(\hat{A}^{\left(u-1,1\right)}_{\theta^{\left(u-1\right)}_{\text{p}},t_k}+\hat{V}\left(t_k,X^{\left(u-1,1\right)}_{t_k};\theta_{\text{v}}^{\left(u-1\right)}\right)-\hat{V}\left(t_k,X^{\left(u-1,1\right)}_{t_k};\theta_{\text{v}}\right)\right)^2\right.\\
&\;\left.+\sum_{e=2}^{E^{\left(u\right)}-1}\sum_{k=0}^{\tilde{n}^{\left(e\right)}-1}\left(\hat{A}^{\left(u-1,e\right)}_{\theta^{\left(u-1\right)}_{\text{p}},t_k}+\hat{V}\left(t_k,X^{\left(u-1,e\right)}_{t_k};\theta_{\text{v}}^{\left(u-1\right)}\right)-\hat{V}\left(t_k,X^{\left(u-1,e\right)}_{t_k};\theta_{\text{v}}\right)\right)^2\right.\\
&\;\left.+\sum_{k=0}^{K_f^{\left(u\right)}-1}\left(\hat{A}^{\left(u-1,E^{\left(u\right)}\right)}_{\theta^{\left(u-1\right)}_{\text{p}},t_k}+\hat{V}\left(t_k,X^{\left(u-1,E^{\left(u\right)}\right)}_{t_k};\theta_{\text{v}}^{\left(u-1\right)}\right)-\hat{V}\left(t_k,X^{\left(u-1,E^{\left(u\right)}\right)}_{t_k};\theta_{\text{v}}\right)\right)^2\right].
\end{align*}
\end{itemize}

Finally, to encourage the RL agent exploring the MDP environment, the third component in the surrogate performance measure of PPO is the entropy bonus. Based on the Gaussian density function, define
\begin{itemize}
\item if $E^{\left(u\right)}=1$,
\begin{equation*}
L_{\text{EN}}^{\left(u-1\right)}\left(\theta_{\text{p}}\right)=\mathbb{E}\left[\sum_{k=K_s^{\left(u\right)}}^{K_s^{\left(u\right)}+K-1}\ln d\left(X^{\left(u-1\right)}_{t_k};\theta_{\text{p}}\right)\right];
\end{equation*}
\item if $E^{\left(u\right)}=2,3,\dots$,
\begin{align*}
&\;L_{\text{EN}}^{\left(u-1\right)}\left(\theta_{\text{p}}\right)=\mathbb{E}\left[\sum_{k=K_s^{\left(u\right)}}^{\tilde{n}^{\left(1\right)}-1}\ln d\left(X^{\left(u-1,1\right)}_{t_k};\theta_{\text{p}}\right)+\sum_{e=2}^{E^{\left(u\right)}-1}\sum_{k=0}^{\tilde{n}^{\left(e\right)}-1}\ln d\left(X^{\left(u-1,e\right)}_{t_k};\theta_{\text{p}}\right)\right.\\&\;\left.\quad\quad\quad\quad\quad\quad\quad\quad+\sum_{k=0}^{K_f^{\left(u\right)}-1}\ln d\left(X^{\left(u-1,E^{\left(u\right)}\right)}_{t_k};\theta_{\text{p}}\right)\right].
\end{align*}
\end{itemize}

Therefore, the surrogate performance measure of PPO is given by:
\begin{equation}
\mathcal{J}^{\left(u-1\right)}\left(\theta\right)=L_{\text{CLIP}}^{\left(u-1\right)}\left(\theta_{\text{p}}\right)-c_1L_{\text{VF}}^{\left(u-1\right)}\left(\theta_{\text{v}}\right)+c_2L_{\text{EN}}^{\left(u-1\right)}\left(\theta_{\text{p}}\right),
\label{eq:PPO_measure}
\end{equation}
where the hyperparameters $c_1,c_2\in\left[0,1\right]$ are the loss coefficients of the RL agent. Its estimated gradient, based on the $K$ realizations, is then computed via automatic differentiation; see, for example, \cite{Baydin_2018}.

\section{Illustrative Example Revisited: Training Phase}\label{sec:baseline_results}

Recall that, in the training phase, the insurer constructs a model of the market environment for an MDP training environment, while the RL agent, which does not know any specifics of this MDP environment, observes states and receives the anchor-hedging reward signals in \eqref{eq:reward_1} from it, and hence gradually learns the hedging strategy by the PPO algorithm reviewed in the last section. This section revisits the illustrative example in Section \ref{sec:pit_revisit} via the two-phase RL approach in the training phase.



\subsection{Markov Decision Process Training Environment}\label{sec:MDP_training}
The model of the market environment is the BS and the CFM in the financial and the actuarial parts. However, unlike the model following the market environment to write a single contract to a single policyholder, for effective training, the insurer writes identical contracts to $N$ homogeneous policyholders in the training environment. Because of the homogeneity of the contracts and the policyholders, for all $i=1,2,\dots,N$, $x_i=x$, $\rho^{\left(i\right)}=\rho$, $m^{\left(i\right)}=m$, $G_M^{\left(i\right)}=G_M$, $G_D^{\left(i\right)}=G_D$, $m_e^{\left(i\right)}=m_e$, and $F_t^{\left(i\right)}=F_t=\rho S_te^{-mt}$, for $t\in\left[0,T\right]$.

At any time $t\in\left[0,T\right]$, the future gross liability of the insurer accumulated to the maturity is thus $
\left(G_M-F_T\right)_+\sum_{i = 1}^{N}J_T^{\left(i\right)} +\sum_{i = 1}^{N} e^{r\left(T - T_{x}^{\left(i\right)} \right)}\left(G_D-F_{T_{x}^{\left(i\right)}}\right)_+\mathds{1}_{\{T_{x}^{\left(i\right)} < T\}}J_{t}^{\left(i\right)},$ and its time-$t$ discounted value is
\begin{align*}
V_t^{\text{GL}}&=e^{-r\left(T-t\right)}\mathbb{E}^{\mathbb{Q}}\left[\left(G_M-F_T\right)_+\sum_{i=1}^{N}J_T^{\left(i\right)}\Big\vert\mathcal{F}_t\right]+\mathbb{E}^{\mathbb{Q}}\left[\sum_{i=1}^{N}e^{-r\left(T_{x}^{\left(i\right)}-t\right)}\left(G_D-F_{T_{x}^{\left(i\right)}}\right)_+\mathds{1}_{\{T_{x}^{\left(i\right)} < T\}}J_{t}^{\left(i\right)}\Big\vert\mathcal{F}_t\right]\\
&=e^{-r\left(T-t\right)}\mathbb{E}^{\mathbb{Q}}\left[\left(G_M-F_T\right)_+\vert\mathcal{F}_t\right]\sum_{i=1}^{N}\mathbb{E}^{\mathbb{Q}}\left[J_T^{\left(i\right)}\big\vert\mathcal{F}_t\right]+\sum_{i=1}^{N}J_t^{\left(i\right)}\mathbb{E}^{\mathbb{Q}}\left[e^{-r\left(T_{x}^{\left(i\right)}-t\right)}\left(G_D-F_{T_{x}^{\left(i\right)}}\right)_+\mathds{1}_{\{T_{x}^{\left(i\right)} < T\}}\Big\vert\mathcal{F}_t\right],
\end{align*}
where the probability measure $\mathbb{Q}$ defined on $\left(\Omega,\mathcal{F}\right)$ is an equivalent martingale measure with respect to $\mathbb{P}$. Herein, the probability measure $\mathbb{Q}$ is chosen to be the product measure of each individual equivalent martingale measure in the actuarial or financial part, which implies the independence among the Brownian motion $W$ and the future lifetime $T_x^{\left(1\right)},T_x^{\left(2\right)},\dots,T_x^{\left(N\right)}$, clarifying the first term in the second equality above. The second term in that equality is due to the fact that, for $i=1,2,\dots,N$, the single-jump process $J^{\left(i\right)}$ is $\mathbb{F}$-adapted. Under the probability measure $\mathbb{Q}$, all future lifetime are identically distributed and have a CFM $\nu>0$, which are the same as those under the probability measure $\mathbb{P}$ in Section \ref{sec:pit_revisit}. Therefore, for any $i=1,2,\dots,N$, and for any $0\leq t\leq s\leq T$, the conditional survival probability $\mathbb{Q}\left(T_x^{\left(i\right)}>s\vert T_x^{\left(i\right)}>t\right)=e^{-\nu\left(s-t\right)}$
. For each policyholder $i=1,2,\dots,N$, by the independence and the Markov property, for any $0\leq t\leq s\leq T$,
\begin{equation}
\mathbb{E}^{\mathbb{Q}}\left[J_s^{\left(i\right)}\big\vert\mathcal{F}_t\right]=\mathbb{E}^{\mathbb{Q}}\left[J_s^{\left(i\right)}\big\vert J_t^{\left(i\right)}\right]=
\begin{cases}
\mathbb{Q}\left(T_x^{\left(i\right)}>s\vert T_x^{\left(i\right)}\leq t\right)=0 & \text{if}\quad T_x^{\left(i\right)}\left(\omega\right)\leq t\\
\mathbb{Q}\left(T_x^{\left(i\right)}>s\vert T_x^{\left(i\right)}>t\right)=e^{-\nu\left(s-t\right)} & \text{if}\quad T_x^{\left(i\right)}\left(\omega\right)>t\\
\end{cases}.
\label{eq:conditional_prob_J}
\end{equation}
Moreover, under the probability measure $\mathbb{Q}$, for any $t\in\left[0,T\right]$, $dF_t=\left(r-m\right)F_tdt+\sigma F_tdW_t^{\mathbb{Q}}$, where $W^{\mathbb{Q}}=\left\{W^{\mathbb{Q}}_t\right\}_{t\in\left[0,T\right]}$ is the standard Brownian motion under the probability measure $\mathbb{Q}$. Hence, the time-$t$ value of the discounted future gross liability, for $t\in\left[0,T\right]$, is given by

\begin{align*}
V_t^{\text{GL}}=&\;e^{-\nu\left(T-t\right)}\left(G_Me^{-r\left(T-t\right)}\Phi\left(-d_2\left(t,G_M\right)\right)-F_te^{-m\left(T-t\right)}\Phi\left(-d_1\left(t,G_M\right)\right)\right)\sum_{i=1}^{N}J_t^{\left(i\right)}\\
&\;+\int_{t}^{T}\left(G_De^{-r\left(T-s\right)}\Phi\left(-d_2\left(s,G_D\right)\right)-F_te^{-m\left(T-s\right)}\Phi\left(-d_1\left(s,G_D\right)\right)\right)\nu e^{-\nu\left(s-t\right)}ds\sum_{i=1}^{N}J_t^{\left(i\right)},
\end{align*}
where, for $s \in \left[0,T\right)$ and $G > 0$, $d_1\left(s,G\right)=\frac{\ln\left(\frac{F_s}{G}\right)+\left(r-m+\frac{\sigma^2}{2}\left(T-s\right)\right)}{\sigma\sqrt{T-s}}$, $d_2\left(s,G\right)=d_1\left(s,G\right)-\sigma\sqrt{T-s}$, $d_1\left(T,G\right) = \lim_{s\rightarrow T^{-}}d_1\left(s,G\right)$, $d_2\left(T,G\right) = d_1\left(T,G\right)$, and $\Phi\left(\cdot\right)$ is the standard Gaussian distribution function. Note that $\sum_{i=1}^{N}J_t^{\left(i\right)}$ represents the number of surviving policyholders at time $t\in\left[0,T\right]$.

As for the cumulative future rider charge to be collected by the insurer from any time $t\in\left[0,T\right]$ onward, it is given by $\sum_{i=1}^{N}\int_{t}^{T}m_eF_sJ_s^{\left(i\right)}e^{r(T-s)}ds$, and its time-$t$ discounted value is
\begin{equation*}
V_t^{\text{RC}}=e^{-r\left(T-t\right)}\mathbb{E}^{\mathbb{Q}}\left[\sum_{i=1}^{N}\int_{t}^{T}m_eF_sJ_s^{\left(i\right)}e^{r(T-s)}ds\Big\vert\mathcal{F}_t\right]=\sum_{i=1}^{N}\int_{t}^{T}m_ee^{-r\left(s-t\right)}\mathbb{E}^{\mathbb{Q}}\left[F_s\vert F_t\right]\mathbb{E}^{\mathbb{Q}}\left[J_s^{\left(i\right)}\big\vert J_t^{\left(i\right)}\right]ds,
\end{equation*}
where the second equality is again due to the independence and the Markov property. Under the probability measure $\mathbb{Q}$, $\mathbb{E}^{\mathbb{Q}}\left[F_s\vert F_t\right]=e^{\left(r-m\right)\left(s-t\right)}F_t$. Together with \eqref{eq:conditional_prob_J},
\begin{equation*}
V_t^{\text{RC}}=\frac{1-e^{-\left(m+\nu\right)\left(T-t\right)}}{m+\nu}m_eF_t\sum_{i=1}^{N}J_t^{\left(i\right)}.
\end{equation*}

Therefore, the time-$t$ net liability of the insurer, for $t\in\left[0,T\right]$, is given by
\begin{equation}
\begin{aligned}
L_t=V_t^{\text{GL}}-V_t^{\text{RC}}=&\;\Bigg(e^{-\nu\left(T-t\right)}\left(G_Me^{-r\left(T-t\right)}\Phi\left(-d_2\left(t,G_M\right)\right)-F_te^{-m\left(T-t\right)}\Phi\left(-d_1\left(t,G_M\right)\right)\right)\\
&\;\quad+\int_{t}^{T}\left(G_De^{-r\left(T-s\right)}\Phi\left(-d_2\left(s,G_D\right)\right)-F_te^{-m\left(T-s\right)}\Phi\left(-d_1\left(s,G_D\right)\right)\right)\nu e^{-\nu\left(s-t\right)}ds \\
&\;\quad-\frac{1-e^{-\left(m+\nu\right)\left(T-t\right)}}{m+\nu}m_eF_t\Bigg)\sum_{i=1}^{N}J_t^{\left(i\right)},
\end{aligned}
\label{eq:net_liability}
\end{equation}
which contributes parts of the reward signals in \eqref{eq:reward_1}. 
The time-$t$ value of the insurer's hedging portfolio, for $t\in\left[0,T\right]$, as in \eqref{eq:hedging_port_value}, is given by: $P_0=0$, and if $t\in\left(t_k,t_{k+1}\right]$, for some $k=0,1,\dots,n-1$,
\begin{equation}
P_t=\left(P_{t_k}-H_{t_k}S_{t_k}\right)e^{r\left(t-t_k\right)}+H_{t_k}S_{t}+m_e\int_{t_k}^{t}F_se^{r\left(t-s\right)}\sum_{i=1}^{N}J_s^{\left(i\right)}ds-\sum_{i = 1}^{N}e^{r\left(t-T_{x}^{\left(i\right)}\right)}\left(G_D-F_{T_{x}^{\left(i\right)}}\right)_+\mathds{1}_{\{t_k<T_{x}^{\left(i\right)}\leq t<T\}},
\label{eq:baseline_hedging_portfolio}
\end{equation}
which is also supplied to the reward signals in \eqref{eq:reward_1}.

At each time $t_k$, where $k=0,1,\dots,n$, the RL agent is given to observe four features from this MDP environment; these four features are summarized in the state vector
\begin{equation}\label{eq:state_vector}
X_{t_k}=\left(\ln{F_{t_k}},\frac{P_{t_k}}{N},\frac{\sum_{i=1}^{N}J^{\left(i\right)}_{t_k}}{N},T-t_k\right).
\end{equation}
The first feature is the natural logarithm of the segregated account value of the policyholder. The second feature is the hedging portfolio value of the insurer, being normalized by the initial number of policyholders. The third feature is the ratio of the number of surviving policyholders with respect to the initial number of policyholders. These features are either log-transformed or normalized to prevent the RL agent from exploring and learning from features with high variability. The last feature is the term to maturity. In particular, when either the third or the last feature first hits zero, i.e. at time $t_{\tilde{n}}$, an episode is terminated. The state space $\mathcal{X}=\mathbb{R}\times\mathbb{R}\times\left[0,1/N, 2/N,\dots, 1\right]\times\left\{0,t_1,t_2,\dots,T\right\}$.

Recall that, at each time $t_k$, where $k=0,1,\dots,\tilde{n}-1$, with the state vector \eqref{eq:state_vector} being the input, the output of the policy network in \eqref{eq:policy_network} is the mean $c\left(X_{t_k};\theta_{\text{p}}\right)$ and the variance $d^2\left(X_{t_k};\theta_{\text{p}}\right)$ of a Gaussian measure; herein, the Gaussian measure represents the distribution of the average number of shares of the risky asset being held by the insurer at the time $t_k$ for each surviving policyholder. Hence, for $k=0,1,\dots,\tilde{n}-1$, the hedging strategy $H_{t_k}$ in \eqref{eq:baseline_hedging_portfolio} is given by $H_{t_k}=\overline{H}_{t_k}\sum_{i=1}^{N}J^{\left(i\right)}_{t_k}$, where $\overline{H}_{t_k}$ is sampled from the Gaussian measure. Since the hedging strategy is assumed to be Markovian with respect to the state vector, it can be shown, albeit tedious, that the state vector, in \eqref{eq:state_vector}, and the hedging strategy together, satisfy the Markov property in \eqref{eq:markov_property}.

Also recall that the infant RL agent is trained in the MDP environment with multiple homogeneous policyholders. The RL agent should then effectively update the ANN weights $\theta$, and learn the hedging strategies, via a more direct inference on the force of mortality from the third feature in the state vector. The RL agent hedges daily, so that the difference between the consecutive discrete hedging time is $\delta t_{k}=t_{k+1}-t_{k}=\frac{1}{252}$, for $k=0,1,\dots,n-1$. In this MDP training environment, the parameters of the model are given in Table \ref{table:param_financial_actuarial}, but with $N=500$.

\subsection{Building Reinforcement Learning Agent}\label{sec:training_RL_agent}


After constructing this MDP training environment, the insurer builds the RL agent which implements the PPO, which was reviewed in Section \ref{sec:PPO}. Table \ref{sub_table:params_PPO} summarizes all hyperparameters of the implemented PPO, in which three of them are determined via grid search\footnote{The grid search was performed using the Hardware-Accelerated Learning cluster in the National Center for Supercomputing Applications; see \cite{Kindratenko_2020}.}, while the remaining two are fixed a priori since they alter the surrogate performance measure itself, and thus should not be based on grid search. Table \ref{sub_table:params_ANN} outlines the hyperparameters of the ANN architecture in Section \ref{sec:network_art}, which are all pre-specified, in which ReLU stands for Rectified Linear Unit; that is, the componentwise activation function is given by, for any $z\in\mathbb{R}$, $\psi\left(z\right)=\max\left\{z,0\right\}$.



\begin{table}[!htb]
\begin{subtable}{1\linewidth}
\centering
\caption{Hyperparameters for Proximal Policy Optimization}
\makebox[\textwidth][c]{
\begin{tabular}{lcc|lcc}
\toprule
\multicolumn{3}{c|}{\textbf{Grid-Searched}} & \multicolumn{3}{c}{\textbf{Pre-Specified}} \\
Hyperparameter & & Value & Hyperparameter & & Value  \\ \midrule
Learning rate $\alpha$ & & $0.07$  & Coefficient of value function & & \multirow{2}{*}{$0.25$}\\
Batch size $K$ & & $2048$ & approximation loss $c_1$ & & \\
Clip factor $\epsilon$ & & $0.18$ & Coefficient of entropy bonus $c_2$ & & $0.01$\\
\bottomrule
\end{tabular}}
\label{sub_table:params_PPO}
\end{subtable}%
\bigbreak
\begin{subtable}{1\linewidth}
\centering
\caption{Hyperparameters for Neural Network}
\begin{tabular}{@{}lc@{}}
\toprule
Hyperparameter & Value(s) 
\\ \midrule
Number of layers in policy network $N_{\text{p}}$ &   $6$\\
Number of layers in value function network $N_{\text{v}}$ &   $6$\\
Number of shared layers $N_{\text{s}}$ & $3$\\
Dimension of hidden layers in policy network $d_{\text{p}}^{\left(l\right)}$ &   $\left[32,64,128,64,32\right]$   \\
Dimension of hidden layers in value function network $d_{\text{v}}^{\left(l\right)}$ &   $\left[32,64,128,64,32\right]$    \\
Activation function $\psi\left(\cdot\right)$ & ReLU\\
\bottomrule   
\end{tabular}
\label{sub_table:params_ANN}
\end{subtable} 
\caption{Hyperparameters setting of Proximal Policy Optimization and neural network}
\label{table:param_rl}
\end{table}
\subsection{Training of Reinforcement Learning Agent}\label{sec:training_RL_agent_2}

With all these being set up, the insurer assigns the RL agent experiencing this MDP training environment, in order to observe the state, decide, as well as revise, the hedging strategy, and collect the anchor-hedging reward signal based on \eqref{eq:reward_1}, as much as possible. Let $\mathcal{U}\in\mathbb{N}$ be the number of update steps in the training environment on the ANN weights. Hence, the policy of the experienced RL agent is given by $\pi\left(\cdot;\theta^{\left(\mathcal{U}\right)}\right)=\pi\left(\cdot;\theta_{\text{p}}^{\left(\mathcal{U}\right)}\right)$.

\begin{figure}[H]  
\centering
\begin{subfigure}[t]{.5\textwidth}
\centering
\includegraphics[width=1\linewidth]{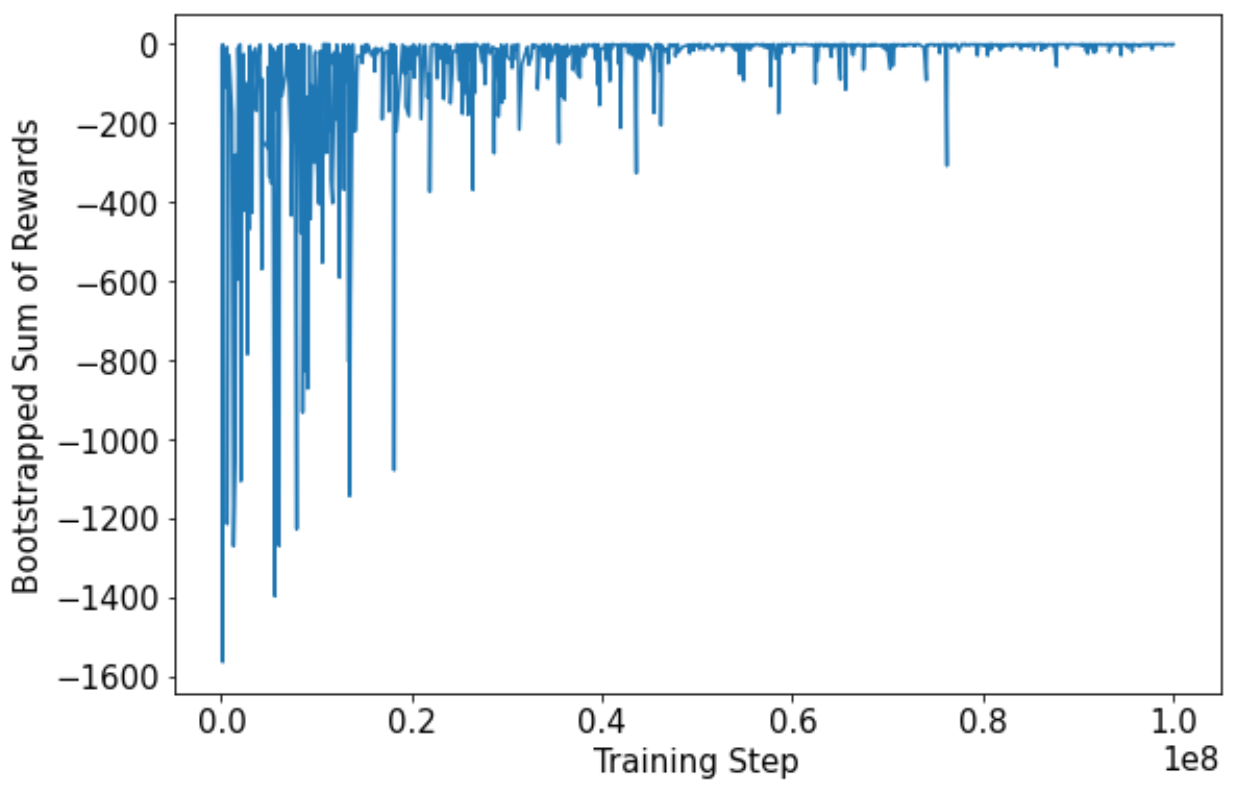}
\caption{Bootstrapped sum of rewards}
\label{fig:episode_reward}
\end{subfigure}%
\begin{subfigure}[t]{.5\textwidth}
\centering
\includegraphics[width=1\linewidth]{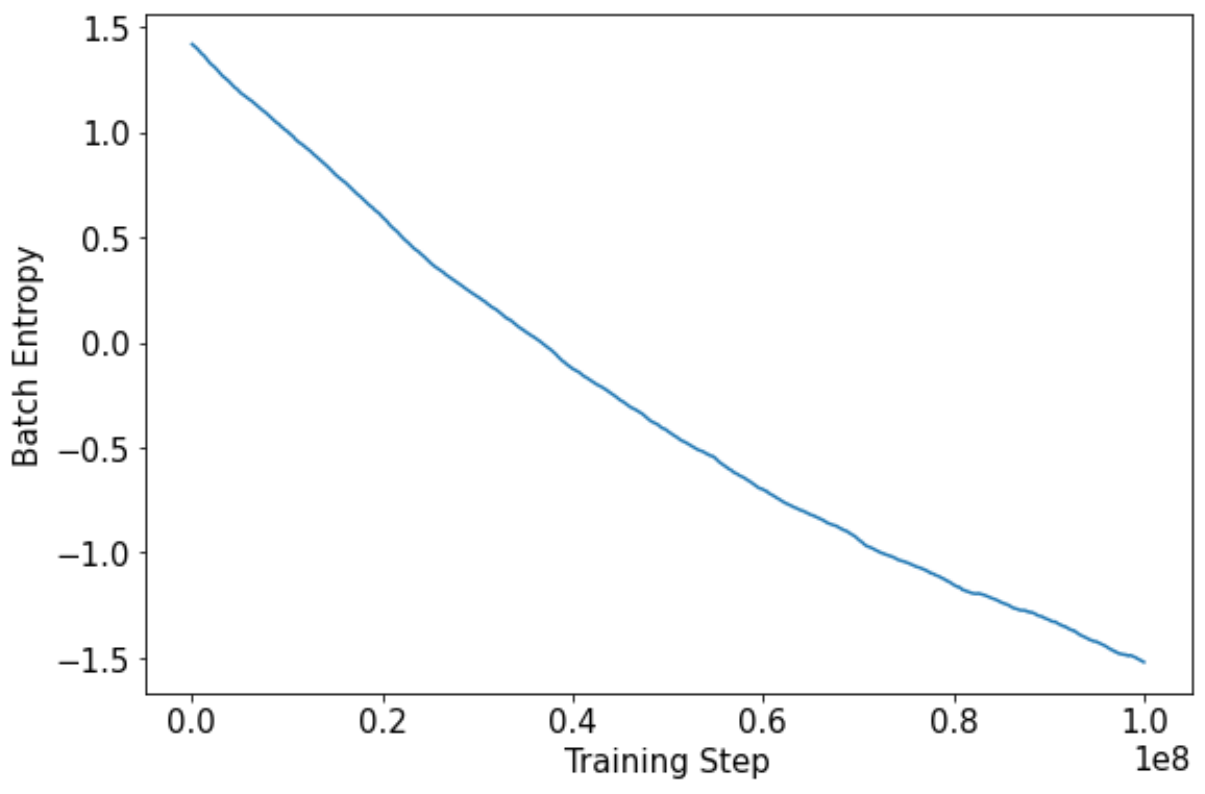}
\caption{Batch entropy}
\label{fig:entropy_loss}
\end{subfigure}%
\caption{Training log in terms of bootstrapped sum of rewards and batch entropy}
\label{fig:tensorboard}
\end{figure}

Figure \ref{fig:tensorboard} depicts the training log of the RL agent in terms of bootstrapped sum of rewards and batch entropy. In particular, Figure \ref{fig:episode_reward} shows that the value function in \eqref{eq:hedging} reduces to almost zero after around $10^8$ training timesteps, which is equivalent to around $48828$ update steps for the ANN weights; within the same number of training timesteps, Figure \ref{fig:entropy_loss} illustrates a gradual depletion on the batch entropy, and hence the Gaussian measure gently becomes more concentrating around its mean, which implies that the RL agent {\it progressively diminishes} the degree of {\it exploration} on the MDP training environment, while {\it increases} the degree of {\it exploitation} on the learned ANN weights.

\subsection{Baseline Hedging Performance}\label{sec:baseline_eval}

In the final step of the training phase, the trained RL agent is assigned to hedge in simulated scenarios from the same MDP training environment, except that $N=1$ which is in line with hedging in the market environment. The trained RL agent takes the deterministic action $c\left(\cdot;\theta_{\text{p}}^{\left(\mathcal{U}\right)}\right)$ which is the mean of the Gaussian measure.

The number of simulated scenarios is $5000$. For each scenario, the insurer documents the realized terminal P\&L, i.e. $P_{t_{\tilde{n}}}-L_{t_{\tilde{n}}}$. After all scenarios are experienced by the trained RL agent, the insurer examines the baseline hedging performance via the empirical distribution and the summary statistics of the realized terminal P\&Ls. The baseline hedging performance of the RL agent is also benchmarked with those by other methods, namely, the classical Deltas and the DH; see Appendix \ref{app:DHT} for the implemented hyperparameters of the DH training. The following four classical Deltas are implemented in the simulated scenarios from the training environment, in which the (in)correctness of the Deltas are with respect to the training environment:
\begin{itemize}
\item (correct) Delta of the CFM actuarial and BS financial models with the model parameters as in Table \ref{table:param_financial_actuarial};
\vspace{-3mm}
\item (incorrect) Delta of the increasing force of mortality (IFM) actuarial and BS financial models, where, for any $i=1,2,\dots,N$, if $T<\overline{b}$, the conditional survival probability $\mathbb{Q}\left(T_x^{\left(i\right)}>s\vert T_x^{\left(i\right)}>t\right)=\frac{\overline{b}-s}{\overline{b}-t}$, for any $0\leq t\leq s\leq T<\overline{b}$, while if $\overline{b}\leq T$, the conditional survival probability $\mathbb{Q}\left(T_x^{\left(i\right)}>s\vert T_x^{\left(i\right)}>t\right)=\frac{\overline{b}-s}{\overline{b}-t}$, for any $0\leq t\leq s<\overline{b}\leq T$, and $\mathbb{Q}\left(T_x^{\left(i\right)}>s\vert T_x^{\left(i\right)}>t\right)=0$, for any $0\leq t\leq\overline{b}\leq s\leq T$ or $0\leq\overline{b}\leq t\leq s\leq T$, with the model parameters as in Tables \ref{sub_table:params_fin} and \ref{table:params_ifm_act};
\vspace{-3mm}
\item (incorrect) Delta in the CFM actuarial and Heston financial models, where, for any $t\in\left[0,T\right]$, $dS_t=\mu S_tdt+\sqrt{\Sigma_t}S_tdW_t^{\left(1\right)}$, $d\Sigma_t=\kappa\left(\overline{\Sigma}-\Sigma_t\right)dt+\eta\sqrt{\Sigma_t}dW_t^{\left(2\right)}$, and $\left\langle W^{(1)},W^{(2)}\right\rangle_t=\phi t$, with the model parameters as in Tables \ref{sub_table:params_act} and \ref{table:params_heston_fin};
\vspace{-3mm}
\item (incorrect) Delta in the IFM actuarial and Heston financial models with the model parameters as in Tables \ref{table:params_ifm_act} and \ref{table:params_heston_fin}.
\end{itemize}

\begin{table}[!htb]
\begin{center}
\begin{tabular}{@{}lc@{}}
\toprule
Parameter & Value 
\\ \midrule
Initial number of policyholder $N$ & $1$\\
Initial age of policyholder $x$ & $20$\\
Lower bound of uniformly distributed lifetime $\underline{b}$ & $0$ \\
Upper bound of uniformly distributed lifetime $\overline{b}$ & $50$\\
Investment strategy of policyholders $\rho$ & $1.19$\\
\bottomrule   
\end{tabular}
\end{center}
\caption{Parameters setting of increasing force of mortality actuarial model for Delta}
\label{table:params_ifm_act}
\end{table}

\begin{table}[!htb]
\begin{center}
\begin{tabular}{@{}lc@{}}
\toprule
Parameter & Value  \\
\midrule
Risk-free interest rate $r$ & $0.02$\\
Risky asset initial price $S_0$ & $100$\\
Risky asset drift $\mu$ & $0.08$\\
Variance initial value $\Sigma_0$ & $0.04$ \\
Variance mean reversion rate $\kappa$ & $0.2$ \\
Variance long-run average $\overline{\Sigma}$ & $0.04$\\
Variance volatility $\eta$ & $0.1$ \\
Brownian motions correlation $\phi$ & $-0.5$\\
\bottomrule
\end{tabular}
\end{center}
\caption{Parameters setting of Heston financial model for Delta}
\label{table:params_heston_fin}
\end{table}

\begin{figure}[H]  
\centering
\begin{subfigure}[t]{.5\textwidth}
\centering
\includegraphics[width=1\linewidth]{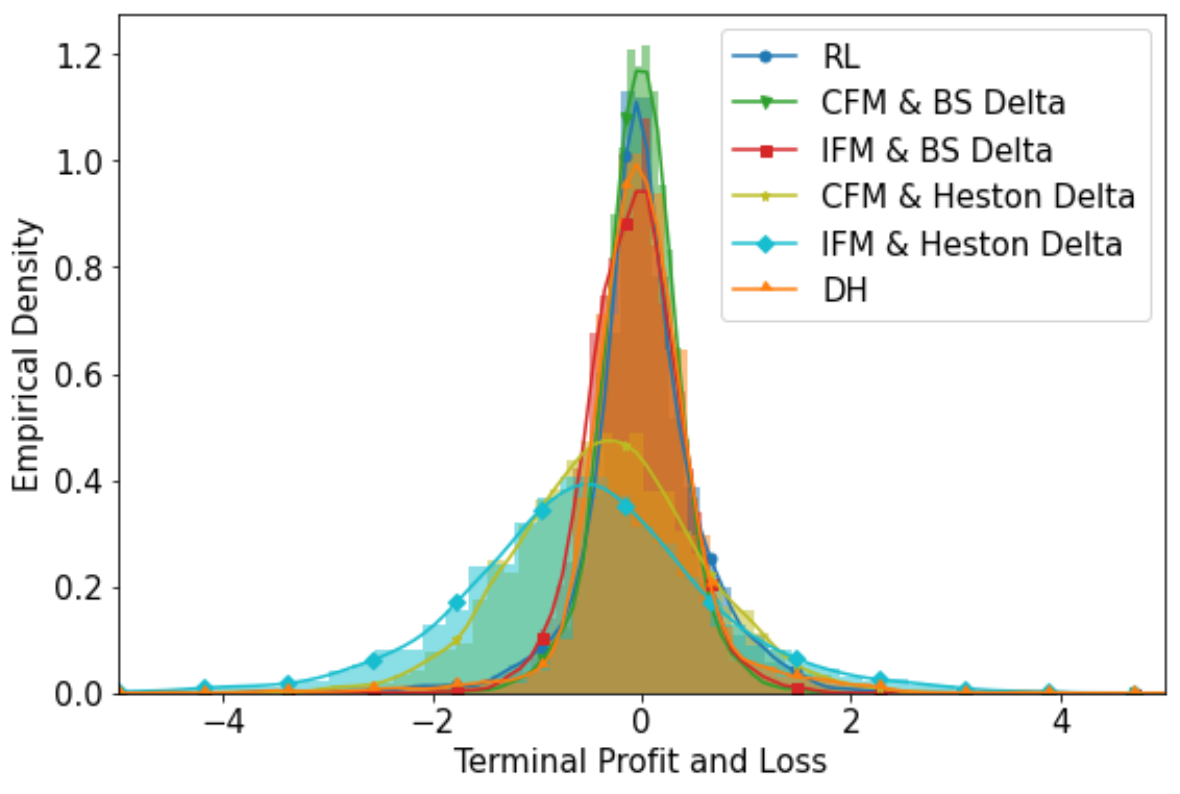}
\caption{Empirical density}
\label{fig:mf_dist}
\end{subfigure}%
\begin{subfigure}[t]{.5\textwidth}
\centering
\includegraphics[width=1\linewidth]{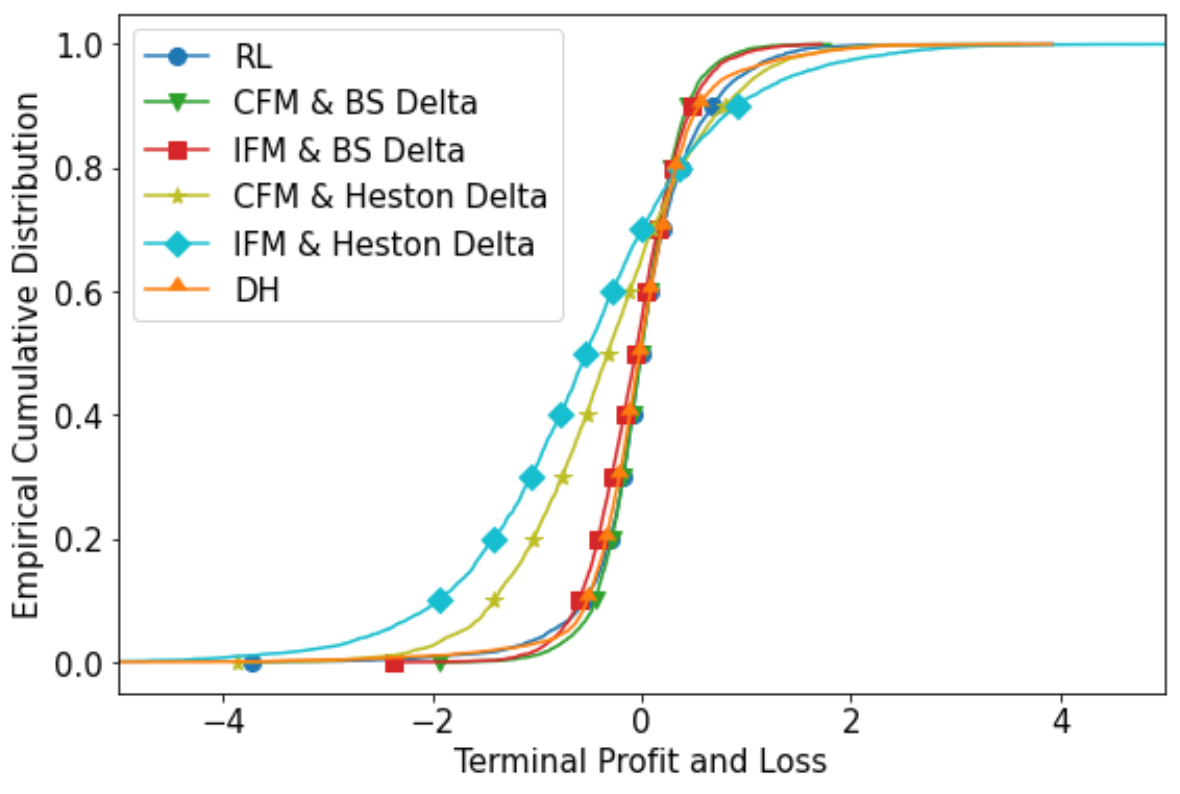}
\caption{Empirical cumulative distribution}
\label{fig:mf_cdf}
\end{subfigure}%
\caption{Empirical density and cumulative distribution functions of realized terminal P\&Ls by the approaches of reinforcement learning, classical Deltas, and deep hedging}
\label{fig:baseline_plots}
\end{figure}

\begin{figure}[H]
\centering

\begin{subfigure}{0.45\columnwidth}
\centering
\includegraphics[width=0.8\textwidth]{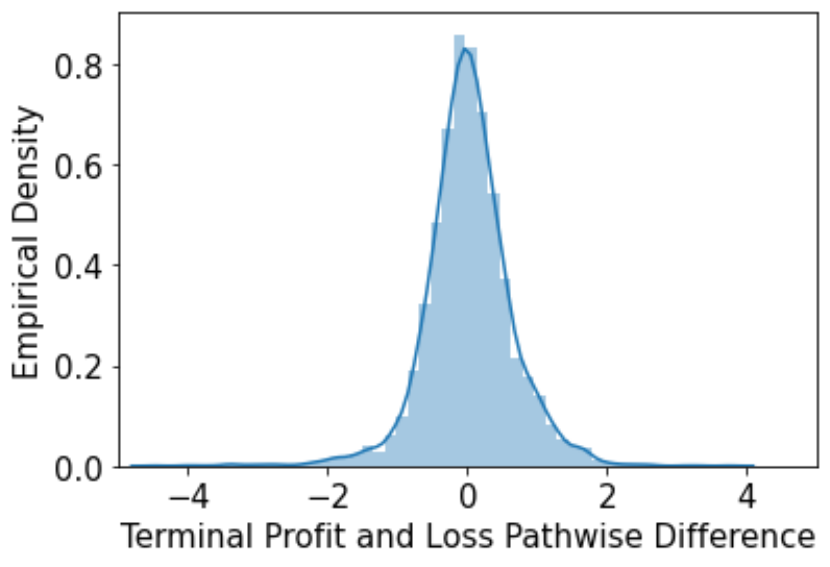}
\caption{Reinforcement learning versus Delta in constant force of mortality and Black-Scholes models}
\label{fig:true_delta}
\end{subfigure}\hfill
\begin{subfigure}{0.45\columnwidth}
\centering
\includegraphics[width=0.8\textwidth]{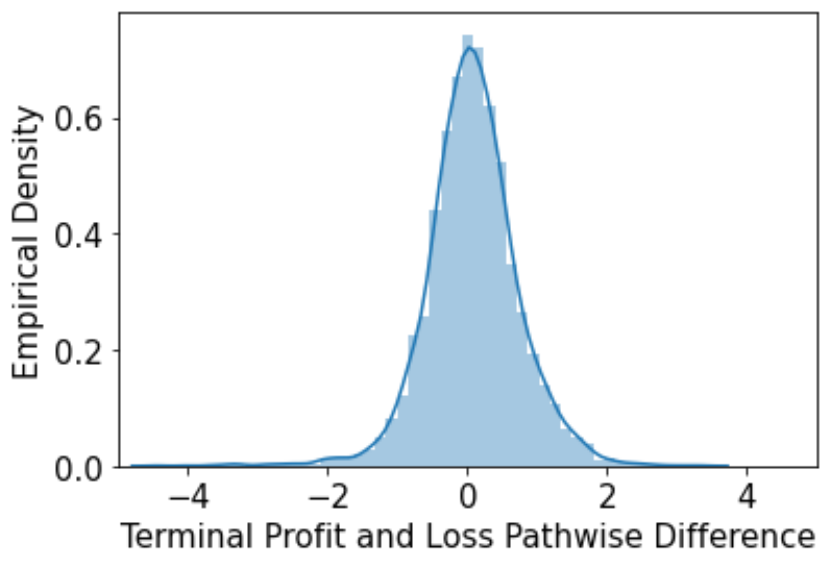}
\caption{Reinforcement learning versus Delta in increasing force of mortality and Black-Scholes models}
\label{fig:bs_incres}
\end{subfigure}

\medskip

\begin{subfigure}{0.45\columnwidth}
\centering
\includegraphics[width=0.8\textwidth]{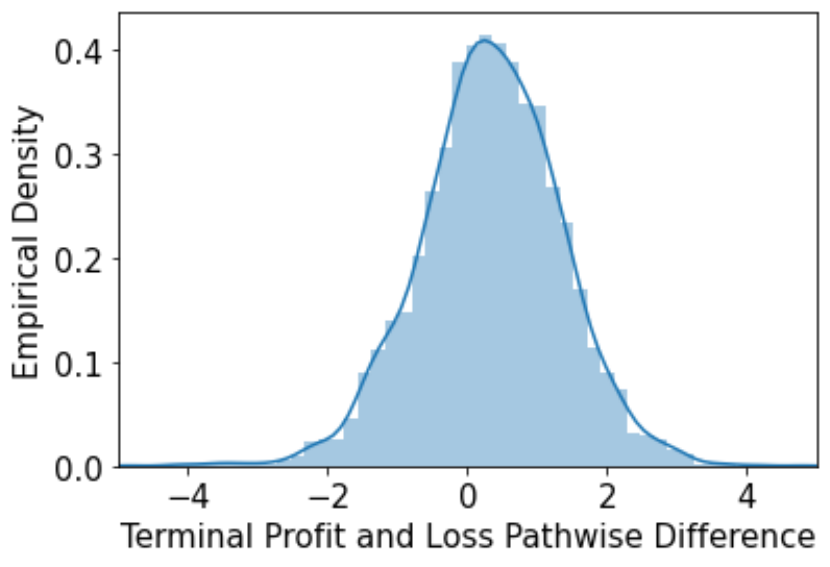}
\caption{Reinforcement learning versus Delta in constant force of mortality and Heston models}
\label{fig:heston_const}
\end{subfigure}\hfill
\begin{subfigure}{0.45\columnwidth}
\centering
\includegraphics[width=0.8\textwidth]{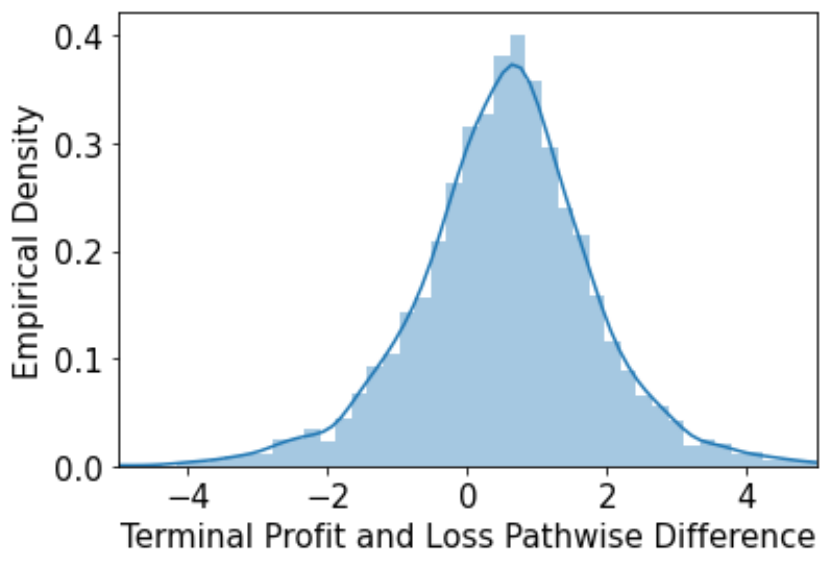}
\caption{Reinforcement learning versus Delta in increasing force of mortality and Heston models}
\label{fig:heston_uncon}
\end{subfigure}

\medskip

\begin{subfigure}{0.45\columnwidth}
\centering
\includegraphics[width=0.8\textwidth]{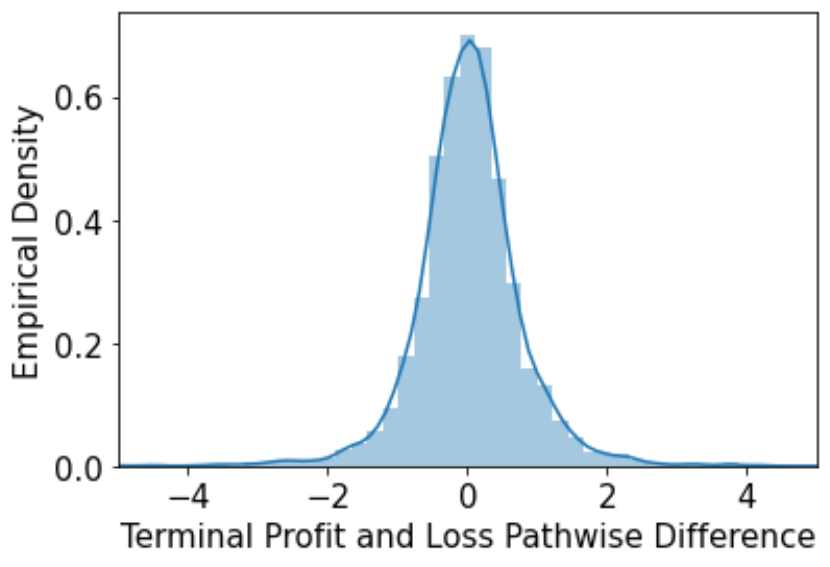}
\caption{Reinforcement learning versus deep hedging}
\label{fig:rl_sl}
\end{subfigure}

\caption{Empirical density functions of realized pathwise differences of terminal P\&Ls comparing with the approaches of classical Deltas and deep hedging}
\label{fig:pathwise_diff}
\end{figure}

Figure \ref{fig:baseline_plots} shows the empirical density and cumulative distribution functions via the $5000$ realized terminal P\&Ls by each hedging approach, while Table \ref{table:baseline_stats} outlines the summary statistics of these empirical distributions. To clearly illustrate the comparisons, Figure \ref{fig:pathwise_diff} depicts the empirical density functions via the $5000$ pathwise differences of the realized terminal P\&Ls between the RL agent and each of the other approaches, while Table \ref{table:pathwise_diff} lists the summary statistics of the empirical distributions; for example, comparing with the DH approach, the pathwise difference of the realized terminal P\&Ls for the $e$-th simulated scenario, for $e=1,2,\dots,5000$, is calculated by $\left(P_{t_{\tilde{n}}}^{\text{RL}}\left(\omega_e\right)-L_{t_{\tilde{n}}}^{\text{RL}}\left(\omega_e\right)\right)-\left(P_{t_{\tilde{n}}}^{\text{DH}}\left(\omega_e\right)-L_{t_{\tilde{n}}}^{\text{DH}}\left(\omega_e\right)\right)$.

\begin{table}[H]
\centering
\begin{tabular}{@{}rccccccccc@{}}
\toprule
Terminal P\&L of & \multirow{2}{*}{Mean} & \multirow{2}{*}{Median} & \multirow{2}{*}{Std. Dev.} & \multirow{2}{*}{$\text{VaR}_{90}$} & \multirow{2}{*}{$\text{VaR}_{95}$} & \multirow{2}{*}{$\text{TVaR}_{90}$} & \multirow{2}{*}{$\text{TVaR}_{95}$} & \multirow{2}{*}{$\widehat{\text{RMSE}}$}
\\
Hedging Approach & &  &  &  &  &  &  & 
\\\midrule
Reinforcement Learning & $0.02$ &  $-0.01$ & $0.58$ & $-0.54$ & $-0.87$ & $-1.05$ & $-1.43$  & $0.58$\\
CFM \& BS Delta  & $-0.01$ &  $0.00$ & $0.38$ & $-0.44$ & $-0.63$ & $-0.70$ & $-0.89$ & $0.38$ \\
IFM \& BS Delta  & $-0.06$ & $-0.06$ & $0.45$ & $-0.60$ & $-0.77$ & $-0.85$ & $-1.02$ & $0.45$ \\
CFM \& Heston Delta  & $-0.32$ &  $-0.33$ & $0.87$ & $-1.41$ & $-1.73$ & $-1.85$ & $-2.17$ & $0.93$ \\
IFM \& Heston Delta  & $-0.53$ &  $-0.53$ & $1.20$ & $-1.94$ & $-2.48$ & $-2.70$ & $-3.23$ & $1.31$ \\
Deep Hedging & $-0.01$ &  $-0.02$ & $0.60$ & $-0.52$ & $-0.71$ & $-1.04$ & $-1.49$ & $0.60$\\
\bottomrule                     
\end{tabular}
\caption{Summary statistics of empirical distributions of realized terminal P\&Ls by the approaches of reinforcement learning, classical Deltas, and deep hedging}
\label{table:baseline_stats}
\end{table}

\begin{table}[H]
\centering
\begin{tabular}{rcccc}
\toprule
Pathwise Difference of & \multirow{2}{*}{Mean} & \multirow{2}{*}{Median} & \multirow{2}{*}{Std. Dev.} & Probability of
\\
Terminal P\&Ls Comparing With & &  &  & Non-Negativity\\ \midrule
CFM \& BS Delta  & $0.02$ & $0.01$ & $0.62$ & $50.6\%$\\
IFM \& BS Delta & $0.08$ & $0.07$ & $0.66$ & $54.7\%$\\
CFM \& Heston Delta  & $0.34$ &  $0.34$ & $1.01$ & $64.3\%$ \\
IFM \& Heston Delta  & $0.54$ & $0.58$ & $1.29$ & $70.0\%$\\
Deep Hedging  & $0.02$ & $0.01$ & $0.75$ & $51.3\%$ \\
\bottomrule                     
\end{tabular}
\caption{Summary statistics of empirical distributions of realized pathwise differences of terminal P\&Ls comparing with the approaches of classical Deltas and deep hedging}
\label{table:pathwise_diff}
\end{table}

As expected, the baseline hedging performance of the trained RL agent in this training environment is comparable with those by, the correct CFM and BS Delta, as well as the DH approach. Moreover, the RL agent outperforms all the other three incorrect Deltas, which are based on either incorrect IFM actuarial or Heston financial model, or both.

\section{Online Learning Phase}\label{sec:online_learning}
Given the satisfactory baseline hedging performance of the experienced RL agent in the MDP training environment, the insurer finally assigns the agent to interact and learn from the market environment.

To distinguish them from the simulated time in the training environment, let $\tilde{t}_k$, for $k=0,1,2,\dots$, be the real time when the RL agent decides the hedging strategy in the market environment, such that $0=\tilde{t}_{0}<\tilde{t}_{1}<\tilde{t}_{2}<\cdots$, and $\delta\tilde{t}_k=\tilde{t}_{k+1}-\tilde{t}_k=\frac{1}{252}$. Note that the current time $t=\tilde{t}_0=0$ and the RL agent shall hedge daily on behalf of the insurer. At the current time $0$, the insurer writes a variable annuity contract with the GMMB and GMDB riders to the first policyholder. When this first contract terminates, due to either the death of the first policyholder or the expiration of the contract, the insurer shall write an identical contract, i.e. contract with the same characteristics, to the second policyholder. And so on. These contract re-establishments ensure that the insurer shall hold only one written variable annuity contract with the GMMB and GMDB riders at a time, and the RL agent shall solely hedge the contract being effective at that moment.

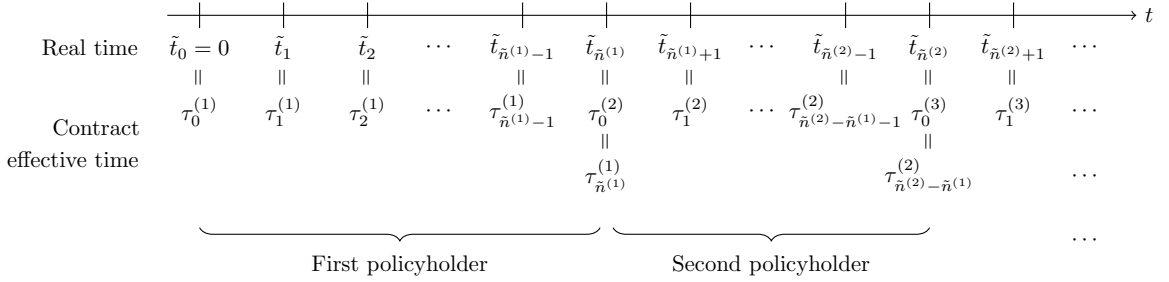
\begin{figure}[ht!]
\centering
\begin{tikzpicture}[scale=0.85, transform shape]
\draw[->] (0,0) -- (15,0);
\draw (15.2,0) node {$t$};

\draw (-1.2,-0.5) node {Real time};
\draw (-1.1,-1.75) node {Contract};
\draw (-1.46,-2.25) node {effective time};
\draw (0.5,0.2) -- (0.5,-0.2);
\draw (0.5,-0.5) node {$\tilde{t}_0=0$};
\draw (0.5,-1) node {$\verteq$};
\draw (0.5,-1.5) node {$\tau^{\left(1\right)}_0$};

\draw (1.8,0.2) -- (1.8,-0.2);
\draw (1.8,-0.5) node {$\tilde{t}_1$};
\draw (1.8,-1) node {$\verteq$};
\draw (1.8,-1.5) node {$\tau^{\left(1\right)}_1$};

\draw (3.1,0.2) -- (3.1,-0.2);
\draw (3.1,-0.5) node {$\tilde{t}_2$};
\draw (3.1,-1) node {$\verteq$};
\draw (3.1,-1.5) node {$\tau^{\left(1\right)}_2$};

\draw (4.2,-0.5) node {$\cdots$};
\draw (4.2,-1.5) node {$\cdots$};

\draw (5.5,0.2) -- (5.5,-0.2);
\draw (5.5,-0.5) node {$\tilde{t}_{\tilde{n}^{\left(1\right)}-1}$};
\draw (5.5,-1) node {$\verteq$};
\draw (5.5,-1.5) node {$\tau^{\left(1\right)}_{\tilde{n}^{\left(1\right)}-1}$};

\draw (6.8,0.2) -- (6.8,-0.2);
\draw (6.8,-0.5) node {$\tilde{t}_{\tilde{n}^{\left(1\right)}}$};
\draw (6.8,-1) node {$\verteq$};
\draw (6.8,-1.5) node {$\tau^{\left(2\right)}_{0}$};
\draw (6.8,-2) node {$\verteq$};
\draw (6.8,-2.5) node {$\tau^{\left(1\right)}_{\tilde{n}^{\left(1\right)}}$};

\draw (8.1,0.2) -- (8.1,-0.2);
\draw (8.1,-0.5) node {$\tilde{t}_{\tilde{n}^{\left(1\right)}+1}$};
\draw (8.1,-1) node {$\verteq$};
\draw (8.1,-1.5) node {$\tau^{\left(2\right)}_{1}$};

\draw (9.2,-0.5) node {$\cdots$};
\draw (9.2,-1.5) node {$\cdots$};

\draw (10.5,0.2) -- (10.5,-0.2);
\draw (10.5,-0.5) node {$\tilde{t}_{\tilde{n}^{\left(2\right)}-1}$};
\draw (10.5,-1) node {$\verteq$};
\draw (10.5,-1.5) node {$\tau^{\left(2\right)}_{\tilde{n}^{\left(2\right)}-\tilde{n}^{\left(1\right)}-1}$};

\draw (11.8,0.2) -- (11.8,-0.2);
\draw (11.8,-0.5) node {$\tilde{t}_{\tilde{n}^{\left(2\right)}}$};
\draw (11.8,-1) node {$\verteq$};
\draw (11.8,-1.5) node {$\tau^{\left(3\right)}_{0}$};
\draw (11.8,-2) node {$\verteq$};
\draw (11.8,-2.5) node {$\tau^{\left(2\right)}_{\tilde{n}^{\left(2\right)}-\tilde{n}^{\left(1\right)}}$};


\draw (13.1,0.2) -- (13.1,-0.2);
\draw (13.1,-0.5) node {$\tilde{t}_{\tilde{n}^{\left(2\right)}+1}$};
\draw (13.1,-1) node {$\verteq$};
\draw (13.1,-1.5) node {$\tau^{\left(3\right)}_{1}$};

\draw (14.2,-0.5) node {$\cdots$};
\draw (14.2,-1.5) node {$\cdots$};
\draw (14.2,-2.5) node {$\cdots$};

\draw [decorate,decoration={brace,amplitude=5pt,mirror,raise=4ex}]
(0.5,-2.6) -- (6.7,-2.6) node[midway,yshift=-13mm]{First policyholder};

\draw [decorate,decoration={brace,amplitude=5pt,mirror,raise=4ex}]
(6.9,-2.6) -- (11.8,-2.6) node[midway,yshift=-13mm]{Second policyholder};
\draw (14.2,-3.5) node {$\cdots$};
\end{tikzpicture}
\caption{An illustrative timeline with the real time and the contract effective time in the online learning phase}
\label{fig:timeline}
\end{figure}

To this end, iteratively, for the $\iota$-th policyholder, where $\iota\in\mathbb{N}$, let $\tilde{t}_{\tilde{n}^{\left(\iota\right)}}$ be the first time (right) after the $\iota$-th policyholder dies or the contract expires, for some $\tilde{n}^{\left(\iota\right)}=\tilde{n}^{\left(\iota-1\right)}+1,\tilde{n}^{\left(\iota-1\right)}+2,\dots,\tilde{n}^{\left(\iota-1\right)}+n$; that is $\tilde{t}_{\tilde{n}^{\left(\iota\right)}}=\min\left\{\tilde{t}_{k},k=\tilde{n}^{\left(\iota-1\right)}+1,\tilde{n}^{\left(\iota-1\right)}+2,\dots,\tilde{n}^{\left(\iota-1\right)}+n:\tilde{t}_{k}-\tilde{t}_{\tilde{n}^{\left(\iota-1\right)}}\geq T^{\left(\iota\right)}_{x_{\iota}}\wedge T\right\}$, where, by convention, $\tilde{n}^{\left(0\right)}=0$. Therefore, the contract effective time for the $\iota$-th policyholder $\tau_{k}^{\left(\iota\right)}=\tilde{t}_{\tilde{n}^{\left(\iota-1\right)}+k}$, where $\iota\in\mathbb{N}$ and $k=0,1,\dots,\tilde{n}^{\left(\iota\right)}-\tilde{n}^{\left(\iota-1\right)}$; in particular, $\tau_{0}^{\left(\iota\right)}=\tilde{t}_{\tilde{n}^{\left(\iota-1\right)}}$ is the contract inception time for the $\iota$-th policyholder. Figure \ref{fig:timeline} depicts one of the possible realizations for clearly illustrating the real time and the contract effective time.
\vspace{3mm}


In the online learning phase, the trained RL agent carries on with the PPO of policy gradient methods in the market environment. That is, as in Section \ref{sec:PPO}, starting from the ANN weights $\theta^{\left(\mathcal{U}\right)}$ at the current time $0$, and via interacting with the market environment to observe the states and collect the reward signals, the RL agent further updates the ANN weights by a batch of $\tilde{K}\in\mathbb{N}$ realizations and the (stochastic) gradient ascent in \eqref{eq:sga} with the surrogate performance measure in \eqref{eq:PPO_measure}, at each update step.

However, there are subtle differences of applying the PPO in the market environment from that in the training environment. At each further update step $v=1,2,\dots$, based on the ANN weights $\theta^{\left(\mathcal{U}+v-1\right)}$, and thus the policy $\pi\left(\cdot;\theta_{\text{p}}^{\left(\mathcal{U}+v-1\right)}\right)$, the RL agent hedges each effective contract of $\tilde{E}^{\left(v\right)}\in\mathbb{N}$ realized policyholders for the $\tilde{K}\in\mathbb{N}$ realizations. Indeed, the concept of episodes in the training environment, by the state re-initiation when one episode ends, should be replaced by sequential policyholders in the real-time market environment, via the contract re-establishment when one policyholder dies or contract expires.

\begin{itemize}
\item If $\tilde{E}^{\left(v\right)}=1$, which is when $\left(v-1\right)\tilde{K},v\tilde{K}\in\left[\tilde{n}^{\left(\iota-1\right)},\tilde{n}^{\left(\iota\right)}\right]$, for some $\iota\in\mathbb{N}$, the batch of $\tilde{K}$ realizations is collected solely from the $\iota$-th policyholder. The realizations are given by
\begin{align*}
&\;\left\{\dots,x_{\tau^{\left(\iota\right)}_{\tilde{K}_s^{\left(v\right)}}}^{\left(v-1,\iota\right)},h_{\tau^{\left(\iota\right)}_{\tilde{K}_s^{\left(v\right)}}}^{\left(v-1,\iota\right)},x_{\tau^{\left(\iota\right)}_{\tilde{K}_s^{\left(v\right)}+1}}^{\left(v-1,\iota\right)},r_{\tau^{\left(\iota\right)}_{\tilde{K}_s^{\left(v\right)}+1}}^{\left(v-1,\iota\right)},h_{\tau^{\left(\iota\right)}_{\tilde{K}_s^{\left(v\right)}+1}}^{\left(v-1,\iota\right)},\right.\\&\;\left.\quad\dots,x_{\tau^{\left(\iota\right)}_{\tilde{K}_s^{\left(v\right)}+\tilde{K}-1}}^{\left(v-1,\iota\right)},r_{\tau^{\left(\iota\right)}_{\tilde{K}_s^{\left(v\right)}+\tilde{K}-1}}^{\left(v-1,\iota\right)},h_{\tau^{\left(\iota\right)}_{\tilde{K}_s^{\left(v\right)}+\tilde{K}-1}}^{\left(v-1,\iota\right)},x_{\tau^{\left(\iota\right)}_{\tilde{K}_s^{\left(v\right)}+\tilde{K}}}^{\left(v-1,\iota\right)},r_{\tau^{\left(\iota\right)}_{\tilde{K}_s^{\left(v\right)}+\tilde{K}}}^{\left(v-1,\iota\right)},\dots\right\},
\end{align*}
where $\tilde{K}_s^{\left(v\right)}=0,1,\dots,\tilde{n}^{\left(\iota\right)}-\tilde{n}^{\left(\iota-1\right)}-1$, such that the time $\tau^{\left(\iota\right)}_{\tilde{K}_s^{\left(v\right)}}$ is when the first state is observed for the $\iota$-th policyholder in this update; 
necessarily, $\tilde{n}^{\left(\iota\right)}-\tilde{n}^{\left(\iota-1\right)}-\tilde{K}_s^{\left(v\right)}\geq\tilde{K}$.

\item If $\tilde{E}^{\left(v\right)}=2,3,\dots$, which is when $\left(v-1\right)\tilde{K}\in\left[\tilde{n}^{\left(\iota-1\right)},\tilde{n}^{\left(\iota\right)}\right]$ and $v\tilde{K}\in\left[\tilde{n}^{\left(j-1\right)},\tilde{n}^{\left(j\right)}\right]$, for some $\iota,j\in\mathbb{N}$ such that $\iota<j$, the batch of $\tilde{K}$ realizations is collected from the $\iota$-th, $\left(\iota+1\right)$-th, $\dots$, and $j$-th policyholders; that is, $\tilde{E}^{\left(v\right)}=j-\iota+1$. The realizations are given by
\begin{align*}
&\;\left\{\dots,x_{\tau^{\left(\iota\right)}_{\tilde{K}_s^{\left(v\right)}}}^{\left(v-1,\iota\right)},h_{\tau^{\left(\iota\right)}_{\tilde{K}_s^{\left(v\right)}}}^{\left(v-1,\iota\right)},x_{\tau^{\left(\iota\right)}_{\tilde{K}_s^{\left(v\right)}+1}}^{\left(v-1,\iota\right)},r_{\tau^{\left(\iota\right)}_{\tilde{K}_s^{\left(v\right)}+1}}^{\left(v-1,\iota\right)},h_{\tau^{\left(\iota\right)}_{\tilde{K}_s^{\left(v\right)}+1}}^{\left(v-1,\iota\right)},\right.\\&\;\left.\quad\dots,x_{\tau^{\left(\iota\right)}_{\tilde{n}^{\left(\iota\right)} - \tilde{n}^{\left(i - 1\right)} -1}}^{\left(v-1,\iota\right)},r_{\tau^{\left(\iota\right)}_{\tilde{n}^{\left(\iota\right)} - \tilde{n}^{\left(i - 1\right)} -1}}^{\left(v-1,\iota\right)},h_{\tau^{\left(\iota\right)}_{\tilde{n}^{\left(\iota\right)} - \tilde{n}^{\left(i - 1\right)} -1}}^{\left(v-1,\iota\right)},x_{\tau^{\left(\iota\right)}_{\tilde{n}^{\left(\iota\right)} - \tilde{n}^{\left(i - 1\right)}}}^{\left(v-1,\iota\right)},r_{\tau^{\left(\iota\right)}_{\tilde{n}^{\left(\iota\right)} - \tilde{n}^{\left(i - 1\right)}}}^{\left(v-1,\iota\right)}\right\},\\
&\;\left\{x_{\tau^{\left(\iota+1\right)}_{0}}^{\left(v-1,\iota+1\right)},h_{\tau^{\left(\iota+1\right)}_{0}}^{\left(v-1,\iota+1\right)},x_{\tau^{\left(\iota+1\right)}_{1}}^{\left(v-1,\iota+1\right)},r_{\tau^{\left(\iota+1\right)}_{1}}^{\left(v-1,\iota+1\right)},h_{\tau^{\left(\iota+1\right)}_{1}}^{\left(v-1,\iota+1\right)},\right.\\&\;\left.\quad\dots,x_{\tau^{\left(\iota+1\right)}_{\tilde{n}^{\left(\iota+1\right)} - \tilde{n}^{\left(\iota\right)} -1}}^{\left(v-1,\iota+1\right)},r_{\tau^{\left(\iota+1\right)}_{\tilde{n}^{\left(\iota+1\right)} - \tilde{n}^{\left(\iota\right)} -1}}^{\left(v-1,\iota+1\right)},h_{\tau^{\left(\iota+1\right)}_{\tilde{n}^{\left(\iota+1\right)} - \tilde{n}^{\left(\iota\right)} -1}}^{\left(v-1,\iota+1\right)},x_{\tau^{\left(\iota+1\right)}_{\tilde{n}^{\left(\iota+1\right)} - \tilde{n}^{\left(\iota\right)}}}^{\left(v-1,\iota+1\right)},r_{\tau^{\left(\iota+1\right)}_{\tilde{n}^{\left(\iota+1\right)} - \tilde{n}^{\left(\iota\right)}}}^{\left(v-1,\iota+1\right)}\right\},\\
&\;\dots,\\
&\;\left\{x_{\tau^{\left(j-1\right)}_{0}}^{\left(v-1,j-1\right)},h_{\tau^{\left(j-1\right)}_{0}}^{\left(v-1,j-1\right)},x_{\tau^{\left(j-1\right)}_{1}}^{\left(v-1,j-1\right)},r_{\tau^{\left(j-1\right)}_{1}}^{\left(v-1,j-1\right)},h_{\tau^{\left(j-1\right)}_{1}}^{\left(v-1,j-1\right)},\right.\\&\;\left.\quad\dots,x_{\tau^{\left(j-1\right)}_{\tilde{n}^{\left(j-1\right)} - \tilde{n}^{\left(j-2\right)} -1}}^{\left(v-1,j-1\right)},r_{\tau^{\left(j-1\right)}_{\tilde{n}^{\left(j-1\right)} - \tilde{n}^{\left(j-2\right)} -1}}^{\left(v-1,j-1\right)},h_{\tau^{\left(j-1\right)}_{\tilde{n}^{\left(j-1\right)} - \tilde{n}^{\left(j-2\right)} -1}}^{\left(v-1,j-1\right)},x_{\tau^{\left(j-1\right)}_{\tilde{n}^{\left(j-1\right)} - \tilde{n}^{\left(j-2\right)}}}^{\left(v-1,j-1\right)},r_{\tau^{\left(j-1\right)}_{\tilde{n}^{\left(j-1\right)} - \tilde{n}^{\left(j-2\right)}}}^{\left(v-1,j-1\right)}\right\},\\
&\;\left\{x_{\tau^{\left(j\right)}_{0}}^{\left(v-1,j\right)},h_{\tau^{\left(j\right)}_{0}}^{\left(v-1,j\right)},x_{\tau^{\left(j\right)}_{1}}^{\left(v-1,j\right)},r_{\tau^{\left(j\right)}_{1}}^{\left(v-1,j\right)},h_{\tau^{\left(j\right)}_{1}}^{\left(v-1,j\right)},\right.\\&\;\left.\quad\dots,x_{\tau^{\left(j\right)}_{\tilde{K}_f^{\left(v\right)}-1}}^{\left(v-1,j\right)},r_{\tau^{\left(j\right)}_{\tilde{K}_f^{\left(v\right)}-1}}^{\left(v-1,j\right)},h_{\tau^{\left(j\right)}_{\tilde{K}_f^{\left(v\right)}-1}}^{\left(v-1,j\right)},x_{\tau^{\left(j\right)}_{\tilde{K}_f^{\left(v\right)}}}^{\left(v-1,j\right)},r_{\tau^{\left(j\right)}_{\tilde{K}_f^{\left(v\right)}}}^{\left(v-1,j\right)},\dots\right\},
\end{align*}
where $\tilde{K}_f^{\left(v\right)}=1,2,\dots,\tilde{n}^{\left(j\right)}-\tilde{n}^{\left(j-1\right)}$, such that the time $\tau^{\left(j\right)}_{\tilde{K}_f^{\left(v\right)}}$ is when the last state is observed for the $j$-th policyholder in this update; necessarily, $\tilde{n}^{\left(j-1\right)} - \tilde{n}^{\left(i - 1\right)} + \tilde{K}_f^{\left(v\right)} - \tilde{K}_s^{\left(v\right)} = \tilde{K}$.

\end{itemize}

Moreover, the first two features in the state vector \eqref{eq:state_vector} are based on the real-time risky asset price realization from the market, while all features depend on a particular effective policyholder. For $\iota\in\mathbb{N}$ and $k=0,1,\dots,\tilde{n}^{\left(\iota\right)}-\tilde{n}^{\left(\iota-1\right)}$,
\begin{equation}
X_{\tau^{\left(\iota\right)}_{k}}^{\left(v-1,\iota\right)}=
\begin{cases}
\left(\ln F^{\left(\iota\right)}_{\tau^{\left(\iota\right)}_{k}},P^{\left(\iota\right)}_{\tau^{\left(\iota\right)}_{k}},1,T-\left(\tau^{\left(\iota\right)}_{k}-\tau^{\left(\iota\right)}_{0}\right)\right)&\text{if}\; k=0,1,\dots,\tilde{n}^{\left(\iota\right)}-\tilde{n}^{\left(\iota-1\right)}-1\\
\left(\ln F^{\left(\iota\right)}_{\tau^{\left(\iota\right)}_{k}},P^{\left(\iota\right)}_{\tau^{\left(\iota\right)}_{k}},0,T-\left(\tau^{\left(\iota\right)}_{k}-\tau^{\left(\iota\right)}_{0}\right)\right)&\text{if}\; k=\tilde{n}^{\left(\iota\right)}-\tilde{n}^{\left(\iota-1\right)}\text{ and }T^{\left(\iota\right)}_{x_{\iota}}\leq T\\
\left(\ln F^{\left(\iota\right)}_{\tau^{\left(\iota\right)}_{k}},P^{\left(\iota\right)}_{\tau^{\left(\iota\right)}_{k}},1,0\right)&\text{if}\; k=\tilde{n}^{\left(\iota\right)}-\tilde{n}^{\left(\iota-1\right)}\text{ and }T^{\left(\iota\right)}_{x_{\iota}}>T
\end{cases},
\label{eq:state_vector_OL}
\end{equation}
where $F^{\left(\iota\right)}_t=\rho^{\left(\iota\right)} S_{t}e^{-m^{\left(\iota\right)}\left(t-\tau^{\left(\iota\right)}_{0}\right)}$, if $t\in\left[\tau^{\left(\iota\right)}_{0},\tilde{t}_{\tilde{n}^{\left(\iota\right)}}\right]$, $P^{\left(\iota\right)}_{\tau^{\left(\iota\right)}_{0}}=0$, and
\begin{align*}
P^{\left(\iota\right)}_{\tau^{\left(\iota\right)}_{k}}=&\;\left(P^{\left(\iota\right)}_{\tau^{\left(\iota\right)}_{k-1}}-H^{\left(\iota\right)}_{\tau^{\left(\iota\right)}_{k-1}}S_{\tau^{\left(\iota\right)}_{k-1}}\right)e^{r\left(\tau^{\left(\iota\right)}_{k}-\tau^{\left(\iota\right)}_{k-1}\right)}+H^{\left(\iota\right)}_{\tau^{\left(\iota\right)}_{k-1}}S_{\tau^{\left(\iota\right)}_{k}}+m^{\left(\iota\right)}_e\int_{\tau^{\left(\iota\right)}_{k-1}}^{\tau^{\left(\iota\right)}_{k}}F^{\left(\iota\right)}_{s}e^{r\left(\tau^{\left(\iota\right)}_{k}-s\right)}J_s^{\left(\iota\right)}ds\\
&\;-\left(G_D-F_{T_{x_\iota}^{\left(\iota\right)}}^{\left(\iota\right)}\right)_+\mathds{1}_{\{\tau^{\left(\iota\right)}_{k-1}<T_{x_\iota}^{\left(\iota\right)}\leq \tau^{\left(\iota\right)}_{k}\}}e^{r\left(\tau^{\left(\iota\right)}_{k}-T_{x_\iota}^{\left(\iota\right)}\right)},
\end{align*}
for $k=1,2,\dots,\tilde{n}^{\left(\iota\right)}-\tilde{n}^{\left(\iota-1\right)}$. Recall also that the reward signals collecting from the market environment should be based on that in \eqref{eq:reward_2}; that is, for $\iota\in\mathbb{N}$ and $k=0,1,\dots,\tilde{n}^{\left(\iota\right)}-\tilde{n}^{\left(\iota-1\right)}$,
\begin{equation*}
R_{\tau^{\left(\iota\right)}_{k}}^{\left(v-1,\iota\right)}=
\begin{cases}
0&\text{if}\quad k=0,1,\dots,\tilde{n}^{\left(\iota\right)}-\tilde{n}^{\left(\iota-1\right)}-1\\
-\left(P^{\left(\iota\right)}_{\tilde{t}_{\tilde{n}^{\left(\iota\right)}}}-L^{\left(\iota\right)}_{\tilde{t}_{\tilde{n}^{\left(\iota\right)}}}\right)^2&\text{if}\quad k=\tilde{n}^{\left(\iota\right)}-\tilde{n}^{\left(\iota-1\right)}
\end{cases},
\end{equation*}
in which $L^{\left(\iota\right)}_{\tilde{t}_{\tilde{n}^{\left(\iota\right)}}}=0$ if $T^{\left(\iota\right)}_{x_{\iota}}\leq T$, and $L^{\left(\iota\right)}_{\tilde{t}_{\tilde{n}^{\left(\iota\right)}}}=\left(G_M-F^{\left(\iota\right)}_{\tau^{\left(\iota\right)}_{0}+T}\right)_+$ if $T^{\left(\iota\right)}_{x_{\iota}}>T$.



Table \ref{table:new_ppo_parameters} summarizes all hyperparameters of the implemented PPO in the market environment, while the hyperparameters of the ANN architecture are still given in Table \ref{sub_table:params_ANN}. In the online learning phase, the insurer should choose a smaller batch size $\tilde{K}$ comparing to that in the training phase; this yields a higher updating frequency by the PPO to ensure that the experienced RL agent could revise the hedging strategy within a reasonable amount of time. However, fewer realizations in the batch cause less credible updates; hence, the insurer should also tune down the learning rate $\tilde{\alpha}$, from that in the training phase, to reduce the reliance on each further update step.

\begin{table}[H]
\centering
\begin{tabular}{lcc|lcc}
\toprule
Hyperparameter & & Value & Hyperparameter & & Value  \\ \midrule
{\bf Learning rate $\tilde{\alpha}$} & & {\bf $0.001$}  & Coefficient of value function & & \multirow{2}{*}{$0.25$}\\
{\bf Batch size $\tilde{K}$} & & {\bf $30$} & approximation loss $c_1$ & & \\
Clip factor $\epsilon$ & & $0.18$ & Coefficient of entropy bonus $c_2$ & & $0.01$\\
\bottomrule
\end{tabular}
\caption{Hyperparameters setting of Proximal Policy Optimization for online learning with bolded hyperparameters being different from those for training}
\label{table:new_ppo_parameters}
\end{table}

\section{Illustrative Example Revisited: Online Learning Phase}\label{sec:illustrative}
This section revisits the illustrative example in Section \ref{sec:pit_revisit} via the two-phase RL approach in the online learning phase. In the market environment, the policyholders being sequentially written of the contracts with both GMMB and GMDB riders are homogeneous. Due to contract re-establishments to these sequential homogeneous policyholders, the number and age of policyholders shall be reset to the values as in Table \ref{sub_table:params_act} at each contract inception time. Furthermore, via the approach discussed in Section \ref{sec:net_lia}, to determine the fee structures of each contract at its inception time, the insurer relies on the parameters of the model of the market environment in Table \ref{table:param_financial_actuarial}, except that now the risky asset initial price therein is replaced by the risky asset price observed at the contract inception time. Note that the fee structures of the first contract are still given as in Table \ref{sub_table:params_fee}, since the risky asset price observed at $t=0$ is exactly the risky asset initial price.

Let $\mathcal{V}\in\mathbb{N}$ be the number of further update steps in the market environment on the ANN weights. In order to showcase the result that, (RLw/OL) the further trained RL agent with the online learning phase, could gradually revise the hedging strategy, from the nearly optimal one in the training environment, to the one in the market environment, we evaluate the hedging performance of RLw/OL on a rolling-basis. That is, right after each further update step $v=1,2,\dots,\mathcal{V}$, we first simulate $\tilde{M}=500$ market scenarios stemming from the real-time realized state vector $x_{\tau^{\left(j\right)}_{\tilde{K}_f^{\left(v\right)}}}^{\left(v-1,j\right)}$ and by implementing the hedging strategy from the updated policy $\pi\left(\cdot;\theta_{\text{p}}^{\left(\mathcal{U}+v\right)}\right)$, i.e. the further trained RL agent takes the deterministic action $c\left(\cdot;\theta_{\text{p}}^{\left(\mathcal{U}+v\right)}\right)$ which is the mean of the Gaussian measure; we then document the realized terminal P\&L, for each of the $500$ simulated scenarios, i.e. $P^{\text{RLw/OL}}_{t}\left(\omega_e\right)-L_{t}\left(\omega_e\right)$, for $e=1,2,\dots,500$, where $t=\tilde{t}_{\tilde{n}^{\left(j\right)}}\left(\omega_e\right)$ if $\tau^{\left(j\right)}_{\tilde{K}_f^{\left(v\right)}}<\tilde{t}_{\tilde{n}^{\left(j\right)}}$, and $t=\tilde{t}_{\tilde{n}^{\left(j+1\right)}}\left(\omega_e\right)$ if $\tau^{\left(j\right)}_{\tilde{K}_f^{\left(v\right)}}=\tilde{t}_{\tilde{n}^{\left(j\right)}}$.

Since the state vector $x_{\tau^{\left(j\right)}_{\tilde{K}_f^{\left(v\right)}}}^{\left(v-1,j\right)}$ is realized in real time, the realized terminal P\&L in fact depends on, not only the simulated scenarios after each update, but also the actual realization in the market environment. To this end, from the current time $0$, we simulate $M=1000$ future trajectories in the market environment; for each future trajectory $f=1,2,\dots,1000$, the aforementioned realized terminal P\&Ls are obtained as $P^{\text{RLw/OL}}_{t}\left(\omega_f,\omega_e\right)-L_{t}\left(\omega_f,\omega_e\right)$, for $e=1,2,\dots,500$, where $t=\tilde{t}_{\tilde{n}^{\left(j\right)}}\left(\omega_f,\omega_e\right)$ if $\tau^{\left(j\right)}_{\tilde{K}_f^{\left(v\right)}}\left(\omega_f\right)<\tilde{t}_{\tilde{n}^{\left(j\right)}}\left(\omega_f\right)$, and $t=\tilde{t}_{\tilde{n}^{\left(j+1\right)}}\left(\omega_f,\omega_e\right)$ if $\tau^{\left(j\right)}_{\tilde{K}_f^{\left(v\right)}}\left(\omega_f\right)=\tilde{t}_{\tilde{n}^{\left(j\right)}}\left(\omega_f\right)$.

The rolling-basis hedging performance of RLw/OL is benchmarked with those by, (RLw/oOL) the trained RL agent without the online learning phase, (CD) the correct Delta based on the market environment, and (ID) the incorrect Delta based on the training environment. For the same set of future trajectories $\omega_f$, for $f=1,2,\dots,1000$, and the same sets of simulated scenarios $\omega_e$, for $e=1,2,\dots,500$, the realized terminal P\&Ls are also obtained, by implementing each of these benchmark strategies starting from the current time $0$, which does not need to be updated throughout; denote the realized terminal P\&L as $P^{\mathcal{S}}_{t}\left(\omega_f,\omega_e\right)-L_{t}\left(\omega_f,\omega_e\right)$, where $\mathcal{S}=\text{RLw/OL},\text{RLw/oOL},\text{CD},\text{ or }\text{ID}$.

This example considers $\mathcal{V}=25$ further update steps of RLw/OL, for each future trajectory $\omega_f$, where $f=1,2,\dots,1000$; as the batch size in the online learning phase $\tilde{K}=30$, this is equivalent to $750$ trading days, which is just less than $3$ years (assuming that non-trading days are uniformly spread across a year). For each $f=1,2,\dots,1000$, and $v=1,2,\dots,25$, let $\mu^{\left(v,j\right)}_{\mathcal{S}}\left(\omega_f\right)$ be the expected terminal P\&L, right after the $v$-th further update step implementing the hedging strategy $\mathcal{S}$ for the future trajectory $\omega_f$:
\begin{equation*}
\mu^{\left(v,j\right)}_{\mathcal{S}}\left(\omega_f\right) = \mathbb{E} \left[P^{\mathcal{S}}_{t}\left(\omega_f,\cdot\right)-L_{t}\left(\omega_f,\cdot\right)\Big\vert X_{\tau^{\left(j\right)}_{\tilde{K}_f^{\left(v\right)}}}^{\left(v-1,j\right)} =  X_{\tau^{\left(j\right)}_{\tilde{K}_f^{\left(v\right)}}}^{\left(v-1,j\right)}\left(\omega_f\right)\right],
\end{equation*}
which is a conditional expectation taking with respect to the scenarios from the time $\tau^{\left(j\right)}_{\tilde{K}_f^{\left(v\right)}}$ forward; let $\hat{\mu}^{\left(v,j\right)}_{\mathcal{S}}\left(\omega_f\right)$ be the sample mean of the terminal P\&L based on the simulated scenarios:
\begin{equation}
\hat{\mu}^{\left(v,j\right)}_{\mathcal{S}}\left(\omega_f\right) = \frac{1}{500}\sum_{e = 1}^{500}\left(P_{t}^{\mathcal{S}}\left(\omega_f,\omega_e\right)-L_{t}\left(\omega_f,\omega_e\right)\right).
\label{eq:sample_mean}
\end{equation}

\begin{figure}[H]
\centering
\includegraphics[scale=0.5]{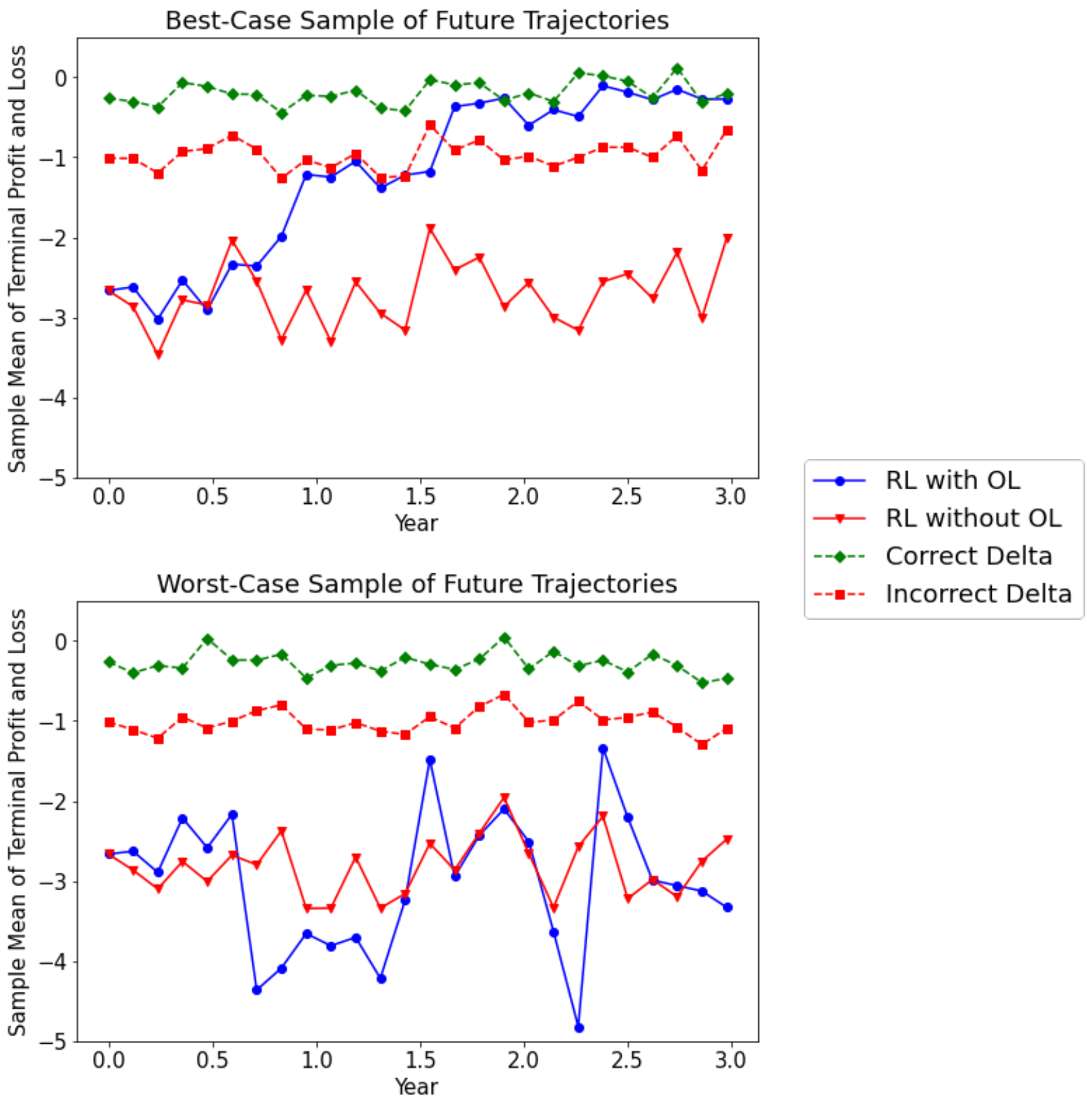}
\caption{Best-case and worst-case samples of future trajectories for rolling-basis evaluation of reinforcement learning agent with online learning phase, and comparisons with classical Deltas and reinforcement learning agent without online learning phase}
\label{fig:best_worst}
\end{figure}

Figure \ref{fig:best_worst} plots the sample means of the terminal P\&L in \eqref{eq:sample_mean}, right after each further update step and implementing each hedging strategy, in two future trajectories. Firstly, notice that, in both future trajectories, the average hedging performance of RLw/oOL is even worse than that of ID. Secondly, the average hedging performances of RLw/OL between the two future trajectories are substantially different. In the best-case future trajectory, the RLw/OL is able to swiftly self-revise the hedging strategy, and hence quickly catch up the average hedging performance of ID by simply twelve further updates on the ANN weights, as well as that of CD in around two years; however, in the {\it worst-case} future trajectory, within $3$ years, the RLw/OL is not able to improve the average hedging performance to even the level of ID, let alone to that of CD.

\begin{figure}[H]
\centering
\includegraphics[width = 0.5\linewidth]{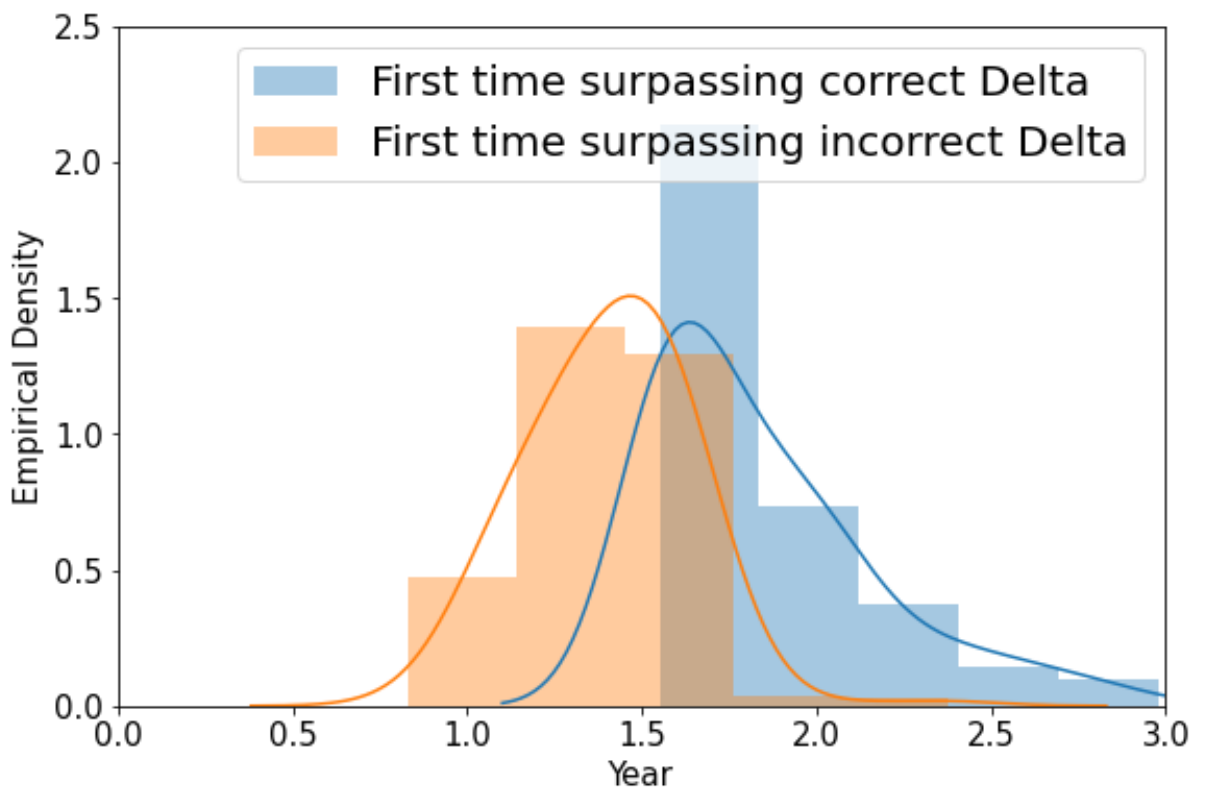}
\caption{Empirical conditional density functions of first surpassing times conditioning on reinforcement learning agent with online learning phase exceeding correct Delta in terms of sample means of terminal P\&L within $3$ years}
\label{fig:hitting_time}
\end{figure}

In view of the second observation above, the hedging performance of RLw/OL should not be concluded for each future trajectory alone; instead, it should be studied among the future trajectories. To this end, for each $f=1,2,\dots,1000$, define
\begin{equation*}
v_{\text{CD}}\left(\omega_f\right) = \min\left\{v = 1, 2, \dots, 25: \hat{\mu}^{\left(v,j\right)}_{\text{RLw/OL}}\left(\omega_f\right)>\hat{\mu}^{\left(v,j\right)}_{\text{CD}}\left(\omega_f\right)\right\}
\end{equation*}
as the first further update step such that the sample mean of the terminal P\&L by RLw/OL is strictly greater than that by CD, for the future trajectory $\omega_f$; herein, let $\min\emptyset=26$, and also define $t_{\text{CD}}\left(\omega_f\right)=v_{\text{CD}}\left(\omega_f\right)\times\frac{\tilde{K}}{252}$ as the corresponding number of years. Therefore, the estimated proportion of the future trajectories, where RLw/OL is able to exceed the average hedging performance of CD within $3$ years, is given by
\begin{equation*}
\frac{1}{1000}\sum_{f = 1}^{1000}\mathds{1}_{\left\{t_{\text{CD}}\left(\omega_f\right)\leq 3\right\}}= 95.4\%.
\end{equation*}
For each $f=1,2,\dots,1000$, define $v_{\text{ID}}\left(\omega_f\right)$ and $t_{\text{ID}}\left(\omega_f\right)$ similarly for comparing RLw/OL with ID. Figure \ref{fig:hitting_time} shows the empirical conditional density functions of $t_{\text{CD}}$ and $t_{\text{ID}}$, both subject to that RLw/OL exceeds the average hedging performance of CD within $3$ years. Table \ref{table:sum_catchup} lists the summary statistics of the empirical conditional distributions.

\begin{table}[H]
\centering
\begin{tabular}{@{}rccccccccc@{}}
\toprule
Reinforcement Learning Agent & \multirow{3}{*}{Mean} & \multirow{3}{*}{Median} & \multirow{3}{*}{Std. Dev.} & \multirow{3}{*}{$\text{VaR}_{90}$} & \multirow{3}{*}{$\text{VaR}_{95}$} & \multirow{3}{*}{$\text{TVaR}_{90}$} & \multirow{3}{*}{$\text{TVaR}_{95}$} &
\\
with Online Learning Phase & &  &  &  &  &  &  & 
\\
First Surpassing Time to & &  &  &  &  &  &  & 
\\\midrule
Correct Delta & $1.84$ & $1.79$ & $0.32$ & $2.38$ & $2.50$ & $2.66$ & $2.73$  \\
Incorrect Delta  & $1.41$ &  $1.43$ & $0.22$ & $1.67$ & $1.67$ & $2.05$ & $2.05$ \\
\bottomrule                     
\end{tabular}
\caption{Summary statistics of empirical conditional distributions of first surpassing times conditioning on reinforcement learning agent with online learning phase exceeding correct Delta in terms of sample means of terminal P\&L within $3$ years}
\label{table:sum_catchup}
\end{table}


The above analysis obviously neglected the variance, due to the simulated scenarios, of hedging performance by each hedging strategy. In the following, for each future trajectory, we define a refined first further update step such that the expected terminal P\&L by RLw/OL is statistically significant to be strictly greater than that by CD. To this end, for each $f=1,2,\dots,1000$, and $v=1,2,\dots,25$, consider the following null and alternative hypotheses:
\begin{equation*}
H_{0,\mathcal{S}}^{\left(v,j\right)}\left(\omega_f\right):\mu^{\left(v,j\right)}_{\text{RLw/OL}}\left(\omega_f\right)\leq\mu^{\left(v,j\right)}_{\mathcal{S}}\left(\omega_f\right)\quad\text{versus}\quad H_{1,\mathcal{S}}^{\left(v,j\right)}\left(\omega_f\right):\mu^{\left(v,j\right)}_{\text{RLw/OL}}\left(\omega_f\right)>\mu^{\left(v,j\right)}_{\mathcal{S}}\left(\omega_f\right),
\end{equation*}
where $\mathcal{S}=\text{CD or ID}$; the analysis before supports this choice of the alternative hypothesis. Define respectively the test statistics and the p-value by
\begin{equation*}
\mathcal{T}^{\left(v,j\right)}_{\mathcal{S}}\left(\omega_f\right) = \frac{\hat{\mu}^{\left(v,j\right)}_{\text{RLw/OL}}\left(\omega_f\right) - \hat{\mu}^{\left(v,j\right)}_{\mathcal{S}}\left(\omega_f\right)}{\sqrt{\frac{\hat{\sigma}^{\left(v,j\right)}_{\text{RLw/OL}}\left(\omega_f\right)^2}{500}+\frac{\hat{\sigma}^{\left(v,j\right)}_{\mathcal{S}}\left(\omega_f\right)^2}{500}}}\quad\text{and}\quad
p^{\left(v,j\right)}_{\mathcal{S}}\left(\omega_f\right) = \mathbb{P}\left(T_{\mathcal{S}}\left(\omega_f\right)>\mathcal{T}^{\left(v,j\right)}_{\mathcal{S}}\left(\omega_f\right)\right),
\end{equation*}
where the random variable $T_{\mathcal{S}}\left(\omega_f\right)$ follows a Student's t-distribution with the degree of freedom
\begin{equation*}
\text{df}^{\left(v,j\right)}_{\mathcal{S}}\left(\omega_f\right) = \frac{\left(\frac{\hat{\sigma}^{\left(v,j\right)}_{\text{RLw/OL}}\left(\omega_f\right)^2}{500}+\frac{\hat{\sigma}^{\left(v,j\right)}_{\mathcal{S}}\left(\omega_f\right)^2}{500}\right)^2}{\frac{\left(\hat{\sigma}^{\left(v,j\right)}_{\text{RLw/OL}}\left(\omega_f\right)^2/500\right)^2}{500-1}+\frac{\left(\hat{\sigma}^{\left(v,j\right)}_{\mathcal{S}}\left(\omega_f\right)^2/500\right)^2}{500-1}},
\end{equation*}
and the sample variance $\hat{\sigma}^{\left(v,j\right)}_{\mathcal{S}}\left(\omega_f\right)^2$ of the terminal P\&L based on the simulated scenarios is given by
\begin{equation*}
\hat{\sigma}^{\left(v,j\right)}_{\mathcal{S}}\left(\omega_f\right)^2= \frac{1}{499}\sum_{e = 1}^{500}\left(\left(P_{t}^{\mathcal{S}}\left(\omega_f,\omega_e\right)-L_{t}\left(\omega_f,\omega_e\right)\right)-\hat{\mu}^{\left(v,j\right)}_{\mathcal{S}}\left(\omega_f\right)\right)^2.
\end{equation*}
For a fixed level of significance $\alpha^*\in\left(0,1\right)$, if $p^{\left(v,j\right)}_{\mathcal{S}}\left(\omega_f\right)<\alpha^*$, then the expected terminal P\&L by RLw/OL is statistically significant to be strictly greater than that by $\mathcal{S}=\text{CD or ID}$.

In turn, for each $f=1,2,\dots,1000$, and for any $\alpha^*\in\left(0,1\right)$, define \begin{equation*}
v_{\mathcal{S}}^{\text{p}}\left(\omega_f;\alpha^*\right) = \min\left\{v = 1, 2, \dots, 25: p^{\left(v,j\right)}_{\mathcal{S}}\left(\omega_f\right)<\alpha^*\right\}
\end{equation*}
as the first further update step such that the expected terminal P\&L by RLw/OL is statistically significant to be strictly greater than that by $\mathcal{S}=\text{CD or ID}$, for the future trajectory $\omega_f$ at the level of significance $\alpha^*$; again, herein, let $\min\emptyset=26$, and define $t_{\mathcal{S}}^{\text{p}}\left(\omega_f;\alpha^*\right)=v_{\mathcal{S}}^{\text{p}}\left(\omega_f;\alpha^*\right)\times\frac{\tilde{K}}{252}$ as the corresponding number of years. Table \ref{table:catchup_prop} lists the estimated proportion of the future trajectories, where RLw/OL is statistically significant to be able to exceed the expected terminal P\&L of $\mathcal{S}$ within $3$ years, which is given by $\sum_{f = 1}^{1000}\mathds{1}_{\left\{t_{\mathcal{S}}^{\text{p}}\left(\omega_f;\alpha^*\right)\leq 3\right\}}/1000$, with various levels of significance.

\begin{table}[H]
\centering
\begin{tabular}{@{}rccccccccc@{}}
\toprule
Estimated Proportion & \multirow{2}{*}{$\alpha^* = 0.20$} & \multirow{2}{*}{$\alpha^* = 0.15$} & \multirow{2}{*}{$\alpha^* = 0.10$} & \multirow{2}{*}{$\alpha^* = 0.05$} & \multirow{2}{*}{$\alpha^* = 0.01$} &
\\
of Exceeding & &  &  &  &  &  &  & 
\\\midrule
Correct Delta & $55.7\%$ & $52.1\%$ & $47.6\%$  & $35.9\%$ & $21.8\%$\\
Incorrect Delta  &  $96.9\%$ & $95.1\%$ & $85.0\%$ & $70.6\%$ & $64.6\%$\\
\bottomrule                     
\end{tabular}
\caption{Estimated proportions of future trajectories where reinforcement learning agent with online learning phase is statistically significant to be exceeding correct Delta and incorrect Delta within $3$ years with various levels of significance}
\label{table:catchup_prop}
\end{table}


When the level of significance $\alpha^*$ gradually decreases from $0.20$ to $0.01$, both estimated proportions, of the future trajectories for RLw/OL being statistically significant to be exceeding CD or ID within $3$ years, decline. This is because, for any $\alpha^*_1,\alpha^*_2\in\left(0,1\right)$ with $\alpha^*_1\leq\alpha^*_2$, and for any $\omega_f$, for $f=1,2,\dots,1000$, $t_{\mathcal{S}}^{\text{p}}\left(\omega_f;\alpha^*_1\right)\leq 3$ implies that $t_{\mathcal{S}}^{\text{p}}\left(\omega_f;\alpha^*_2\right)\leq 3$, and thus $\mathds{1}_{\left\{t_{\mathcal{S}}^{\text{p}}\left(\omega_f;\alpha^*_1\right)\leq 3\right\}}\leq\mathds{1}_{\left\{t_{\mathcal{S}}^{\text{p}}\left(\omega_f;\alpha^*_2\right)\leq 3\right\}}$, which leads to that $\sum_{f = 1}^{1000}\mathds{1}_{\left\{t_{\mathcal{S}}^{\text{p}}\left(\omega_f;\alpha^*_1\right)\leq 3\right\}}/1000\leq\sum_{f = 1}^{1000}\mathds{1}_{\left\{t_{\mathcal{S}}^{\text{p}}\left(\omega_f;\alpha^*_2\right)\leq 3\right\}}/1000$; indeed, since $t_{\mathcal{S}}^{\text{p}}\left(\omega_f;\alpha^*_1\right)\leq 3$, or equivalently $v_{\mathcal{S}}^{\text{p}}\left(\omega_f;\alpha^*_1\right) \leq 25$, we have $p^{\left(v_{\mathcal{S}}^{\text{p}}\left(\omega_f;\alpha^*_1\right),j\right)}_{\mathcal{S}}\left(\omega_f\right) < \alpha^*_1\leq\alpha^*_2$, and thus
\begin{equation*}
v_{\mathcal{S}}^{\text{p}}\left(\omega_f;\alpha^*_2\right) = \min\left\{v = 1, 2, \dots, 25: p^{\left(v,j\right)}_{\mathcal{S}}\left(\omega_f\right)<\alpha^*_2\right\}\leq v_{\mathcal{S}}^{\text{p}}\left(\omega_f;\alpha^*_1\right)\leq 25,
\end{equation*}
or equivalently $t_{\mathcal{S}}^{\text{p}}\left(\omega_f;\alpha^*_2\right) \leq t_{\mathcal{S}}^{\text{p}}\left(\omega_f;\alpha^*_1\right)\leq 3$. However, notably, the declining rate of the estimated proportion for exceeding CD is greater than that for exceeding ID.

\begin{figure}[H]
\centering
\includegraphics[width = 0.5\linewidth]{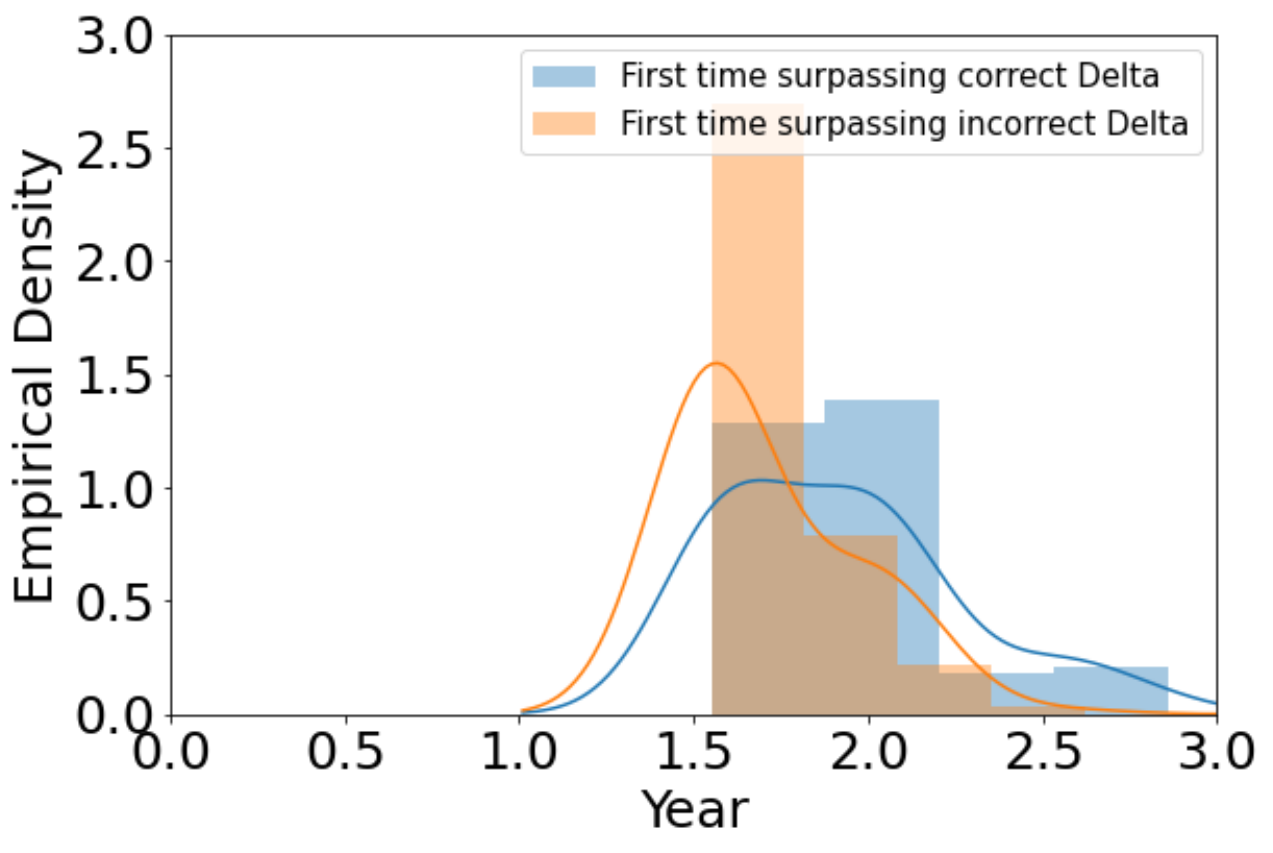}
\caption{Empirical conditional density functions of first statistically significant surpassing times conditioning on reinforcement learning agent with online learning phase being statistically significant to be exceeding correct Delta within $3$ years for $0.1$ level of significance}
\label{fig:hitting_time_p_0.9}
\end{figure}

Similar to Figure \ref{fig:hitting_time} and Table \ref{table:sum_catchup}, one can depict the empirical conditional density functions and list the summary statistics of $t_{\text{CD}}^{\text{p}}\left(\cdot;\alpha^*\right)$ and $t_{\text{ID}}^{\text{p}}\left(\cdot;\alpha^*\right)$, for each level of significance $\alpha^*$, subject to that RLw/OL is statistically significant to be exceeding CD within $3$ years. For example, with $\alpha^*=0.1$, Figure \ref{fig:hitting_time_p_0.9} and Table \ref{table:sum_catchup_p_0.9} illustrate that, comparing with Figure \ref{fig:hitting_time} and Table \ref{table:sum_catchup}, the distributions are right-shifted as well as more spread, and the summary statistics are all increased.

\begin{table}[H]
\centering
\begin{tabular}{@{}rccccccccc@{}}
\toprule
Reinforcement Learning Agent & \multirow{3}{*}{Mean} & \multirow{3}{*}{Median} & \multirow{3}{*}{Std. Dev.} & \multirow{3}{*}{$\text{VaR}_{90}$} & \multirow{3}{*}{$\text{VaR}_{95}$} & \multirow{3}{*}{$\text{TVaR}_{90}$} & \multirow{3}{*}{$\text{TVaR}_{95}$} &
\\
with Online Learning Phase & &  &  &  &  &  &  & 
\\
First Surpassing Time to & &  &  &  &  &  &  & 
\\\midrule
Correct Delta & $1.92$ &  $1.90$ & $0.34$ & $2.50$ & $2.62$ & $2.61$ & $2.70$  \\
Incorrect Delta  & $1.70$ & $1.55$ & $0.24$ & $2.02$ & $2.14$ & $2.07$ & $2.20$ \\
\bottomrule                     
\end{tabular}
\caption{Summary statistics of empirical conditional distributions of first statistically significant surpassing times conditioning on reinforcement learning agent with online learning phase being statistically significant to be exceeding correct Delta within $3$ years for $0.1$ level of significance}
\label{table:sum_catchup_p_0.9}
\end{table}

\begin{figure}[H]
\centering
\begin{adjustbox}{minipage=\linewidth,scale=0.9}
\begin{subfigure}{0.5\columnwidth}
\centering
\includegraphics[width=0.8\textwidth]{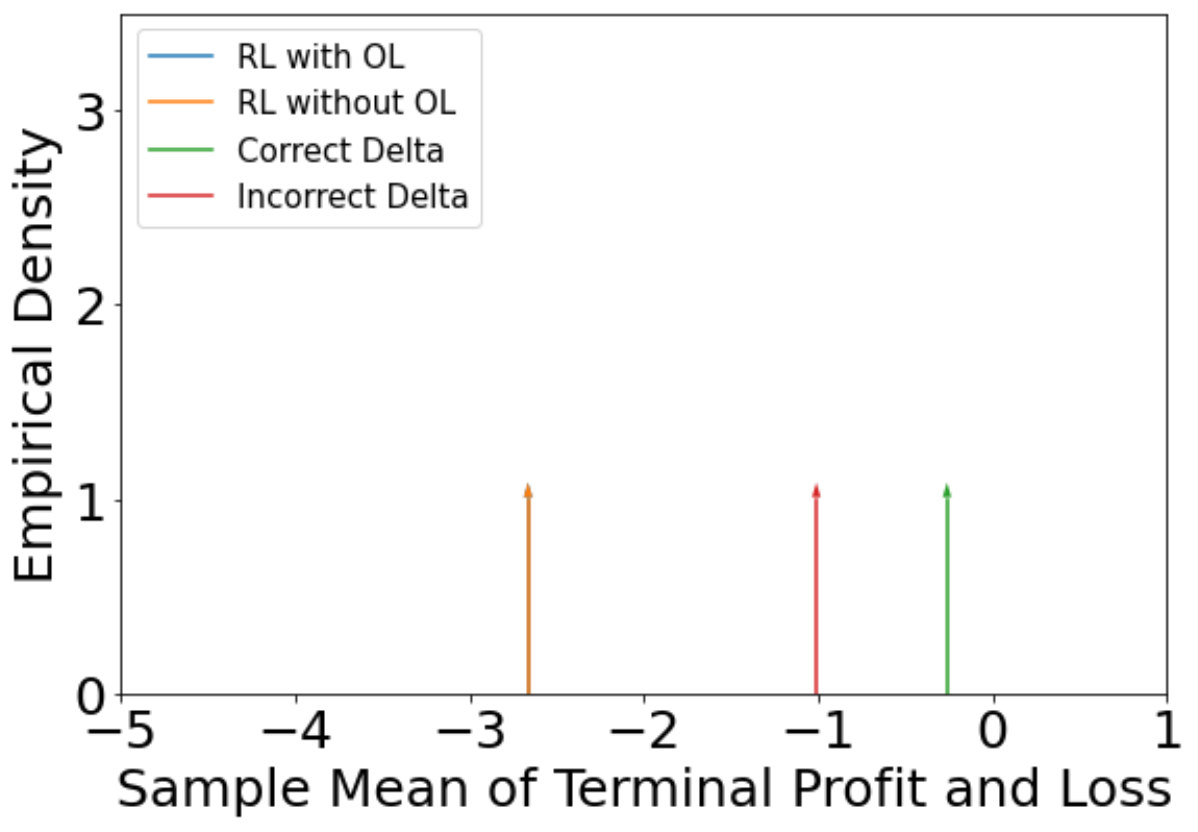}
\caption{$t = 0$}
\label{fig:eval_0}
\end{subfigure}\hfill
\begin{subfigure}{0.5\columnwidth}
\centering
\includegraphics[width=0.8\textwidth]{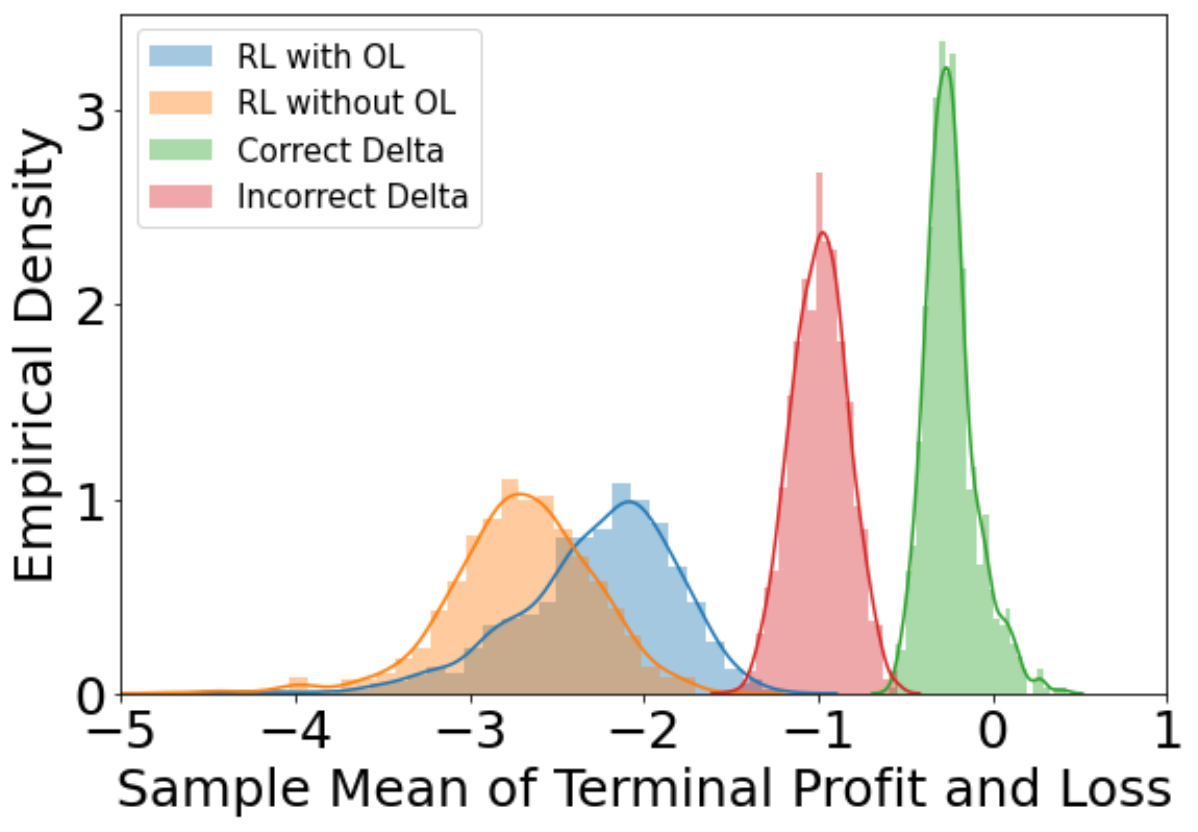}
\caption{$t = 0.6$}
\label{fig:eval_5}
\end{subfigure}

\begin{subfigure}{0.5\columnwidth}
\centering
\includegraphics[width=0.8\textwidth]{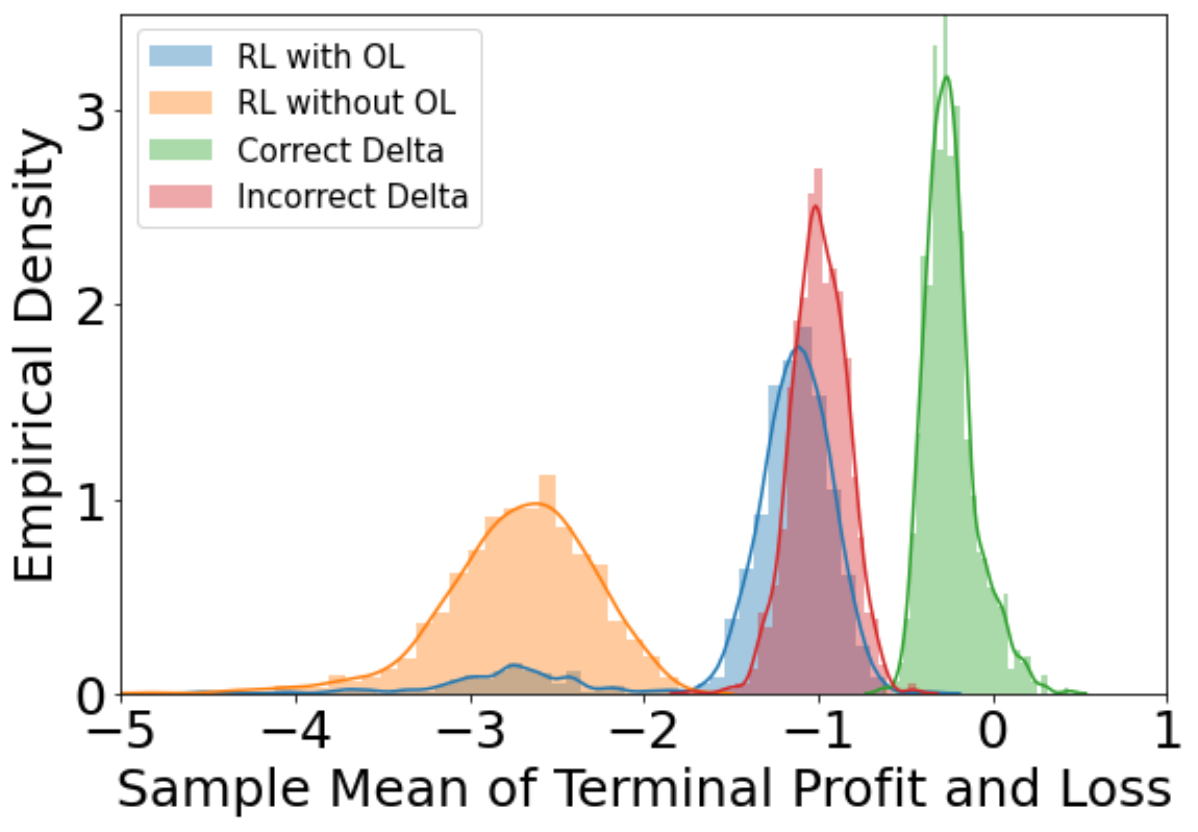}
\caption{$t = 1.2$}
\label{fig:eval_10}
\end{subfigure}\hfill
\begin{subfigure}{0.5\columnwidth}
\centering
\includegraphics[width=0.8\textwidth]{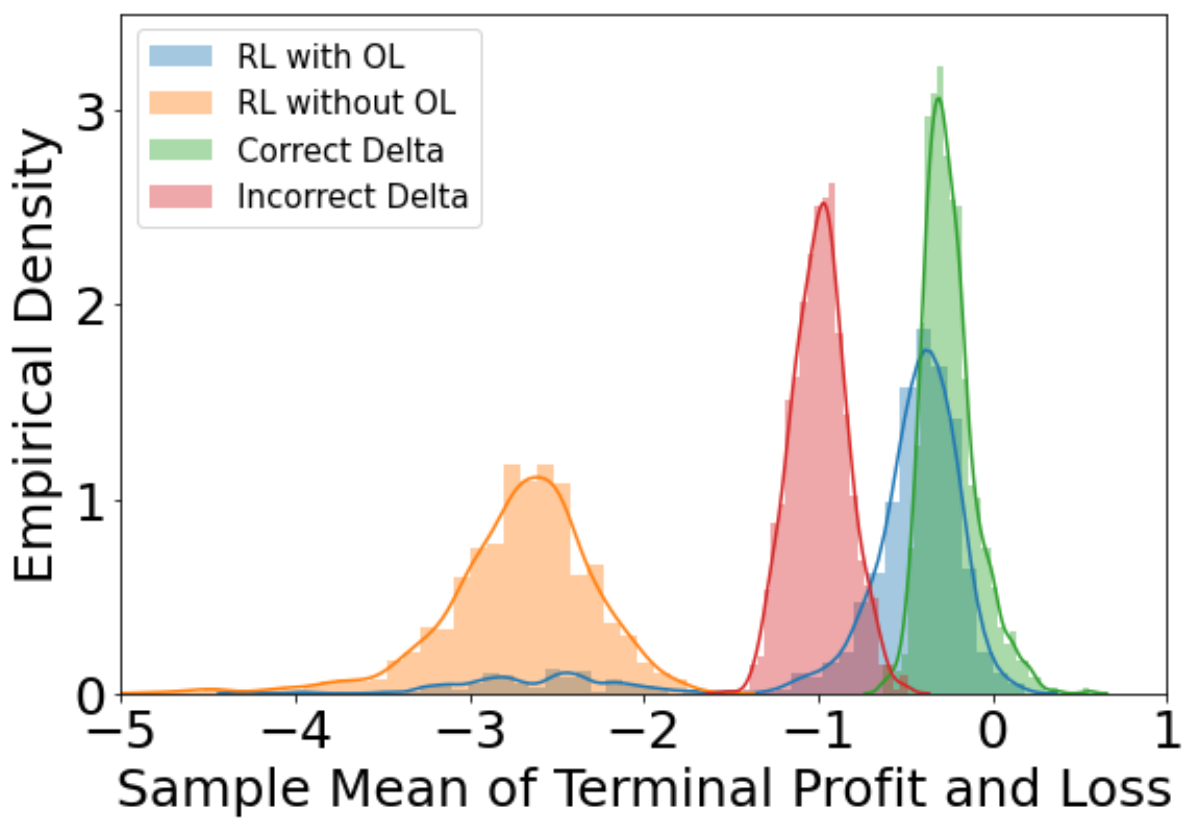}
\caption{$t = 1.8$}
\label{fig:eval_15}
\end{subfigure}
\begin{subfigure}{0.5\columnwidth}
\centering
\includegraphics[width=0.8\textwidth]{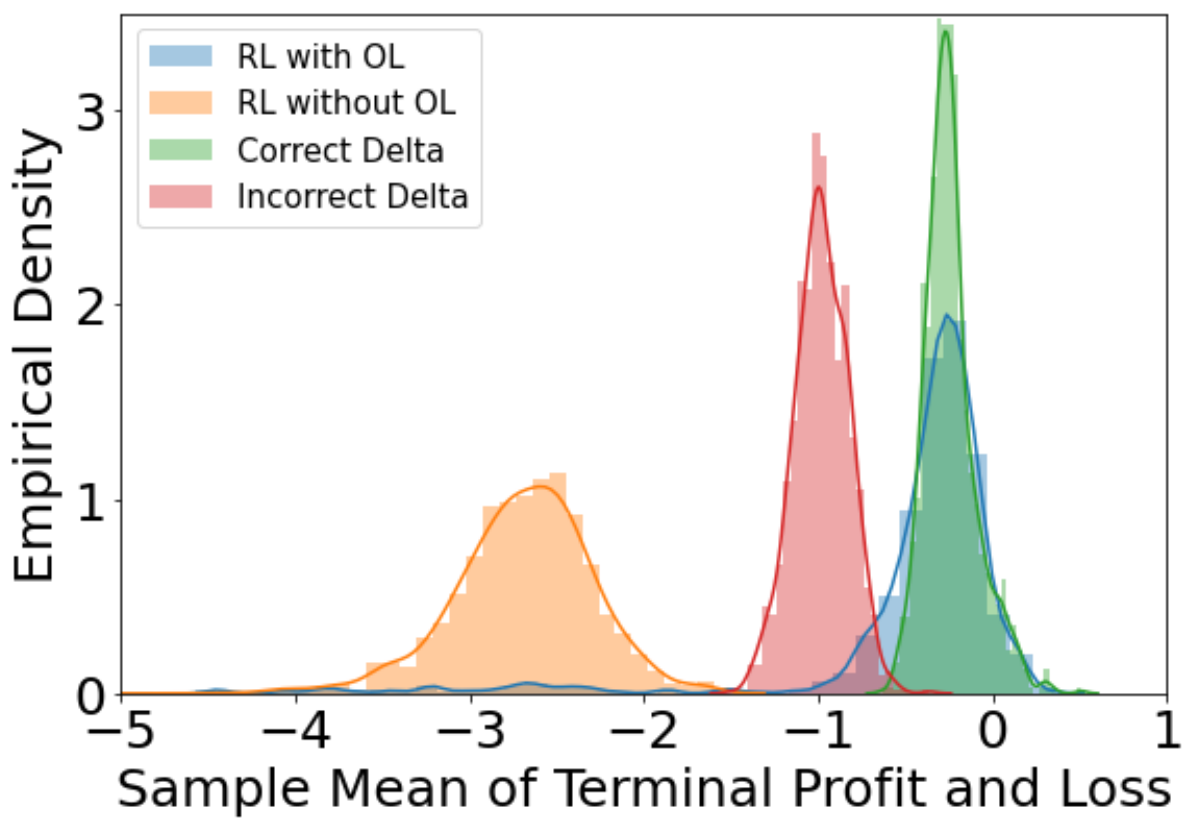}
\caption{$t = 2.4$}
\label{fig:eval_20}
\end{subfigure}\hfill
\begin{subfigure}{0.5\columnwidth}
\centering
\includegraphics[width=0.8\textwidth]{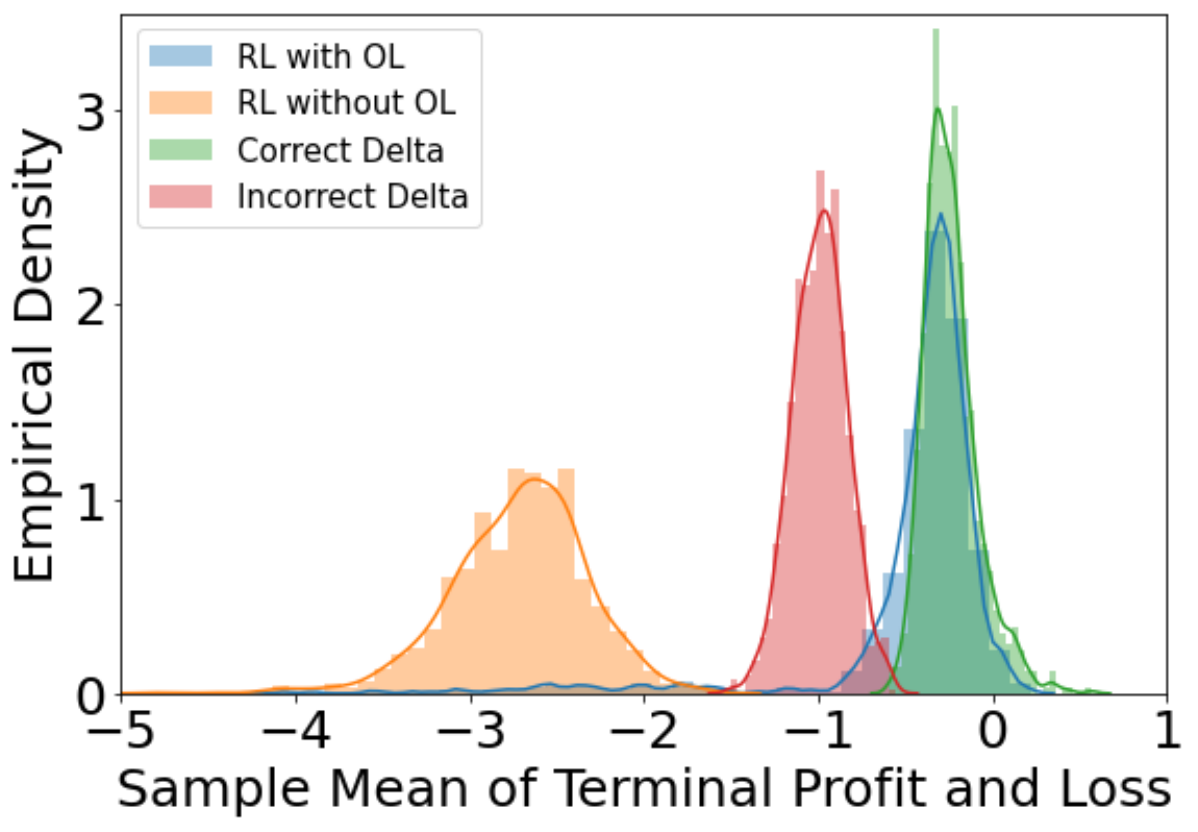}
\caption{$t = 3$}
\label{fig:eval_25}
\end{subfigure}
\end{adjustbox}
\caption{Snapshots of empirical density functions of sample mean of terminal P\&L by reinforcement learning agent with online learning phase, reinforcement learning agent without online learning phase, correct Delta, and incorrect Delta at different time points}
\label{fig:shift_of_online}
\end{figure}

Finally, to further examine the hedging performance of RLw/OL in terms of the sample mean of the terminal P\&L in \eqref{eq:sample_mean}, as well as take the random future trajectories into account, Figure \ref{fig:shift_of_online} shows the snapshots of the empirical density functions, among the future trajectories, of the sample mean by each hedging strategy over time at $t=0,0.6,1.2,1.8,2.4\text{ and }3$; Table \ref{table:mean_stats} outlines their summary statistics. Note that, at the current time $t=0$, since none of the future trajectories has been realized yet, the empirical density functions are given by Dirac delta at the corresponding sample mean by each hedging strategy, which only depends on the simulated scenarios. As the time progresses, one can observe that the empirical density function by RLw/OL is gradually shifting to the right, substantially passing the one by ID and almost catching up the one by CD at $t=1.8$. This sheds light on the high probability that RLw/OL is able to self-revise the hedging strategy from a very sub-optimal one to a nearly optimal one close to the CD.

\begin{table}[H]
\centering
\begin{subtable}{1\linewidth}
\centering
\begin{tabular}{@{}rccccccccc@{}}
\toprule
Sample Mean of & \multirow{2}{*}{Mean} & \multirow{2}{*}{Median} & \multirow{2}{*}{Std. Dev.} & \multirow{2}{*}{$\text{VaR}_{90}$} & \multirow{2}{*}{$\text{VaR}_{95}$} & \multirow{2}{*}{$\text{TVaR}_{90}$} & \multirow{2}{*}{$\text{TVaR}_{95}$} &
\\
Terminal P\&L by & &  &  &  &  &  &  & 
\\\midrule
RL with OL  & $-2.66$ &  $-2.66$ & $0$ & $-2.66$ & $-2.66$ & $-2.66$ & $-2.66$ \\
RL without OL  & $-2.66$ &  $-2.66$ & $0$ & $-2.66$ & $-2.66$ & $-2.66$ & $-2.66$\\
Correct Delta  & $-0.26$ &  $-0.26$ & $0$ & $-0.26$ & $-0.26$ & $-0.26$ & $-0.26$ \\
Incorrect Delta  & $-1.01$ & $-1.01$ & $0$ & $-1.01$ & $-1.01$ & $-1.01$ & $-1.01$\\
\bottomrule                     
\end{tabular}
\caption{$t = 0$}
\end{subtable}
\medskip
\begin{subtable}{1\linewidth}
\centering
\begin{tabular}{@{}rccccccccc@{}}
\toprule
Sample Mean of & \multirow{2}{*}{Mean} & \multirow{2}{*}{Median} & \multirow{2}{*}{Std. Dev.} & \multirow{2}{*}{$\text{VaR}_{90}$} & \multirow{2}{*}{$\text{VaR}_{95}$} & \multirow{2}{*}{$\text{TVaR}_{90}$} & \multirow{2}{*}{$\text{TVaR}_{95}$} &
\\
Terminal P\&L by & &  &  &  &  &  &  & 
\\\midrule
RL with OL  & $-2.27$ &  $-2.19$ & $0.45$ & $-2.86$ & $-3.11$ & $-3.17$ & $-3.38$ \\
RL without OL  & $-2.71$ &  $-2.69$ & $0.42$ & $-3.20$ & $-3.39$ & $-3.52$ & $-3.76$\\
Correct Delta  & $-0.24$ &  $-0.26$ & $0.15$ & $-0.40$ & $-0.45$ & $-0.46$ & $-0.49$ \\
Incorrect Delta  & $-0.99$ &  $-0.99$ & $0.16$ & $-1.20$ & $-1.26$ & $-1.27$ & $-1.31$\\
\bottomrule                     
\end{tabular}
\caption{$t = 0.6$}
\end{subtable}
\medskip
\begin{subtable}{1\linewidth}
\centering
\begin{tabular}{@{}rccccccccc@{}}
\toprule
Sample Mean of & \multirow{2}{*}{Mean} & \multirow{2}{*}{Median} & \multirow{2}{*}{Std. Dev.} & \multirow{2}{*}{$\text{VaR}_{90}$} & \multirow{2}{*}{$\text{VaR}_{95}$} & \multirow{2}{*}{$\text{TVaR}_{90}$} & \multirow{2}{*}{$\text{TVaR}_{95}$} &
\\
Terminal P\&L by & &  &  &  &  &  &  & 
\\\midrule
RL with OL  & $-1.29$ &  $-1.14$ & $0.55$ & $-1.83$ & $-2.75$ & $-2.80$ & $-3.10$ \\
RL without OL & $-2.71$ & $-2.68$ & $0.42$ & $-3.22$ & $-3.45$ & $-3.54$ & $-3.78$ \\
Correct Delta  & $-0.24$ & $-0.26$ & $0.14$ & $-0.41$ & $-0.44$ & $-0.45$ & $-0.48$ \\
Incorrect Delta  & $-0.99$ &  $-0.99$ & $0.16$ & $-1.19$ & $-1.25$ & $-1.27$ & $-1.33$\\
\bottomrule                     
\end{tabular}
\caption{$t = 1.2$}
\end{subtable}
\medskip
\begin{subtable}{1\linewidth}
\centering
\begin{tabular}{@{}rccccccccc@{}}
\toprule
Sample Mean of & \multirow{2}{*}{Mean} & \multirow{2}{*}{Median} & \multirow{2}{*}{Std. Dev.} & \multirow{2}{*}{$\text{VaR}_{90}$} & \multirow{2}{*}{$\text{VaR}_{95}$} & \multirow{2}{*}{$\text{TVaR}_{90}$} & \multirow{2}{*}{$\text{TVaR}_{95}$} &
\\
Terminal P\&L by & &  &  &  &  &  &  & 
\\\midrule
RL with OL  & $-0.63$ &  $-0.43$ & $0.69$ & $-1.14$ & $-2.50$ & $-2.52$ & $-2.94$ \\
RL without OL & $-2.70$ &  $-2.67$ & $0.43$ & $-3.22$ & $-3.42$ & $-3.58$ & $-3.86$ \\
Correct Delta  & $-0.25$ &  $-0.27$ & $0.15$ & $-0.41$ & $-0.45$ & $-0.47$ & $-0.50$ \\
Incorrect Delta  & $-0.99$ &  $-0.99$ & $0.15$ & $-1.20$ & $-1.25$ & $-1.27$ & $-1.31$\\
\bottomrule                     
\end{tabular}
\caption{$t = 1.8$}
\end{subtable}
\medskip
\begin{subtable}{1\linewidth}
\centering
\begin{tabular}{@{}rccccccccc@{}}
\toprule
Sample Mean of & \multirow{2}{*}{Mean} & \multirow{2}{*}{Median} & \multirow{2}{*}{Std. Dev.} & \multirow{2}{*}{$\text{VaR}_{90}$} & \multirow{2}{*}{$\text{VaR}_{95}$} & \multirow{2}{*}{$\text{TVaR}_{90}$} & \multirow{2}{*}{$\text{TVaR}_{95}$} &
\\
Terminal P\&L by & &  &  &  &  &  &  & 
\\\midrule
RL with OL  & $-0.46$ &  $-0.33$ & $0.71$ & $-0.72$ & $-2.24$ & $-2.13$ & $-3.24$ \\
RL without OL & $-2.69$ &  $-2.66$ & $0.41$ & $-3.18$ & $-3.40$ & $-3.48$ & $-3.69$ \\
Correct Delta  & $-0.24$ &  $-0.26$ & $0.15$ & $-0.40$ & $-0.45$ & $-0.46$ & $-0.50$ \\
Incorrect Delta  & $-0.98$ &  $-0.98$ & $0.15$ & $-1.18$ & $-1.24$ & $-1.26$ & $-1.30$\\
\bottomrule                     
\end{tabular}
\caption{$t = 2.4$}
\end{subtable}
\begin{subtable}{1\linewidth}
\centering
\begin{tabular}{@{}rccccccccc@{}}
\toprule
Sample Mean of & \multirow{2}{*}{Mean} & \multirow{2}{*}{Median} & \multirow{2}{*}{Std. Dev.} & \multirow{2}{*}{$\text{VaR}_{90}$} & \multirow{2}{*}{$\text{VaR}_{95}$} & \multirow{2}{*}{$\text{TVaR}_{90}$} & \multirow{2}{*}{$\text{TVaR}_{95}$} &
\\
Terminal P\&L by & &  &  &  &  &  &  & 
\\\midrule
RL with OL  & $-0.45$ &  $-0.33$ & $0.56$ & $-0.66$ & $-1.66$ & $-1.75$ & $-2.59$ \\
RL without OL & $-2.71$ &  $-2.68$ & $0.41$ & $-3.24$ & $-3.38$ & $-3.49$ & $-3.68$ \\
Correct Delta  & $-0.24$ &  $-0.26$ & $0.15$ & $-0.40$ & $-0.44$ & $-0.46$ & $-0.49$ \\
Incorrect Delta  & $-0.99$ & $-0.99$ & $0.15$ & $-1.19$ & $-1.24$ & $-1.26$ & $-1.31$\\
\bottomrule                     
\end{tabular}
\caption{$t = 3$}
\end{subtable}
\caption{Summary statistics of empirical distributions of sample mean of terminal P\&L by reinforcement learning agent with online learning phase, reinforcement learning agent without online learning phase, correct Delta, and incorrect Delta at different time points}
\label{table:mean_stats}
\end{table}

\section{Methodological Assumptions and Implications in Practice}\label{sec:assumption_practice}

To apply the proposed two-phase RL approach to a hedging problem of contingent claims, there are at least four assumptions to be satisfied. This section discusses these assumptions and elaborates their implications in practice.


\subsection{Observable, Sufficient, Relevant, and Transformed Features in State}\label{sec:state_assumption}
One of the crucial components in an MDP environment of the training phase or the online learning phase is the state, in which the features provide information from the environment to the RL agent. First, the features must be observable by the RL agent for learning. For instance, in our proposed state vectors \eqref{eq:state_vector} and \eqref{eq:state_vector_OL}, all the four features, namely, the segregated account value, the hedging portfolio value, the number of surviving policyholders, and the term to maturity, are observable. Any unobservable, albeit desirable, features cannot be included in the state, such as insider information which could provide a better inference on the future value of a risky asset, or exact health condition of a policyholder. Second, the observable features in the state should be sufficient for the RL agent to learn. For example, due to the dual-risk bearing nature of the contract in this paper, the proposed state vectors \eqref{eq:state_vector} and \eqref{eq:state_vector_OL} incorporate both financial and actuarial features; also, the third and the fourth features in the state vectors \eqref{eq:state_vector} and \eqref{eq:state_vector_OL} would inform the RL agent to halt its hedging at the terminal time. However, incorporating sufficient observable features in the state does not imply that every observable feature in the environment should be included; the observable features in the state need to be relevant for learning efficiently. Since the segregated account value and the term to maturity have already been included in the state vectors \eqref{eq:state_vector} and \eqref{eq:state_vector_OL} as features, the risky asset value and the hedging time are respective similar information from the environment, and thus are redundant features to be contained in the state. Finally, the features in the state which have high variance might be appropriately transformed for reducing the volatility due to exploration. For instance, the segregated account value in the state vectors \eqref{eq:state_vector} and \eqref{eq:state_vector_OL} is log-transformed in both phases.

\subsection{Reward Engineering}
Another crucial component in an MDP environment is the reward, which supplies signals to the RL agent to evaluate its actions, i.e. the hedging strategy, for learning. First, the reward signals, if available, should suggest the local hedging performance. For example, in this paper, the RL agent is provided by the sequential anchor-hedging reward, given in \eqref{eq:reward_1}, in the training phase; through the net liability value in the MDP training environment, the RL agent often receives a positive (resp. negative) signal for encouragement (resp. punishment), which is more informative than collecting the zero reward. However, any informative reward signals need to be computable from an MDP environment. In this paper, since the insurer does not know the MDP market environment, the RL agent could not be supplied the sequential anchor-hedging reward signals, which consist of the net liability values, in the online learning phase, even though they are more informative; instead, the RL agent is given the less informative single terminal reward, given in \eqref{eq:reward_2}, in the online learning phase which can be computed from the market environment.

\subsection{Markov Property in State and Action}
In an MDP environment of the training phase or the online learning phase, the state and action pair needs to satisfy the Markov property as in \eqref{eq:markov_property}. In the training phase, since the MDP training environment is constructed, the Markov property can be verified theoretically for the state, with the included features in line with Section \ref{sec:state_assumption}, and the action, which is the hedging strategy. For example, in this paper, with the model of the market environment being the BS and the CFM, the state vector in \eqref{eq:state_vector} and the Markovian hedging strategy satisfy the Markov property in the training phase. Since the illustrative example in this paper assumes that the market environment also follows the BS and the CFM, the state vector in \eqref{eq:state_vector_OL} and the Markovian hedging strategy satisfy the Markov property in the online learning phase as well. However, in general, as the market environment is unknown, the Markov property for the state and action pair would need to be checked statistically in the online phase as follows.

After the training phase and before an RL agent proceeding to the online learning phase, historical state and action sequences in a time frame are derived by hypothetically writing identical contingent claims and using the historical realizations from the market environment. For instance, historical values of risky assets are publicly available, or an insurer retrieves historical survival status of its policyholders with similar demographic information and medical history as the policyholder being actually written. These historical samples of the state and action pair are then used to conduct hypothesis testing on whether the Markov property in \eqref{eq:markov_property} holds for the pair in the market environment, by, for example, the test statistics proposed in \cite{Chen_2012}. If the Markov property holds statistically, the RL agent could begin the online learning phase. Yet, if the property does not hold statistically, the state and action pair should be revised and then the training phase should be revisited; since the hedging strategy is the action in a hedging problem, only the state could be amended by including more features from the environment. Moreover, during the online learning phase, right after each further update step, new historical state and action sequences in a shifted time frame of the same duration are obtained together with the most recent historical realizations from the market environment and using the action samples being drawn from the updated policy. These regularly new samples should be applied to statistically verify the Markov property on a rolling basis. If the property fails to hold at any time, the state needs to be revised and the RL agent must be re-trained before resuming the online learning.

\subsection{Re-Establishment of Contingent Claims in Online Learning Phase}
Any contingent claims must have a finite terminal time realization. On one hand, in the training phase, that would be the time when an episode ends and the state is re-initialized so that the RL agent can be trained in the training environment as long as possible. On the other hand, in the online learning phase, the market environment, and hence the state, could not be re-initialized; instead, at each terminal time realization, the seller re-establishes identical contingent claims of the same contract characteristics and writing on (more or less) the same assets so that the RL agent can be trained in the market environment successively. In this paper, the terms to maturity and the minimum guarantees of all variable annuity contracts in the online learning phase are the same. Moreover, all re-established contracts therein write on the same financial risky asset, though the initial values of the asset are given by the real-time realizations in the market environment. Finally, while a new policyholder is written at each contract inception time, these policyholders have similar, if not identical, distributions of their random future lifetimes via examining their demographic information and medical history.

\section{Concluding Remarks and Future Directions}\label{sec:conclusion}
This paper proposed the two-phase deep RL approach which can tackle practically common model miscalibration in hedging variable annuity contracts with both GMMB and GMDB riders in the BS financial and CFM actuarial market environments. The approach is composed of the training phase and the online learning phase. While the satisfactory hedging performance of the trained RL agent in the training environment was anticipated, the performance by the further trained RL agent in the market environment via the illustrative example should be highlighted. First, by comparing their sample means of terminal P\&L from simulated scenarios, in most future trajectories, within a reasonable amount of time, the further trained RL agent was able to exceed the hedging performance by the correct Delta from the market environment and the incorrect Delta from the training environment. Second, through a more delicate hypothesis testing analysis, similar conclusions can be drawn in a fair amount of future trajectories. Finally, snapshots of empirical density functions, among the future trajectories, of the sample means of terminal P\&L from simulated scenarios by each hedging strategy, shed light on the high probability that, the further trained RL agent is indeed able to self-revise the hedging strategy.

There should be at least two future directions derived from this paper. (I) The market environment in the illustrative example of this paper was assumed to be the BS financial and CFM actuarial models, which turned out to be the same as designed by the insurer for the training environment, with different parameters though. Moreover, the policyholders were assumed to be homogeneous that their survival probabilities and investment behaviors are all the same, with even identical contracts of the same minimum guarantee and maturity. In the market environment, the agent only had to hedge one contract at a time, instead of a portfolio of contracts. Obviously, if any of these is to be relaxed, the trained RL agent from the current training environment should not be able to produce satisfactory hedging performance in a market environment. Therefore, the training environment will certainly need to be substantially extended in terms of its sophistication, in order for the trained RL agent to be able to further learn and hedge well in any realistic market environments. (II) Beyond this, an even more ambitious question needs to be addressed is that how much similar do the training and market environments have to be, such that the online learning for self-revision on hedging strategy is possible, if not efficient. This second future direction is related to the transfer learning being adapted to the variable annuities hedging problem, and shall be investigated carefully in the future.

\newpage
\begin{appendices}
\section{Deep Hedging Approach}\label{app:dh_agent_1}
In this section, we provide a brief review of the DH approach adapted from \cite{Buhler_2019}. In particular, the hedging objective of the insurer is still given as $\sqrt{\mathbb{E}\left[\left(P_{t_{\tilde{n}}}-L_{t_{\tilde{n}}}\right)^2\right]}$, with Equation \eqref{eq:hedging} being the optimal (discrete) hedging strategy. The hedging agent built by the insurer using the DH algorithm shall be called the DH agent hereafter.
\subsection{Deterministic Action}
Different from Section \ref{sec:stoc_action}, in which the RL agent takes a stochastic action which is sampled from the policy for the exploration in the MDP environment, the DH agent only deploys a deterministic action $H^{\text{DH}}: \mathcal{X} \to \mathcal{A}$, which is a direct mapping from the state space to the action space. Specifically, at each time $t_k$, where $k = 0,1, \dots, n-1$, given the current state $X_{t_k} \in \mathcal{X}$, the DH agent takes an action $H^{\text{DH}}\left(X_{t_k}\right) \in \mathcal{A}$. In this case, the objective of the DH agent is to solve for the optimal hedging strategy $H^{\text{DH},*}\left(\cdot\right)$ that minimizes $\sqrt{\mathbb{E}\left[\left(P_{t_{\tilde{n}}}-L_{t_{\tilde{n}}}\right)^2\right]}$, or equivalently minimizes $\mathbb{E}\left[\left(P_{t_{\tilde{n}}}-L_{t_{\tilde{n}}}\right)^2\right]$. 

\subsection{Action Approximation and Parameterization}
The deterministic action mapping $H^{\text{DH}}: \mathcal{X} \to \mathcal{A}$ is then approximated and parameterized by an ANN with weights $\upsilon_{\text{a}}$. 
The construction of such ANN $\mathcal{N}_{\text{a}}\left(\cdot;\upsilon_{\text{a}}\right)$ is similar to that in Section \ref{sec:pc_network}, except that $\mathcal{N}_{\text{a}}\left(x;\upsilon_{\text{a}}\right) \in \mathbb{R}$ for any $x \in \mathbb{R}^{p}$; that is, $\mathcal{N}_{\text{a}}\left(\cdot;\upsilon_{\text{a}}\right)$ takes a state vector $x\in \mathbb{R}^p$ as the input, and directly outputs a deterministic action $a\left(x;\upsilon_{\text{a}}\right) \in \mathbb{R}$, instead of the Gaussian mean-variance tuple $\left(c\left(x;\upsilon_{\text{a}}\right),d^2\left(x;\upsilon_{\text{a}}\right)\right) \in \mathbb{R} \times \mathbb{R}^+$ in the RL approach, which then samples an action from the Gaussian measure. Hence, in the DH approach, solving the optimal hedging strategy $H^{\text{DH},*}\left(\cdot\right)$ boils down to finding the optimal weights  $\upsilon_{a}^*$.

\subsection{Deep Hedging Method}\label{app:DHM}
The DH agent starts from initial ANN weights $\upsilon_{\text{a}}^{\left(0\right)}$, deploys the hedging strategy to collect terminal P\&Ls, and gradually updates the ANN weights by stochastic gradient ascent as shown in Equation \eqref{eq:sga}, with $\theta$ replaced by $\upsilon$. For the DH agent, at each update step $u = 1, 2, \dots$, the surrogate performance measure is given as 
$$
\mathcal{J}^{\left(u-1\right)}\left(\upsilon_{\text{a}}^{\left(u-1\right)}\right) = -\mathbb{E}\left[\left(P_{t_{\tilde{n}}}^{\left(u-1\right)}-L_{t_{\tilde{n}}}^{\left(u-1\right)}\right)^2\right].
$$
Correspondingly, the gradient of the surrogate performance measure with respect to the ANN weights $\upsilon_{\text{a}}$ is
$$
\nabla_{\upsilon_{\text{a}}}\mathcal{J}^{\left(u-1\right)}\left(\upsilon_{\text{a}}^{\left(u-1\right)}\right) = -2\mathbb{E}\left[\left(P_{t_{\tilde{n}}}^{\left(u-1\right)}-L_{t_{\tilde{n}}}^{\left(u-1\right)}\right)\nabla_{\upsilon_{\text{a}}}P_{t_{\tilde{n}}}^{\left(u-1\right)}\right].
$$
Therefore, based on the \textit{realized} terminal P\&L $p_{t_{\tilde{n}}}^{\left(u-1\right)}$ and $l_{t_{\tilde{n}}}^{\left(u-1\right)}$, the estimated gradient is given as 
$$
\widehat{\nabla_{\upsilon_{\text{a}}}\mathcal{J}^{\left(u-1\right)}\left(\upsilon_{\text{a}}^{\left(u-1\right)}\right)} = -2\left(p_{t_{\tilde{n}}}^{\left(u-1\right)} - l_{t_{\tilde{n}}}^{\left(u-1\right)}\right)\nabla_{\upsilon_{\text{a}}}p_{t_{\tilde{n}}}^{\left(u-1\right)}.
$$
Algorithm \ref{algo:dh} summarizes the DH method above.
\begin{algorithm}[H]
\SetAlgoLined
\textbf{Input} initial ANN model $\mathcal{N}_{\text{a}}\left(\cdot;\upsilon_{\text{a}}^{\left(0\right)}\right)$, total number of updates $\hat{M} \in \mathbb{N}$, learning rate $\hat{\alpha} \in [0,1]$.\\
\For{$u = 1, 2, \cdots, \hat{M}$}{
$\boldsymbol{\cdot}$ Initialize the MDP training environment and observe the initial state vector $x_{t_0}^{\left(u-1\right)}$.\\
$\boldsymbol{\cdot}$ Follow the hedging strategy $\mathcal{N}_{\text{a}}\left(\cdot;\upsilon_{\text{a}}^{\left(u-1\right)}\right)$ to realize an episode and evaluate the terminal P\&L $p_{t_{\tilde{n}}}^{\left(u-1\right)}$ and $l_{t_{\tilde{n}}}^{\left(u-1\right)}$.\\
$\boldsymbol{\cdot}$ Update $\upsilon_{\text{a}}^{\left(u-1\right)}$ as
$$\upsilon_{\text{a}}^{\left(u\right)} = \upsilon_{\text{a}}^{\left(u-1\right)} - 2\hat{\alpha} \left(p_{t_{\tilde{n}}}^{\left(u-1\right)} - l_{t_{\tilde{n}}}^{\left(u-1\right)}\right)\nabla_{\upsilon_{\text{a}}}p_{t_{\tilde{n}}}^{\left(u-1\right)}.$$
}
\textbf{Return} the trained ANN model $\mathcal{N}_{\text{a}}\left(\cdot;\upsilon_{\text{a}}^{\left(\hat{M}\right)}\right)$.
\caption{Pseudo-code for deep hedging method}
\label{algo:dh}
\end{algorithm}

Compared with policy gradient methods introduced in Section \ref{sec:PPO}, the DH method shows two key differences. First, it assumes that the hedging portfolio value $P_{t_{\tilde{n}}}^{\left(u-1\right)}$ is differentiable with respect to $\upsilon_{\text{a}}$ at each update $u = 1, 2,\dots$. Second, the update of ANN weights does not depend on intermediate rewards collected during an episode; that is, to update the weights, the DH agent has to experience a complete episode to realize the terminal P\&L. Therefore, the update frequency of the DH method is lower than that of the RL method with TD feature.

\section{REINFORCE: A Monte Carlo Policy Gradient Method}\label{sec:REINFORCE}



At each update step $u=1,2,\dots$, based on the ANN weights $\theta^{\left(u-1\right)}$, and thus the policy $\pi\left(\cdot;\theta_{\text{p}}^{\left(u-1\right)}\right)$, the RL agent experiences the realized episode:
\begin{equation*}
\left\{x_{t_0}^{\left(u-1\right)},h_{t_0}^{\left(u-1\right)},x_{t_1}^{\left(u-1\right)},r_{t_1}^{\left(u-1\right)},h_{t_1}^{\left(u-1\right)},\dots,x_{t_{\tilde{n}-1}}^{\left(u-1\right)},r_{t_{\tilde{n}-1}}^{\left(u-1\right)},h_{t_{\tilde{n}-1}}^{\left(u-1\right)},x_{t_{\tilde{n}}}^{\left(u-1\right)},r_{t_{\tilde{n}}}^{\left(u-1\right)}\right\},
\end{equation*}
where $h_{t_k}^{\left(u-1\right)}$, for $k=0,1,\dots,\tilde{n}-1$, is the time-$t_k$ realized hedging strategy being sampled from the Gaussian distribution with the mean $c\left(x_{t_k}^{\left(u-1\right)};\theta_{\text{p}}^{\left(u-1\right)}\right)$ and the variance $d^2\left(x_{t_k}^{\left(u-1\right)};\theta_{\text{p}}^{\left(u-1\right)}\right)$. In the following, fix an update step $u=1,2,\dots$.

REINFORCE takes directly the time-$0$ value function $V^{\left(u-1\right)}\left(0,x;\theta_{\text{p}}\right)$, for any $x\in\mathcal{X}$, as a part of the surrogate performance measure:
\begin{equation*}
V^{\left(u-1\right)}\left(0,x;\theta_{\text{p}}\right)=\mathbb{E}\left[\sum_{k=0}^{\tilde{n}-1}R_{t_{k+1}}^{\left(u-1\right)}\Big\vert X^{\left(u-1\right)}_{0}=x\right].
\end{equation*}
In \cite{Williams_1992}, the {\it Policy Gradient Theorem} was proved, which states that
\begin{equation*}
\nabla_{\theta_{\text{p}}}V^{\left(u-1\right)}\left(0,x;\theta_{\text{p}}\right)=\mathbb{E}\left[\sum_{k=0}^{\tilde{n}-1}\left(\sum_{l=k}^{\tilde{n}-1}R_{t_{l+1}}^{\left(u-1\right)}\right)\nabla_{\theta_{\text{p}}}\ln\phi\left(H^{\left(u-1\right)}_{t_k};X^{\left(u-1\right)}_{t_k},\theta_{\text{p}}\right)\Big\vert X^{\left(u-1\right)}_{0}=x\right],
\end{equation*}
where $\phi\left(\cdot;X^{\left(u-1\right)}_{t_k},\theta_{\text{p}}\right)$ is the Gaussian density function with mean $c\left(X^{\left(u-1\right)}_{t_k};\theta_{\text{p}}\right)$ and variance $d^2\left(X^{\left(u-1\right)}_{t_k};\theta_{\text{p}}\right)$. Therefore, based on the realized episode, the estimated gradient of the time-$0$ value function is given by
\begin{equation*}
\widehat{\nabla_{\theta_{\text{p}}}V^{\left(u-1\right)}\left(0,x;\theta_{\text{p}}^{\left(u-1\right)}\right)}=\sum_{k=0}^{\tilde{n}-1}\left(\sum_{l=k}^{\tilde{n}-1}r_{t_{l+1}}^{\left(u-1\right)}\right)\nabla_{\theta_{\text{p}}}\ln\phi\left(h_{t_k}^{\left(u-1\right)};x_{t_k}^{\left(u-1\right)},\theta_{\text{p}}^{\left(u-1\right)}\right).
\end{equation*}



Notice that, thanks to the Policy Gradient Theorem, the gradient of the surrogate performance measure does not depend on the gradient of the reward function, and hence the reward function could be discrete or non-differentiable while the estimated gradient of the surrogate performance measure only needs the numerical reward values. However, in the DH approach of \cite{Buhler_2019}, the gradient of the surrogate performance measure therein does depend on the gradient of the terminal loss function, and thus that approach implicitly requires the differentiability of the hedging portfolio value while the estimated gradient of the surrogate performance requires its numerical gradient values. See Appendix \ref{app:dh_agent_1} for more details.


To reduce the variance of estimated gradient above, \cite{Williams_1992} suggested to introduce an unbiased baseline in this gradient, where a natural choice is the value function:
\begin{equation*}
\nabla_{\theta_{\text{p}}}V^{\left(u-1\right)}\left(0,x;\theta_{\text{p}}\right)=\mathbb{E}\left[\sum_{k=0}^{\tilde{n}-1}\left(\sum_{l=k}^{\tilde{n}-1}R_{t_{l+1}}^{\left(u-1\right)}-V\left(t_k,X^{\left(u-1\right)}_{t_k};\theta_{\text{p}}\right)\right)\nabla_{\theta_{\text{p}}}\ln\phi\left(H^{\left(u-1\right)}_{t_k};X^{\left(u-1\right)}_{t_k},\theta_{\text{p}}\right)\Big\vert X^{\left(u-1\right)}_{0}=x\right];
\end{equation*}
see also Weaver and Tao (2001). Herein, at any time $t_k$, for $k=0,1,\dots,\tilde{n}-1$, $A^{\left(u-1\right)}_{t_k}=\sum_{l=k}^{\tilde{n}-1}R_{t_{l+1}}^{\left(u-1\right)}-V\left(t_k,X^{\left(u-1\right)}_{t_k};\theta_{\text{p}}\right)$ is called an {\it advantage}. Since the true value function is unknown to the RL agent, it is approximated by $\hat{V}\left(t_k,X^{\left(u-1\right)}_{t_k};\theta_{\text{v}}^{\left(u-1\right)}\right)=\mathcal{N}_{\text{v}}\left(X^{\left(u-1\right)}_{t_k};\theta_{\text{v}}^{\left(u-1\right)}\right)$, defined in \eqref{eq:value_function_network}, and in which the ANN weights are evaluated at $\theta_{\text{v}}=\theta_{\text{v}}^{\left(u-1\right)}$ as the gradient of the time-$0$ value function is independent of the ANN weights $\theta_{\text{v}}$; hence, the estimated advantage is given by $\hat{A}^{\left(u-1\right)}_{t_k}=\sum_{l=k}^{\tilde{n}-1}R_{t_{l+1}}^{\left(u-1\right)}-\hat{V}\left(t_k,X^{\left(u-1\right)}_{t_k};\theta_{\text{v}}^{\left(u-1\right)}\right)$.

Due to the value function approximation in the baseline, REINFORCE includes a second component in the surrogate performance measure, which aims to minimize the loss between the sum of reward signals and the approximated value function by the ANN. Therefore, the surrogate performance measure is given by:
\begin{equation*}
\mathcal{J}^{\left(u-1\right)}\left(\theta\right)=V^{\left(u-1\right)}\left(0,x;\theta_{\text{p}}\right)-\mathbb{E}\left[\sum_{k=0}^{\tilde{n}-1}\left(\hat{A}^{\left(u-1\right)}_{\theta^{\left(u-1\right)}_{\text{p}},t_k}+\hat{V}\left(t_k,X^{\left(u-1\right)}_{t_k};\theta_{\text{v}}^{\left(u-1\right)}\right)-\hat{V}\left(t_k,X^{\left(u-1\right)}_{t_k};\theta_{\text{v}}\right)\right)^2\Big\vert X^{\left(u-1\right)}_{0}=x\right],
\end{equation*}
where the estimated advantaged $\hat{A}^{\left(u-1\right)}_{\theta^{\left(u-1\right)}_{\text{p}},t_k}$ is evaluated at $\theta_{\text{p}}=\theta^{\left(u-1\right)}_{\text{p}}$.

Hence, at each update step $u=1,2,\dots$, based on the ANN weights $\theta^{\left(u-1\right)}$, and thus the policy $\pi\left(\cdot;\theta_{\text{p}}^{\left(u-1\right)}\right)$, the estimated gradient of the surrogate performance measure is given by
\begin{align*}
\widehat{\nabla_{\theta}\mathcal{J}^{\left(u-1\right)}\left(\theta^{\left(u-1\right)}\right)}=&\;\sum_{k=0}^{\tilde{n}-1}\left(\sum_{l=k}^{\tilde{n}-1}r_{t_{l+1}}^{\left(u-1\right)}-\hat{V}\left(t_k,x_{t_k}^{\left(u-1\right)};\theta_{\text{v}}^{\left(u-1\right)}\right)\right)\nabla_{\theta_{\text{p}}}\ln\phi\left(h_{t_k}^{\left(u-1\right)};x_{t_k}^{\left(u-1\right)},\theta_{\text{p}}^{\left(u-1\right)}\right)\\&\;+\sum_{k=0}^{\tilde{n}-1}\left(\sum_{l=k}^{\tilde{n}-1}r_{t_{l+1}}^{\left(u-1\right)}-\hat{V}\left(t_k,x_{t_k}^{\left(u-1\right)};\theta_{\text{v}}^{\left(u-1\right)}\right)\right)\nabla_{\theta_{\text{v}}}\hat{V}\left(t_k,x_{t_k}^{\left(u-1\right)};\theta_{\text{v}}^{\left(u-1\right)}\right)\\=&\;\sum_{k=0}^{\tilde{n}-1}\hat{a}^{\left(u-1\right)}_{t_k}\left(\nabla_{\theta_{\text{p}}}\ln\phi\left(h_{t_k}^{\left(u-1\right)};x_{t_k}^{\left(u-1\right)},\theta_{\text{p}}^{\left(u-1\right)}\right)+\nabla_{\theta_{\text{v}}}\hat{V}\left(t_k,x_{t_k}^{\left(u-1\right)};\theta_{\text{v}}^{\left(u-1\right)}\right)\right),
\end{align*}
where $\hat{a}_{t_k}^{\left(u-1\right)}=\sum_{l=k}^{\tilde{n}-1}r_{t_{l+1}}^{\left(u-1\right)}-\hat{V}\left(t_k,x_{t_k}^{\left(u-1\right)};\theta_{\text{v}}^{\left(u-1\right)}\right)$, for $k=0,1,\dots,\tilde{n}-1$, is the realized estimated advantage.

\section{Deep Hedging Training}\label{app:DHT}
The state vector observed by the DH agent is the same as that by the RL agent in Equation \eqref{eq:state_vector}. Table \ref{table:DH_hyper} summarizes the hyperparameters of DH agent training, while Table \ref{table:DH_net_hyper} outlines the hyperparameters of the ANN architecture of DH agent; see Appendix \ref{app:dh_agent_1}.

\begin{table}[!htb]
\centering
\begin{subtable}{.5\linewidth}
\centering
\caption{Hyperparameters of Deep Hedging Training}
\begin{tabular}{@{}lc@{}}
\toprule
Parameter & Value 
\\ \midrule
Number of updates $\hat{M}$ &   $10^8$ \\
Learning rate $\hat{\alpha}$ & $0.0001$\\
$\text{Optimizer}$ & Adam \\
\bottomrule   
\end{tabular}
\label{table:DH_hyper}
\end{subtable}%
\begin{subtable}{.5\linewidth}
\centering
\caption{Hyperparameters for Neural Network}
\begin{tabular}{@{}lc@{}}
\toprule
Parameter & Value(s) 
\\ \midrule
Number of layers  &   $6$\\
Dimension of hidden layers &   $[32,64,128,64,32]$     \\
Activation function & ReLU\\
\bottomrule   
\end{tabular}
\label{table:DH_net_hyper}
\end{subtable} 
\caption{The hyperparameters of deep hedging training and the neural network}
\label{table:param_DH}
\end{table}

\end{appendices}
\end{document}